\documentclass[11pt]{article}
\usepackage{graphicx,ifthen,color,algorithm,palatino}
\usepackage{amsmath,amsthm,amssymb,amsfonts,verbatim}
\usepackage{mathrsfs,accents,setspace}
\usepackage{subcaption}
\usepackage{multirow}
\usepackage{float}
\usepackage{caption}
\newcommand{\AL}[1]{\bA_{#1}}
\newcommand{\ALT}[1]{\bA_{#1}^T}
\newcommand{\AU}[1]{\bA^{#1}}
\newcommand{\AUT}[1]{\bA^{#1\,T}}
\newtheorem{result}{\textbf{Result}}
\def\bib{\vskip12pt\par\noindent\hangindent=1 true cm\hangafter=1}
\def\bigX{{\LARGE\mbox{$\times$}}}
\def\myand{\&\ }
\def\bbeta{\boldsymbol{\beta}}
\def\bmu{\boldsymbol{\mu}}
\def\bLambda{\boldsymbol{\Lambda}}
\def\bomega{\boldsymbol{\omega}}	\def\bOmega{\boldsymbol{\Omega}}
\def\bSigma{\boldsymbol{\Sigma}}
\def\bzeta{\boldsymbol{\zeta}}
\def\btheta{\boldsymbol{\theta}}		\def\bTheta{\boldsymbol{\Theta}}
\def\by{\boldsymbol{y}}
\def\bu{\boldsymbol{u}}
\def\ba{\boldsymbol{a}}			\def\bA{\boldsymbol{A}}
\def\bb{\boldsymbol{b}}			\def\bB{\boldsymbol{B}}
\def\bC{\boldsymbol{C}}
\def\bG{\boldsymbol{G}}
\def\bh{\boldsymbol{h}}			\def\bH{\boldsymbol{H}}
\def\bI{\boldsymbol{I}}
\def\bM{\boldsymbol{M}}
\def\bO{\boldsymbol{O}}
\def\bQ{\boldsymbol{Q}}
\def\bR{\boldsymbol{R}}
\def\bx{\boldsymbol{x}}			\def\bX{\boldsymbol{X}}
\def\bZ{\boldsymbol{Z}}
\def\bzero{\boldsymbol{0}}
\def\pDens{\mathfrak{p}}			\def\qDens{\mathfrak{q}}
\def\Gfull{G_{\mbox{\tiny full}}}		\def\Gdiag{G_{\mbox{\tiny diag}}}
\def\tr{\mbox{tr}}
\def\diag{\mbox{diag}}
\def\stackdum{\mathop{\mbox{\rm stack}}}		
\def\stack#1{\stackdum_{#1}}
\def\blockdiagdum{\mathop{\mbox{\rm blockdiag}}}
\def\blockdiag#1{\blockdiagdum_{#1}}
\def\simind{\stackrel{{\tiny \mbox{ind.}}}{\sim}}
\def\qLone{q_1}				
\def\qLtwo{q_2}
\def\sigsq{\sigma^2}
\def\asigsq{a_{\sigsq}}
\def\nusigsq{\nu_{\sigsq}}
\def\ssigsq{s_{\sigsq}}
\def\tausq{\tau^2}
\def\atausq{a_{\tausq}}
\def\stausq{s_{\tausq}}
\def\bZLone{\bZ^{\mbox{\tiny\rm L1}}}
\def\bZLtwo{\bZ^{\mbox{\tiny\rm L2}}}
\def\buLone{\bu^{\mbox{\tiny\rm L1}}}
\def\buLtwo{\bu^{\mbox{\tiny\rm L2}}}
\def\bSigmaLone{\bSigma^{\mbox{\tiny\rm L1}}}
\def\bSigmaLtwo{\bSigma^{\mbox{\tiny\rm L2}}}
\def\bSigmaLoneMinusOne{(\bSigma^{\mbox{\tiny\rm L1}})^{-1}}
\def\bSigmaLtwoMinusOne{(\bSigma^{\mbox{\tiny\rm L2}})^{-1}}
\def\ASigmaLone{\bA_{\mbox{\tiny$\bSigmaLone$}}}
\def\ASigmaLtwo{\bA_{\mbox{\tiny$\bSigmaLtwo$}}}
\def\AG{\bA_{\mbox{\tiny{$\bG$}}}}
\def\ASigma{\bA_{\mbox{\tiny{$\bSigma$}}}}
\def\nuG{\nu_{\mbox{\tiny{$\bG$}}}}
\def\nuSigma{\nu_{\mbox{\tiny{$\bSigma$}}}}
\def\nuSigmaLone{\nu_{\mbox{\tiny{$\bSigmaLone$}}}}
\def\nuSigmaLtwo{\nu_{\mbox{\tiny{$\bSigmaLtwo$}}}}
\def\sGOne{s_{\mbox{\tiny{$\bG,1$}}}}
\def\sSigmaOne{s_{\mbox{\tiny{$\bSigma,1$}}}}
\def\sGq{s_{\mbox{\tiny{$\bG,q$}}}}
\def\sSigmaq{s_{\mbox{\tiny{$\bSigma,q$}}}}
\def\sSigmaOneLone{s_{\mbox{\tiny{$\bSigmaLone,1$}}}}
\def\sSigmaqLone{s_{\mbox{\tiny{$\bSigmaLone,\qLone$}}}}
\def\sSigmaOneLtwo{s_{\mbox{\tiny{$\bSigmaLtwo,1$}}}}
\def\sSigmaqLtwo{s_{\mbox{\tiny{$\bSigmaLtwo,\qLtwo$}}}}
\def\raneff{\text{R}}
\def\additional{\text{A}}
\def\subjtosel{\text{S}}
\def\muq#1{\mu_{\qDens(#1)}}
\def\bmuq#1{\bmu_{\qDens(#1)}}
\def\bSigmaq#1{\bSigma_{\qDens(#1)}}

\def\xiq#1{\xi_{\qDens(#1)}}
\def\lambdaq#1{\lambda_{\qDens(#1)}}
\def\bMq#1{\bM_{\qDens(#1)}}

\def\bLambdaq#1{\bLambda_{\qDens(#1)}}
\def\SscAlgStreamlinedMFVBTwoLevel{\mathcal{S}_1}
\def\SolveTwoLevelSparseMatrix{\textsc{\footnotesize SolveTwoLevelSparseMatrix}}
\def\SolveTwoLevelSparseLeastSquares{\textsc{\footnotesize SolveTwoLevelSparseLeastSquares}}
\def\SscAlgStreamlinedMFVBThreeLevel{\mathcal{S}_2}
\def\SolveThreeLevelSparseMatrix{\textsc{\footnotesize SolveThreeLevelSparseMatrix}}
\def\SolveThreeLevelSparseLeastSquares{\textsc{\footnotesize SolveThreeLevelSparseLeastSquares}}
\begin{document}

\thispagestyle{empty}

\centerline{\Large\bf Sparse Linear Mixed Model Selection via}
\vskip2mm
\centerline{\Large\bf Streamlined Variational Bayes}
\vskip5mm
\centerline{\normalsize\sc By Emanuele Degani$\null^{\dag,\natural}$, Luca Maestrini$\null^\ddag$,}
\centerline{\normalsize\sc Dorota Toczydłowska$\null^\sharp$ \myand Matt P. Wand$\null^\sharp$}
\vskip5mm
\centerline{\textit{Università degli Studi di Padova$\null^\dag$, Banca d'Italia -- Eurosystem$\null^\natural$,}}
\vskip1mm
\centerline{\textit{The Australian National University$\null^\ddag$ and University of Technology Sydney$\null^\sharp$}}
\vskip6mm
\centerline{\today}
\vskip6mm
\centerline{\large\bf Abstract}
\vskip2mm

Linear mixed models are a versatile statistical tool to study data by accounting for fixed effects and random effects from multiple sources of variability. In many situations, a large number of candidate fixed effects is available and it is of interest to select a parsimonious subset of those being effectively relevant for predicting the response variable. 
Variational approximations facilitate fast approximate Bayesian inference for the parameters of a variety of statistical models, including linear mixed models. However, for models having a high number of fixed or random effects, simple application of standard variational inference principles does not lead to fast approximate inference algorithms, due to the size of model design matrices and inefficient treatment of sparse matrix problems arising from the required approximating density parameters updates.
We illustrate how recently developed streamlined variational inference procedures can be generalized to make fast and accurate inference for the parameters of linear mixed models with nested random effects and global-local priors for Bayesian fixed effects selection. 
Our variational inference algorithms achieve convergence to the same optima of their standard implementations, although with significantly lower computational effort, memory usage and time, especially for large numbers of random effects.
Using simulated and real data examples, we assess the quality of automated procedures for fixed effects selection that are free from hyperparameters tuning and only rely upon variational posterior approximations. Moreover, we show high accuracy of variational approximations against model fitting via Markov Chain Monte Carlo sampling.

\vskip2mm
\noindent

\textit{Keywords:} mean field variational Bayes, multilevel models, longitudinal data analysis, fixed effects selection, global-local shrinkage priors.

\section{Introduction\label{sec:Introduction}}

A variety of statistical models can be formulated as linear regression models incorporating both fixed and random effects in the linear predictor. The former are effects associated with the entire population or repeatable levels of experimental factors; the latter arise from individual experimental units drawn at random. In the statistical literature, models admitting both fixed and random effects are known as mixed-effects models (Pinheiro \myand Bates, 2006). 
These models are employed in an assortment of regression-type studies, including the analysis of classical longitudinal data (e.g. Fitzmaurice \emph{et al.}, 2008), repeated measurements (e.g. Vonesh \myand Chinchilli, 1997), blocked designs (e.g. Lindner \myand Rodger, 1997), multilevel data (e.g. Goldstein, 2010), as well as semi-parametric regression models (e.g. Ruppert \emph{et al.}, 2003) such as those including spatial or spline-type components.

The focus of this work is on Bayesian fitting of linear mixed-effects models with nested random effects structures, which are commonly used for the analysis of longitudinal, multilevel and panel data (e.g. Verbeke \myand Molenberghs, 2000; Baltagi, 2013), or small area estimation (e.g. Rao \myand Molina, 2015). These data are typically collected from experimental units that can be grouped into different levels of nesting, and the interest is in modeling within-group correlations. In areas of application such as genome-wide association studies (e.g. Korte \emph{et al.}, 2012; Sikorska \emph{et al.}, 2013; Li \emph{et al.}, 2015) and medical research (e.g. Brown \myand Prescott, 2014), datasets typically possess a large number of group-invariant predictors of which only a few are effectively relevant. A common misleading strategy is that of including all the predictors as fixed effects in the model specification. This may compromise the parsimony of the model specification and validity of inferential conclusions, especially in sparse covariate settings. Therefore, a proper fixed effects selection procedure is recommended to identify the effectively relevant effects. 

Although many frequentist procedures have been developed to tackle this problem (e.g. Schelldorfer \emph{et al.}, 2011; Fan \myand Li, 2012; Groll \myand Tutz, 2012; Hui \emph{et al.}, 2017; Li \emph{et al.}, 2018), little exists in the Bayesian literature. Bayesian approaches are mostly focused on random effects selection induced by the decomposition of their covariance matrix (e.g. Chen \myand Dunson, 2003; Yang, 2013), or joint fixed and random effects selection (e.g. Kinney \myand Dunson, 2007; Yang \emph{et al.}, 2020). 

The current work focuses on fixed effects selection procedures from a Bayesian perspective. This may be advantageous over frequentist approaches especially in high-dimensional settings when likelihood-based inference is computationally intractable and allow for prior knowledge about the parameters to be incorporated in the model specification. Markov Chain Monte Carlo (MCMC) sampling still represents the reference toolkit for \emph{exact} Bayesian inference and all the aforementioned references on Bayesian approaches for effects selection perform model fitting via MCMC. Although not accounting for selection procedures, the \textsf{brms} package (B{\"u}rkner, 2017) allows to fit Bayesian multilevel models in \textsf{R} (R Core Team, 2022) making use of the popular probabilistic programming language \textsf{Stan} (Carpenter \emph{et al.}, 2017). Using this package, practitioners only have to specify the appropriate model structure, the model is automatically fitted and convergence can be assessed. However, automatic sampling procedures usually generate higher computational times and, in general, proper MCMC procedures necessitate convergence assessment for all the model parameters, which may arise problematics such as poor mixing connected to the model parameterization.
These and other drawbacks have supported the development of variational approximations for linear mixed models to improve the speed of convergence, at the cost of employing an approximation to the true posterior distribution for carrying out inferential conclusions. 

Wang \myand Wand (2011) provide some insights on how to implement variational approximations for approximate Bayesian inference in hierarchical models through \textsf{Infer.NET} (Minka \emph{et al.}, 2018). Although this computational framework is suitable for longitudinal and multilevel models, its computational advantage quickly decreases for high numbers of groups and sub-groups, limiting the usefulness of variational inference. Algorithm 3 of Ormerod \myand Wand (2010), and Algorithms 3 and 5 of Luts \emph{et al.} (2014) allow to implement variational inference for fitting longitudinal and multilevel data; however, they do not perform efficiently for large dimensions, as they include na{\"i}ve updates based on inefficient matrix inversions. 

Lee \myand Wand (2016) developed a streamlined updating scheme for variational inference making efficient use of sub-matrix inversion operations whose number is linear in the size of groups at each level. The streamlined scheme represents an improvement of two orders of magnitude over na{\"i}ve implementations of variational approximations. These results have also been extended to the class of generalized linear mixed-effects models and applied, for instance, to models for multiple longitudinal markers (Hughes \emph{et al.}, 2021).
Nolan \emph{et al.} (2020) took advantage of the sparse matrix results developed in Nolan \myand Wand (2020) for deriving streamlined algorithms and performing efficient Bayesian variational approximations for linear mixed models with two and three-level nested random effects structures. This framework, named \emph{streamlined variational inference}, allows to dramatically reduce computational times when compared to na{\"i}ve implementations of variational inference, although achieving the same approximation. Furthermore, streamlined variational inference allows to efficiently store the matrices needed to perform algorithm updates, hence providing significant memory savings. 

These developments have recently inspired streamlined algorithms for linear mixed-effects models with crossed random effects (Menictas \emph{et al.}, 2022) and group-specific curves (Menictas \emph{et al.}, 2021). 
Many extensions can be envisaged and are motivated by the high demand for fast and accurate processing methods for big amounts of data from clinical studies, psychological experiments or surveys in social sciences.
The current streamlined variational algorithms have been developed and tested using generic uninformative priors over the fixed effects vector. In this work, we introduce streamlined variational inference for models with priors inducing Bayesian posterior shrinkage and study an efficient selection procedure for fixed effects. 

\subsection{Contribution and Article Organization\label{subsec:InnovationsAndOrganization}}

To the best of our knowledge, scalable variational approximation methods for fixed effects selection in linear mixed models are rarely present in literature. Armagan \myand Dunson (2011) propose a sparse variational Bayes analysis of linear mixed models which focuses on random effects shrinkage via decomposition of the random effects vector covariance matrix. A more recent contribution is Tung \emph{et al.} (2019), where the suggested approach performs simultaneous fixed-effect selection and parameter estimation via variational Bayes and Bayesian adaptive lasso. However, the approach is limited to high-dimensional two-level generalized linear mixed models and does not account for any streamlined updating improvements. 

The current work extends the results and algorithms of Nolan \emph{et al.} (2020) by developing streamlined Bayesian variational approximations for multilevel linear mixed models with two or three-level random effects where a subset of fixed effects is subject to selection. The selection is performed by first placing global-local priors over the fixed effects being subject to selection, which ensures good shrinkage properties towards the origin for irrelevant fixed effects marginal posteriors, and then identifying those being relevant via an automated selection procedure free from hyperparameters tuning.

The article is organized as follows. Section \ref{sec:LinearMixedModels} provides an overview of linear mixed models from a Bayesian perspective, with a specific focus on two- and three-level random effects specifications. Section \ref{sec:VariationalBayesInference} explains variational approximations for this class of models, with a particular focus on issues arising from na{\"i}ve implementations of variational algorithms and the benefits of streamlined variational inference. Section \ref{sec:ApproximateVarSelShrinkagePriors} discusses automated approximate Bayesian methods for performing variable selection when global-local shrinkage priors are introduced in a linear regression model. Section \ref{sec:VariationalInferenceLMMwithGLPriors} connects the previous two sections and provides streamlined variational Bayes algorithms for mixed-effects models with global-local priors placed over a subset of fixed effects which are subject to selection. Section \ref{sec:Assessments} provides a detailed simulation study that demonstrates the advantages provided by the methodology proposed in this work. A real data illustration is included in Section \ref{sec:Application}. The article is supported by additional supplementary material containing details on distributions, complementary algorithms and derivations.

\subsection{Notation}

The notation of this article matches the one of Nolan \emph{et al.} (2020). Here we briefly recall some essential notation. Data vectors and design matrices can be combined using \emph{stack} and \emph{blockdiag} operators defined as 
$$
\stack{1\le i \le d}(\bM_i)\equiv\left[\begin{array}{c} \bM_1\\ \vdots \\\bM_d \end{array} \right] \quad\mbox{and}\quad \blockdiag{1\le i \le d}(\bM_i)\equiv \left[ \begin{array}{cccc} \bM_1 & \bO   & \cdots & \bO \\ \bO   & \bM_2 & \cdots & \bO \\ \vdots& \vdots& \ddots & \vdots \\ \bO   & \bO& \cdots &\bM_d \end{array} \right],
$$
for a sequence of matrices $\bM_1,\ldots,\bM_d$. The stack operator requires $\bM_i$ matrices with the same number of columns. If $\bM$ is a square matrix, $\text{diagonal}(\bM)$ is the main diagonal of $\bM$ and $\tr(\bM)$ is the trace of $\bM$. For a vector $\boldsymbol{v}$ of length $d$, $\diag(\boldsymbol{v})$ produces a diagonal matrix having the elements of $\boldsymbol{v}$ as main diagonal. Unless specified otherwise, given two vectors $\boldsymbol{v}_1$ and $\boldsymbol{v}_2$ of same length, $\boldsymbol{v}_1/\boldsymbol{v}_2$ indicates their element-wise division. Element-wise addition, subtraction and multiplication are similarly defined.

We use $\pDens$ and $\qDens$ for density functions. In particular, $\qDens$ is used for densities arising from variational approximations. The letters $p$ and $q$ are used for the dimensions of model vectors and matrices. 

We use $\muq{\theta} \equiv E_\qDens(\theta)$ for a generic parameter $\theta$, $\bmuq{\btheta} \equiv E_\qDens(\btheta)$ for a generic vector of parameters $\btheta$ and $\bMq{\bTheta} \equiv E_\qDens(\bTheta)$ for a generic matrix of parameters $\bTheta$, with $E_\qDens(\cdot)$ denoting the expectation with respect to the probability density function $\qDens$.

\section{Linear Mixed Models\label{sec:LinearMixedModels}}

This article treats linear mixed models with Gaussian responses and homoskedastic independent errors from a Bayesian inference perspective. A general formulation for these models is
\begin{equation}
\label{eq:gauss_heter_lmm}
\begin{array}{c}
\by \vert \bbeta, \bu, \sigsq \sim \text{N}(\bX \bbeta + \bZ \bu, \sigsq \bI),\quad \bu \vert \bG \sim \text{N}(\bzero, \bG), \\[1.5ex]
\bbeta \sim \text{N}(\bmu_{\bbeta}, \bSigma_{\bbeta}),\quad \sigsq\vert \asigsq \sim \text{Inverse-}\chi^2(\nusigsq, 1/\asigsq), \quad \asigsq \sim \text{Inverse-}\chi^2(1, 1/(\nusigsq \ssigsq^2)), \\[1.5ex]
 \bG\vert \AG \sim \pDens(\bG \vert \AG), \quad \AG \sim \pDens(\AG),
\end{array}
\end{equation}
where $\by$ is a vector of observed data, $\bbeta$ and $\bu$ are respectively the vectors of fixed and random effects, $\bX$ and $\bZ$ are the associated fixed and random effects design matrices, $\sigsq$ is the variance of the unit-specific error term and $\bG$ is the random effects covariance matrix. 

A very general prior specification for the parameters of model \eqref{eq:gauss_heter_lmm} is considered. The fixed effects vector $\bbeta$ has a multivariate Normal prior with hyperparameters $\bmu_{\bbeta}$ and $\bSigma_{\bbeta}$. Following Gelman (2006), the hierarchical prior specification on $\sigsq$ generates a Half-$t$ distribution on $\sigma$ with $\nusigsq$ degrees of freedom and scale parameter $\ssigsq$, where larger values of $\ssigsq$ correspond to weaker informativity. A similar hierarchical prior is imposed on the random effects vector covariance matrix $\bG$: if, for instance, $\pDens(\bG \vert \AG)$ is an  $\text{Inverse-G-Wishart}(\Gfull,\nuG+2q-2,\AG^{-1})$ density function and $\pDens(\AG)$ is an $\text{Inverse-G-Wishart}(\Gdiag,1,\bLambda_{\AG})$ density function with $\bLambda_{\AG} \equiv \lbrace \nuG \: \diag(\sGOne^2, \ldots,\sGq^2) \rbrace^{-1}$, then according to Huang \myand Wand (2013) such a prior imposition may induce arbitrarily noninformative priors on the standard deviation parameters for large values of $\sGOne,\ldots,\sGq$ and a Uniform$(-1,1)$ distribution over the correlation parameters. The notations $\Gfull$ and $\Gdiag$ symbolize fully connected and disconnected graphs arising from the structure of $\bG^{-1}$ and $\AG^{-1}$, as explained in Maestrini \myand Wand (2021).

The structures of $\bZ$, $\bu$ and $\bG$ embed a rich ensemble of mixed model specifications (Zhao \emph{et al.}, 2006). Hereafter, we will focus on multilevel models having two-level and three-level random effects specifications.

\subsection{Two-Level Linear Mixed Models\label{subsec:TwoLevelLMM}}

Multilevel models with two-level random effects arise from applications where observations from different units belonging to separate groups are available, and the interest is in capturing the within-group variability. Let $m$ be the number of groups, each composed by $o_i$ units, $1 \le i \le m$. A two-level linear mixed model can be expressed in terms of the observations from the $i$th group as follows:
\begin{equation}
\label{eq:gauss_heter_lmm_2level}
\begin{array}{c}
\by_i \vert \bbeta, \bu_i, \sigsq \simind \text{N}(\bX_i \bbeta + \bZ_i \bu_i, \sigsq \bI), \quad \bu_i \vert \bSigma \simind \text{N}(\bzero, \bSigma), \quad 1 \le i \le m, \\[1.5ex]
\bbeta \sim \text{N}(\bmu_{\bbeta}, \bSigma_{\bbeta}), \quad \sigsq\vert \asigsq \sim \text{Inverse-}\chi^2(\nusigsq, 1/\asigsq), \quad \asigsq \sim \text{Inverse-}\chi^2(1, 1/(\nusigsq \ssigsq^2)), \\[1.5ex]
\bSigma\vert \ASigma \sim \text{Inverse-G-Wishart}(\Gfull,\nuSigma+2q-2,\ASigma^{-1}), \\[1.5ex]
\ASigma \sim \text{Inverse-G-Wishart}(\Gdiag,1,\bLambda_{\ASigma}), \quad \bLambda_{\ASigma} \equiv \lbrace \nuSigma\: \diag(\sSigmaOne^2, \ldots, \sSigmaq^2) \rbrace^{-1}.
\end{array}
\end{equation}
Here $\bSigma$ is the covariance matrix for the group-specific random effects vector $\bu_i$ of length $q$. 
Notice model \eqref{eq:gauss_heter_lmm_2level} is a particular case of model \eqref{eq:gauss_heter_lmm}, with
\begin{equation}
\label{eq:2level-to-general}
\begin{array}{c}
\by = \displaystyle{\stack{1 \le i \le m} (\by_i)}, \quad \bX = \displaystyle{\stack{1 \le i \le m} (\bX_i)}, \quad \bZ = \displaystyle{\blockdiag{1 \le i \le m} (\bZ_i)}, \\[1.5ex]
\bu = \displaystyle{\stack{1 \le i \le m} (\bu_i)}, \quad \bG = \bI_m \otimes \bSigma, \quad \AG = \bI_m \otimes \ASigma.
\end{array}
\end{equation}
The structure of $\bZ$ is such that $\bZ \bu = \sum_{i=1}^m \bZ_i \bu_i$ and notice that as $m$ increases, $\bZ$ becomes sparser with only the $(100/m)\%$ of its cells being non-zero.

\subsection{Three-Level Linear Mixed Models\label{subsec:ThreeLevelLMM}}

Multilevel models with three-level random effects extend two-level models by adding a further hierarchy level.
Such structures are employed when there is interest in capturing both the variability within groups and that within their subgroups.  

Let $m$ denote the number of level 1 ($\mbox{L1}$) groups, $n_i$ be the number of level 2 ($\mbox{L2}$) subgroups belonging to the $i$th group, $1 \le i \le m$, and $o_{ij}$ be the number of units belonging to the $j$th subgroup, $1 \le j \le n_i$, of the $i$th group. A three-level linear mixed model can be defined in terms of the observations from the $j$th subgroup belonging to the $i$th group as follows:
\begin{equation}
\label{eq:gauss_heter_lmm_3level}
\begin{array}{c}
\by_{ij} \vert \bbeta, \buLone_i, \buLtwo_{ij}, \sigsq \simind \text{N}(\bX_{ij} \bbeta + \bZLone_{ij} \buLone_i + \bZLtwo_{ij} \buLtwo_{ij}, \sigsq \bI), \\[1.5ex]
\left[\begin{array}{c}\buLone_i \\[1ex] \buLtwo_{ij} \end{array}\right] \Big\vert\: \bSigmaLone, \bSigmaLtwo \simind \text{N}\left(\left[\begin{array}{c}\bzero \\[1ex] \bzero \end{array}\right], \left[\begin{array}{cc} \bSigmaLone & \bO \\[1ex] \bO & \bSigmaLtwo \end{array} \right]\right), \quad 1 \le i \le m, \: 1 \le j \le n_i, \\[3ex]
\bbeta \sim \text{N}(\bmu_{\bbeta}, \bSigma_{\bbeta}),\quad \sigsq\vert \asigsq \sim \text{Inverse-}\chi^2(\nusigsq, 1/\asigsq), \quad \asigsq \sim \text{Inverse-}\chi^2(1, 1/(\nusigsq \ssigsq^2)),\\[1.5ex]
\bSigmaLone\vert \ASigmaLone \sim \text{Inverse-G-Wishart}(\Gfull,\nuSigmaLone+2\qLone-2,\ASigmaLone^{-1}),\\[1.5ex]
\ASigmaLone \sim \text{Inverse-G-Wishart}(\Gdiag,1,\bLambda_{\ASigmaLone}), \quad \bLambda_{\ASigmaLone} \equiv \lbrace \nuSigmaLone \:\diag(\sSigmaOneLone^2,\ldots,\sSigmaqLone^2) \rbrace^{-1},\\[1.5ex]
\bSigmaLtwo\vert \ASigmaLtwo \sim \text{Inverse-G-Wishart}(\Gfull,\nuSigmaLtwo+2\qLtwo-2,\ASigmaLtwo^{-1}),\\[1.5ex]
\ASigmaLtwo \sim \text{Inverse-G-Wishart}(\Gdiag,1,\bLambda_{\ASigmaLtwo}), \quad \bLambda_{\ASigmaLtwo} \equiv \lbrace \nuSigmaLtwo \:\diag(\sSigmaOneLtwo^2,\ldots,\sSigmaqLtwo^2) \rbrace^{-1}.
\end{array}
\end{equation}
Here $\bSigmaLone$ is the covariance matrix for the group-specific random effects vector $\buLone_i$ of length $\qLone$ and $\bSigmaLtwo$ is that for the subgroup-specific random effects vector $\buLtwo_{ij}$ having length $\qLtwo$. 
Notice model \eqref{eq:gauss_heter_lmm_3level} is a particular case of model \eqref{eq:gauss_heter_lmm}, with
\begin{equation}
\label{eq:3level-to-general}
\begin{array}{c}
\by = \displaystyle{\stack{1 \le i \le m} \left(\stack{1 \le j \le n_i} (\by_{ij})\right)}, \quad \bX = \displaystyle{\stack{1 \le i \le m} \left(\stack{1 \le j \le n_i} (\bX_{ij})\right)}, \\[2.5ex]
\bZ = \displaystyle{\blockdiag{1 \le i \le m} \left(\left[\begin{array}{c|c} \displaystyle{\stack{1 \le j \le n_i} (\bZLone_{ij})} & \displaystyle{\blockdiag{1 \le j \le n_i} (\bZLtwo_{ij})} \end{array}\right]\right)},\\[2.5ex]
\bu = \displaystyle{\stack{1 \le i \le m} \left(\left[\begin{array}{c|c} (\buLone_i)^T & \left(\displaystyle{\stack{1 \le j \le n_i} (\buLtwo_{ij})}\right)^T \end{array}\right]^T \right) }, \\[2.5ex]
\bG = \displaystyle{\blockdiag{1 \le i \le m}} \left(\left[\begin{array}{cc} \bSigmaLone & \bO \\[1ex] \bO & \bI_{n_i} \otimes \bSigmaLtwo \end{array}\right]\right), \quad \AG =  \displaystyle{\blockdiag{1 \le i \le m}} \left(\left[\begin{array}{cc} \ASigmaLone & \bO \\[1ex] \bO & \bI_{n_i} \otimes \ASigmaLtwo \end{array}\right]\right).
\end{array}
\end{equation}
The structure of $\bZ$ is more involved than the one of two-level random effects models and is such that $\bZ \bu = \sum_{i=1}^m \sum_{j=1}^{n_i}  (\bZLone_{ij} \buLone_i + \bZLtwo_{ij} \buLtwo_{ij})$. As $m$ and the $n_i$'s increase, $\bZ$ becomes sparser with only the $\{(\qLone + \qLtwo)/(\qLone m + \qLtwo \sum_{i=1}^m n_i) \times 100\}\%$ of its cells being non-zero.
Notice that in the particular case where $n_i = 1$ for all $1 \le i \le m$ and $\qLone = \qLtwo = q$, the three-level specification corresponds to the two-level one.

\section{Variational Bayesian Inference\label{sec:VariationalBayesInference}}

In this section, we provide a brief overview of mean field variational Bayes, shortly MFVB (see e.g. Bishop, 2006 or Blei \emph{et al.}, 2017), which is the variational approximation technique we employ in this work for fitting two- and three-level linear mixed models. 

\subsection{Overview\label{subsec:Overview}}

Consider a generic model with an observed data vector $\by$ and a parameter vector $\btheta \in \bTheta$. Let $\qDens(\btheta)$ be an arbitrary density function over the parameter space $\bTheta$. Then the logarithm of the marginal likelihood $\pDens(\by)$ satisfies
\begin{equation}
	\log \pDens(\by) = \int_{\bTheta} \qDens(\btheta) \log\left\lbrace\frac{\pDens(\by, \btheta)}{\qDens(\btheta)} \right\rbrace\,\mathrm{d}\btheta + \int_{\bTheta} \qDens(\btheta) \log\left\lbrace\frac{\qDens(\btheta)}{\pDens(\btheta\vert\by)} \right\rbrace\,\mathrm{d}\btheta,
	\label{eq:logliksplit}
\end{equation}
where $\pDens(\btheta \vert \by) \equiv \pDens(\by, \btheta)/\pDens(\by)$ is the posterior density function and $\pDens(\by,\btheta)$ is the model joint density function. The first addend of \eqref{eq:logliksplit} is a lower bound to the marginal log-likelihood and it is maximized when the second addend, that is the Kullback-Leibler divergence $\mbox{KL}\lbrace \qDens(\btheta) \:\lVert\: \pDens(\btheta\vert\by) \rbrace$, is minimized. The lower bound corresponds to $\log\pDens(\by)$ when $\qDens(\btheta) = \pDens(\btheta \vert \by)$, although in practical situations exact computation of the posterior density function is infeasible. 

The central idea of MFVB is to approximate $\pDens(\btheta \vert \by)$ with an approximating density function $\qDens(\btheta)$ that solves the following optimization problem:
\begin{equation}
\qDens^*(\btheta) = \underset{\qDens(\btheta) \in \mathcal{Q}}{\arg\min}\, \mbox{KL}\lbrace \qDens(\btheta) \:\lVert\: \pDens(\btheta \vert \by) \rbrace.
\label{eq:qstar_optim_problem}
\end{equation}
Tractable solutions arise when $\qDens(\btheta)$ is restricted to some convenient product of densities such that $\mathcal{Q} = \lbrace \qDens(\btheta) : \qDens(\btheta) = \prod_{i=1}^M \qDens_i(\btheta_i), \, \mbox{for some partition}\, \lbrace \btheta_1, \ldots, \btheta_M \rbrace \, \mbox{of} \, \btheta \rbrace$. It is possible to show (e.g. Ormerod \myand Wand, 2010) that under this restriction the optimal $\qDens$-density functions satisfy 
\begin{equation}
\qDens^*_i(\btheta_i) \propto \exp\left\lbrace E_{\qDens(-\btheta_i)}\lbrace\log \pDens(\btheta_i \vert \text{rest})\rbrace \right\rbrace, \quad 1 \le i \le M,
\label{eq:optimalqdensity}
\end{equation}
where $E_{\qDens(-\btheta_i)}$ denotes expectation with respect to $\prod_{j\neq i}\qDens_j(\btheta_j) = \qDens(\btheta) / \qDens_i(\btheta_i)$ and $\pDens(\btheta_i \vert \text{rest})$ is the full-conditional density function of $\btheta_i$. An iterative coordinate ascent procedure  can be used to solve \eqref{eq:qstar_optim_problem} and maximize the first addend on the right-hand side of \eqref{eq:logliksplit} through iterative updates for the optimal approximating densities arising from \eqref{eq:optimalqdensity}. Tractability is achieved when the updating steps reduce to updates of the approximating densities parameters and this typically occurs when all the $\qDens_i^*(\btheta_i)$'s belong to known families of parametric distributions. 
Convergence to at least local optima is guaranteed from convexity properties of the lower bound (Boyd \myand Vandenberghe, 2004) and once it is reached inference can be performed employing the $\qDens^*_i(\btheta_i)$ densities in place of the corresponding $\pDens(\btheta_i \vert \by)$ marginal posterior densities.

\subsection{Na{\"i}ve Variational Inference \label{subsec:NaiveUpdatesTwoLevelThreeLevelLMM}}

Assume the posterior density function $\pDens(\bbeta, \bu, \sigsq, \asigsq, \bG, \AG \vert \by)$ of the generic linear mixed model \eqref{eq:gauss_heter_lmm} is approximated by a $\qDens$-density function factorized as follows:
$$
\qDens(\bbeta, \bu, \sigsq, \asigsq, \bG, \AG) = \qDens(\bbeta, \bu)\: \qDens(\sigsq)\: \qDens(\asigsq)\: \qDens(\bG)\: \qDens(\AG).
$$
Using arguments from Section \ref{subsec:Overview}, it is possible to show that the optimal approximating densities that are function of $(\bbeta, \bu)$, $\sigsq$ and $\asigsq$ are the following:
\begin{equation}
\begin{array}{c}
\qDens^*(\bbeta, \bu) \text{ is a } \text{N}(\bmuq{\bbeta, \bu}, \bSigmaq{\bbeta, \bu})\text{ density function },\\[1.5ex]
\qDens^*(\sigsq) \text{ is an }\mbox{Inverse-}\chi^2(\xiq{\sigsq}, \lambdaq{\sigsq})\text{ density function }\\[1.5ex]
\mbox{and}\quad\qDens^*(\asigsq) \text{ is an }\mbox{Inverse-}\chi^2(\xiq{\asigsq}, \lambdaq{\asigsq})\text{ density function }.
\end{array}
\label{eq:qstar_gauss_heter_lmm}
\end{equation}
Let $\bC \equiv [\:\bX \:\vert\: \bZ \:]$ and denote with $n$ the number of its rows. The parameters of these densities can be obtained by iteratively performing the updates
\begin{equation}
\begin{array}{c}
	\bSigmaq{\bbeta,\bu} \longleftarrow \left(\muq{1/\sigsq} \:\bC^T \bC + \left[\begin{array}{cc} \bSigma_{\bbeta}^{-1} & \bO \\[1ex] \bO & E_{\qDens}(\bG^{-1})\end{array}\right]\right)^{-1}, \\[3ex]
	\bmuq{\bbeta,\bu} \longleftarrow \bSigmaq{\bbeta,\bu} \left(\muq{1/\sigsq}\: \bC^T \by + \left[\begin{array}{c} \bSigma_{\bbeta}^{-1} \bmu_{\bbeta} \\[1ex] \bzero \end{array}\right] \right),\\[3ex]
	\xiq{\sigsq} \longleftarrow \nusigsq + n, \qquad
	\lambdaq{\sigsq} \longleftarrow \muq{1/\asigsq} + \lVert \by - \bC \bmuq{\bbeta,\bu} \rVert^2 + \tr\left\lbrace \bSigmaq{\bbeta,\bu} \bC^T \bC \right\rbrace, \\[1.5ex]
	\xiq{\asigsq} \longleftarrow \nusigsq + 1\qquad\mbox{and} \qquad
	\lambdaq{\asigsq} \longleftarrow \muq{1/\sigsq} + 1/(\nusigsq \ssigsq^2)
\end{array}
\label{eq:baseModelUpdates}
\end{equation}
until convergence, where $\muq{1/\sigsq} \longleftarrow \xiq{\sigsq}/\lambdaq{\sigsq}$ and $\muq{1/\asigsq} \longleftarrow \xiq{\asigsq}/\lambdaq{\asigsq}$.

The approximating densities for the matrices $\bG$ and $\AG$ vary according to the random effects structure considered and are related to the structure of the random effects design matrix. 
For the two-level random effects specification, the structures of $\bG$ and $\AG$ are given by the second row of \eqref{eq:2level-to-general}, which involves matrices $\bSigma$ and $\ASigma$. Their approximating densities are the following:
\begin{equation*}
	\begin{array}{c}
		\qDens^*(\bSigma) \text{ is an Inverse-G-Wishart}\left(\Gfull, \xiq{\bSigma}, \bLambdaq{\bSigma}\right) \text{density function }\\[1.5ex]
		\mbox{and}\quad\qDens^*(\ASigma) \text{ is an Inverse-G-Wishart}\left(\Gdiag, \xiq{\ASigma}, \bLambdaq{\ASigma}\right)\text{density function }.
	\end{array}
\end{equation*}
The parameters of these $\qDens$-densities are updated according to:
\begin{equation*}
\begin{array}{c}
	\xiq{\bSigma} \longleftarrow \nuSigma + m + 2 q - 2, \quad \bLambdaq{\bSigma} \longleftarrow \bMq{\ASigma^{-1}} + \displaystyle{\sum_{i=1}^m} \left\lbrace \bmuq{\bu_i} \bmuq{\bu_i}^T + \bSigmaq{\bu_i} \right\rbrace,\\[1.5ex]
	\xiq{\ASigma} \longleftarrow \nuSigma + q \quad\mbox{and}\quad
		\bLambdaq{\ASigma} \longleftarrow \bLambda_{\ASigma} + \diag \left\lbrace \mbox{diagonal}\left(\bMq{\bSigma^{-1}}\right) \right\rbrace,
\end{array}
\end{equation*}
where $\bMq{\ASigma^{-1}} \longleftarrow \xiq{\ASigma} \bLambdaq{\ASigma}^{-1}$, and $\bmuq{\bu_i}$ and $\bSigmaq{\bu_i}$ respectively correspond to the sub-vector of $\bmuq{\bbeta, \bu}$ and sub-matrix of $\bSigmaq{\bbeta,\bu}$ associated with the $i$th group random effects vector $\bu_i$, for $1 \le i \le m$. Also, the update for $E_\qDens(\bG^{-1})$ appearing in the first line of \eqref{eq:baseModelUpdates} is
\begin{equation}
	E_\qDens(\bG^{-1}) \longleftarrow \bI_m \otimes \bMq{\bSigma^{-1}},
\label{eq:EGinv_twolevel}
\end{equation}
with $\bMq{\bSigma^{-1}} \longleftarrow (\xiq{\bSigma} - q + 1) \bLambdaq{\bSigma}^{-1}$.

For the three-level random effects specification, the structures of $\bG$ and $\AG$ are given by the last row of \eqref{eq:3level-to-general}, which involves matrices $\bSigmaLone$, $\bSigmaLtwo$, $\ASigmaLone$ and $\ASigmaLtwo$. Their approximating densities are the following:
\begin{equation*}
	\begin{array}{c}
		\qDens^*(\bSigmaLone) \text{ is an Inverse-G-Wishart}\left(\Gfull, \xiq{\bSigmaLone}, \bLambdaq{\bSigmaLone}\right)\text{density function },\\[1.5ex]
		\qDens^*(\ASigmaLone) \text{ is an Inverse-G-Wishart}\left(\Gdiag, \xiq{\ASigmaLone}, \bLambdaq{\ASigmaLone}\right)\text{density function },\\[1.5ex]
		\qDens^*(\bSigmaLtwo) \text{ is an Inverse-G-Wishart}\left(\Gfull, \xiq{\bSigmaLtwo}, \bLambdaq{\bSigmaLtwo}\right)\text{density function }\\[1.5ex]
		\mbox{and}\quad\qDens^*(\ASigmaLtwo) \text{ is an Inverse-G-Wishart}\left(\Gdiag, \xiq{\ASigmaLtwo}, \bLambdaq{\ASigmaLtwo}\right)\text{density function }.
	\end{array}
\end{equation*}
The parameters of these $\qDens$-densities are updated according to:
\begin{equation*}
\begin{array}{c}
	\xiq{\bSigmaLone} \longleftarrow \nuSigmaLone + m + 2 \qLone - 2, \quad \bLambdaq{\bSigmaLone} \longleftarrow \bMq{\ASigmaLone^{-1}} + \displaystyle{\sum_{i=1}^m} \left\lbrace \bmuq{\buLone_i} \bmuq{\buLone_i}^T + \bSigmaq{\buLone_i} \right\rbrace,\\[1.5ex]
	\xiq{\ASigmaLone} \longleftarrow \nuSigmaLone + \qLone, \quad \bLambdaq{\ASigmaLone} \longleftarrow \bLambda_{\ASigmaLone} + \diag \left\lbrace \mbox{diagonal}\left(\bMq{\bSigmaLoneMinusOne}\right) \right\rbrace,\\[1.5ex]
	\xiq{\bSigmaLtwo} \longleftarrow \nuSigmaLtwo + \displaystyle{\sum_{i=1}^m n_i} + 2 \qLtwo - 2, \quad \bLambdaq{\bSigmaLtwo} \longleftarrow \bMq{\ASigmaLtwo^{-1}} + \sum_{i=1}^m \sum_{j=1}^{n_i} \left\lbrace \bmuq{\buLtwo_{ij}} \bmuq{\buLtwo_{ij}}^T + \bSigmaq{\buLtwo_{ij}} \right\rbrace,\\[1.5ex]
	\xiq{\ASigmaLtwo} \longleftarrow \nuSigmaLtwo + \qLtwo \quad\mbox{and}\quad
		\bLambdaq{\ASigmaLtwo} \longleftarrow \bLambda_{\ASigmaLtwo} + \diag \left\lbrace \mbox{diagonal}\left(\bMq{\bSigmaLtwoMinusOne}\right) \right\rbrace,
\end{array}
\end{equation*}
where $\bMq{\ASigmaLone^{-1}} \longleftarrow \xiq{\ASigmaLone} \bLambdaq{\ASigmaLone}^{-1}$, $\bMq{\ASigmaLtwo^{-1}} \longleftarrow \xiq{\ASigmaLtwo} \bLambdaq{\ASigmaLtwo}^{-1}$, $\bmuq{\buLone_i}$ and $\bSigmaq{\buLone_i}$ respectively correspond to the sub-vector of $\bmuq{\bbeta, \bu}$ and sub-matrix of $\bSigmaq{\bbeta,\bu}$ associated with the $i$th group random effects vector at level 1 $\buLone_i$, and $\bmuq{\buLtwo_{ij}}$ and $\bSigmaq{\buLtwo_{ij}}$ respectively correspond to the sub-vector of $\bmuq{\bbeta, \bu}$ and sub-matrix of $\bSigmaq{\bbeta,\bu}$ associated with the $j$th subgroup of the $i$th group random effects vector at level 2 $\buLtwo_{ij}$, for $1 \le i \le m$ and $1 \le j \le n_i$.
Furthermore, the update for $E_\qDens(\bG^{-1})$ appearing in the first line of \eqref{eq:baseModelUpdates} is
\begin{equation}
	E_\qDens(\bG^{-1}) \longleftarrow \displaystyle{\blockdiag{1 \le i \le m}}\left(\left[\arraycolsep=0.5pt\def\arraystretch{1.25}
	\begin{array}{cc}\bMq{\bSigmaLoneMinusOne} & \bO \\[1ex] \bO & \bI_{n_i} \otimes \bMq{\bSigmaLtwoMinusOne} \end{array}\right]\right),
\label{eq:EGinv_threelevel}
\end{equation}
with $\bMq{\bSigmaLoneMinusOne} \longleftarrow (\xiq{\bSigmaLone} - \qLone + 1) \bLambdaq{\bSigmaLone}^{-1}$ and $\bMq{\bSigmaLtwoMinusOne} \longleftarrow (\xiq{\bSigmaLtwo} - \qLtwo + 1) \bLambdaq{\bSigmaLtwo}^{-1}$.  

The term \emph{na{\"i}ve} is used in this work when the MFVB updates described in this section are implemented without exploiting sparse matrix structures. Note, for example, that the update for $\bSigmaq{\bbeta,\bu}$ in \eqref{eq:baseModelUpdates} involves the inversion of a potentially massive matrix whose sparse structure is induced by those of $\bZ$ and $\bG$. As explained in Sections \ref{subsec:TwoLevelLMM} and \ref{subsec:ThreeLevelLMM}, when the model dimensions increase such matrices may become extremely sparse and the inversion operation can face many complications, both in terms of memory storage and computational efficiency. By taking advantage of the specific random effects structure it is possible to perform efficient \emph{streamlined} variational updates. 

\subsection{Streamlined Variational Inference \label{subsec:StreamlinedUpdatesTwoLevelThreeLevelLMM}}

The concept of streamlined variational inference for linear mixed models first appears in Lee \myand Wand (2016), where the sparse structure of $\bSigmaq{\bbeta,\bu}$ is exploited for efficiently fitting a particular version of model \eqref{eq:gauss_heter_lmm_2level} via MFVB. Nolan \myand Wand (2020) define sparse matrix classes arising from two-level and three-level random effects specifications and provide efficient mathematical solutions to the associated matrix inversion problems in their Theorems 2.2, 2.3, 3.2 and 3.3. Nolan \emph{et al.} (2020) implement such results and develop streamlined MFVB algorithms for linear mixed models having both two-level and three-level random effects specifications. These results are presented as solutions of two- and three-level sparse matrix problems.

Two-level sparse matrix problems are described in Section 2 of 
Nolan \myand Wand (2020). These problems are related to finding the vector $\bx$ such that $\bA\bx=\ba$, where
\begin{equation*}
\bA \equiv
\left[
\arraycolsep=2pt\def\arraystretch{1.5}
\begin{array}{ccccc}
\setstretch{4}
\AL{11}     & \AL{12,1} & \AL{12,2}  &\ \ \cdots\ \ &\AL{12,m} \\
\ALT{12,1} & \AL{22,1} & \bO      & \cdots   & \bO    \\
\ALT{12,2} & \bO     & \AL{22,2}  & \cdots   & \bO    \\ 
\vdots    & \vdots  & \vdots   & \ddots   & \vdots   \\
\ALT{12,m} & \bO     & \bO      & \cdots   &\AL{22,m} \\ 
\end{array}
\right],
\quad
\ba\equiv
\left[
\arraycolsep=2pt\def\arraystretch{1.5}
\begin{array}{c}
\setstretch{4}
\ba_1     \\
\ba_{2,1} \\
\ba_{2,2} \\ 
\vdots    \\
\ba_{2,m} \\ 
\end{array}
\right]
\qquad\mbox{and}\qquad
\bx\equiv
\left[
\arraycolsep=2pt\def\arraystretch{1.5}
\begin{array}{c}
\setstretch{4}
\bx_1     \\
\bx_{2,1} \\
\bx_{2,2} \\ 
\vdots    \\
\bx_{2,m} \\ 
\end{array}
\right],
\end{equation*}
and obtaining the sub-blocks of $\bA^{-1}$ corresponding to the non-zero blocks of $\bA$. The structure of $\bA^{-1}$ is 
\begin{equation*}
\bA^{-1}\equiv
\left[
\arraycolsep=2pt\def\arraystretch{1.5}
\begin{array}{ccccc}
\setstretch{4}
\AU{11}     & \AU{12,1} & \AU{12,2}  &\ \ \cdots\ \ &\AU{12,m} \\
\AUT{12,1} & \AU{22,1} & \bigX      & \cdots   & \bigX    \\
\AUT{12,2} & \bigX     & \AU{22,2}  & \cdots   & \bigX    \\ 
\vdots    & \vdots  & \vdots   & \ddots   & \vdots   \\
\AU{12,m} & \bigX     & \bigX      & \cdots   &\AU{22,m} \\ 
\end{array}
\right].
\end{equation*}
The blocks represented by the $\bigX$ symbol are not of interest. The relevant blocks of $\bA^{-1}$ can be efficiently computed applying Theorem 1 of Nolan \emph{et al.} (2020) and the formulas therein can be used to derive streamlined MFVB updates and achieve fast computation.

Three-level sparse matrix problems are useful for the treatment of three-level random effects models and details about this class of problems are provided in Section \ref{sec:threeLevSMP} of the supplementary material.

Algorithms using streamlined updates achieve the same MFVB approximations obtained with na{\"ive} updates, yet reducing memory usage and performing algebraic steps more efficiently. The former is obtained by circumventing the need of storing the zero sub-blocks of $\bC$ and the sub-blocks of $\bSigmaq{\bbeta,\bu}$ which are not needed for performing the updates. The latter is achieved by computing the useful sub-blocks of $\bSigmaq{\bbeta,\bu}$ with faster lower-dimensional matrix inversions, and the updates of $\bmuq{\bbeta,\bu}$ and $\lambdaq{\sigsq}$ solely relying on the non-zero sub-blocks of $\bC$ and $\bSigmaq{\bbeta,\bu}$.

Excellent performances both in terms of approximation accuracy, computational time and memory saving when compared to na{\"ive} MFVB or efficient MCMC samplers are shown in Nolan \emph{et al.} (2020), especially for large values of $m$. Nevertheless, this reference only treats the generic $\bbeta \sim \text{N}(\bmu_{\bbeta}, \bSigma_{\bbeta})$ prior specification for the fixed effects vector given in \eqref{eq:gauss_heter_lmm}. In this article, we develop streamlined variational inference procedures allowing for more general prior specifications on $\bbeta$ aiding selection of fixed effects.

\section{Approximate Variable Selection with Global-Local Priors\label{sec:ApproximateVarSelShrinkagePriors}}

Regression modeling is often concerned with the problem of selecting an optimal subset of plausible regressors with a significant impact on explaining the variability of the response variable. This is of particular interest in  \emph{sparse} covariate settings, where a large set of regressors is considered but only a small proportion of them is effectively relevant. We refer to O'Hara \myand Sillanp{\"a}{\"a} (2009) and references therein for an exhaustive introductory review on variable selection procedures from a Bayesian perspective.

\subsection{Bayesian Methods for Variable Selection\label{subsec:BayesianMethodsVarSel}}

Most common Bayesian approaches involve placing suitable prior distributions over the parameters subject to selection. Approaches of this type can be essentially subdivided into two main families, based on the so-called \textit{spike-and-slab} priors (Mitchell \myand Beauchamp, 1988; George \myand McCulloch, 1997; Johnstone \myand Silverman, 2005; Ishwaran \myand Rao, 2005; Efron, 2008; Bogdan \emph{et al.}, 2011) and \textit{global-local} shrinkage priors (Carvalho \emph{et al.}, 2009, 2010; Griffin \myand Brown, 2010; Polson \myand Scott, 2011; Armagan \emph{et al.}, 2013). 

Spike-and-slab priors are two-component mixture priors. The first prior component, the \emph{spike}, is a point mass function at zero characterizing the noise, usually given by a Dirac delta function or a Gaussian density function having mean zero and very small variance. The second component, the \emph{slab}, is an absolutely continuous density function representing the signal density of nonzero coefficients associated with relevant covariates. The slab is usually given by Laplace or Gaussian density functions and is typically centered around zero. A weight parameter taking values in the unit interval is used to balance the contribution of the two components. Although being highly appealing and allowing for separate control of the level of sparsity and the size of the signal coefficients, these priors may suffer from computational hurdles in high-dimensions.

Global-local shrinkage priors are absolutely continuous shrinkage priors that are placed on each coefficient $\beta_h$, $1 \le h \le H$, which is subject to selection. These priors admit the following convenient scale mixture representation (Polson \myand Scott, 2011), for proper choices of $\pDens(\tau)$ and $\pDens(\zeta_h)$:
\begin{equation}
	\beta_h \vert \tau, \zeta_h \simind \text{N}\left(0, \tausq/\zeta_h \right), \quad \tau \sim \pDens(\tau), \quad \zeta_h \sim \pDens(\zeta_h),\quad 1 \le h \le H.
\label{eq:globallocal}
\end{equation}
The global variance parameter $\tausq$ is common to all the coefficients and induces shrinkage towards the origin in the associated posterior density; the local variance parameter $\zeta_h$ is coefficient-specific. A more general specification including the model response error variance parameter is proposed in Bhattacharya et al. (2016). Depending on the distributional specifications for $\tau$ and $\zeta_h$ in \eqref{eq:globallocal}, many well-known shrinkage priors arise. Examples are the Horseshoe prior of Carvalho \emph{et al.} (2009, 2010), the Bayesian lasso of Park \myand Casella (2008), the Normal-Gamma prior of Griffin \myand Brown (2010), the Normal-Exponential-Gamma prior of Griffin \myand Brown (2011), the generalized double Pareto prior of Armagan \emph{et al.} (2013), the Dirichlet-Laplace prior of Bhattacharya \emph{et al.} (2015) and the Horseshoe+ prior of Bhadra \emph{et al.} (2017). Longer lists are given in Table 1 of Tang \emph{et al.} (2018) and Table 2 of Bhadra \emph{et al.} (2019).

For spike-and-slab priors, the posterior distributions of negligible effects present a higher weight for the spike: this provides a direct way to detect relevant effects, and therefore to perform the selection. For global-local priors, there is no posterior spike. The posterior density function, instead, is continuous with probability mass highly concentrated around zero, and a direct way for determining relevant effects is usually unavailable.

We employ global-local priors as they may offer substantial computational advantages over spike-and-slab priors due to their convenient representation as Gaussian scale mixtures, which give rise to convenient conjugate updates for all the $\beta_h$'s and $\zeta_h$'s. Bhattacharya \emph{et al.} (2015) also show that such priors exhibit improved posterior concentrations. Furthermore, the estimates of frequentist regularization procedures such as \emph{ridge} (Hoerl \myand Kennard, 1970), \emph{lasso} (Tibshirani, 1996), \emph{bridge} (Frank \myand Friedman, 1993) and \emph{elastic net} (Zou \myand Hastie, 2005) can be recasted as posterior mode estimates from models with global-local priors.

\subsection{Variational Inference with Global-Local Priors\label{subsec:VariationalInferenceLMwithGLPriors}}

Before considering their use within linear mixed model specifications, we illustrate the essential elements of variational inference for the simpler linear regression model. Without loss of generality, we will treat three of the most commonly adopted global-local prior specifications, namely:
\begin{equation}
	\beta_h\vert\tau \simind \mbox{Laplace}(0, \tau), \quad \beta_h\vert\tau \simind \mbox{Horseshoe}(0, \tau) \quad\text{or}\quad \beta_h\vert\tau \simind \mbox{NEG}(0, \tau, \lambda),
	\label{eq:global_local_priors}
\end{equation}
for each model coefficient $\beta_h$, $1 \le h \le H$. Hereafter $\lambda > 0$ is an additional shape parameter that we always assume being user-specified. 
\begin{figure}[t!]
	\centering
	\includegraphics[width=0.65\textwidth]{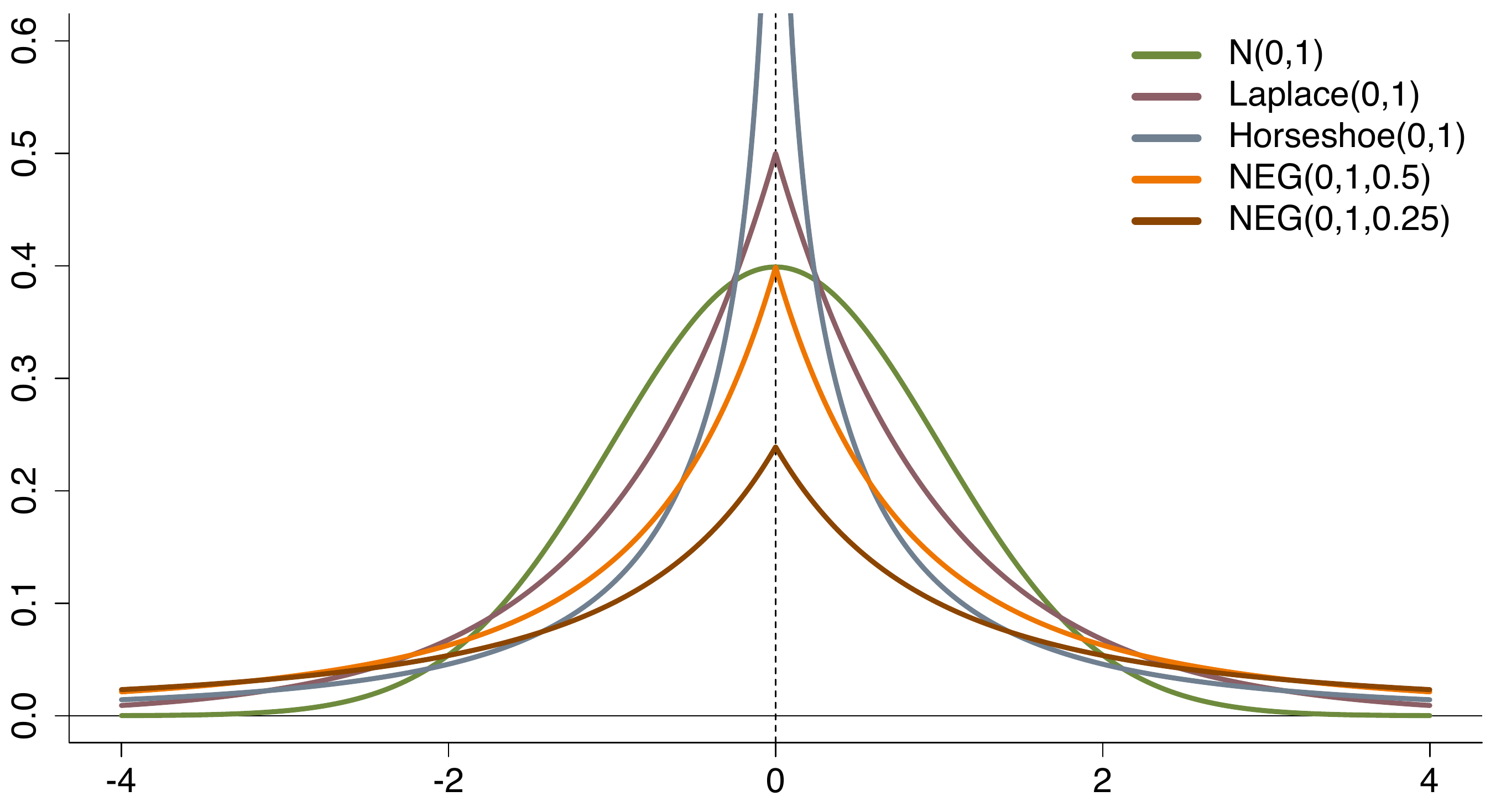}
	\caption{\it Visual comparison of the probability density functions for the Laplace, Horseshoe and Normal-Exponential-Gamma (NEG) distributions with zero mean and unit standard deviation. For easiness of comparison, the standard Gaussian probability density function is also displayed.}
	\label{fig:globallocal_comparison_plot} 
\end{figure}

The Laplace, Horseshoe and Normal-Exponential-Gamma (NEG) distributions account for different degrees of prior shrinkage towards zero and have different tail behaviors, as shown in Figure \ref{fig:globallocal_comparison_plot}. 
\begin{table}[t!]
	\centering
	\resizebox{0.75\textwidth}{!}{%
	\begin{tabular}{l c c c}
		\hline\\[-2ex]
		\emph{Prior specification} $\pDens(\beta_h)$ & $\pDens(\beta_h \vert \zeta_h)$ & $\pDens(\zeta_h \vert a_{\zeta_h})$ & $\pDens(a_{\zeta_h})$ \\[0.5ex] \hline \\[-1.5ex]
		$\mbox{Laplace}(0, \tau)$ & $\mbox{N}\left(0, \tausq/\zeta_h \right)$ & $\mbox{Inverse-}\chi^2(2, 1)$ & $-$ \\[0.5ex]
		$\mbox{Horseshoe}(0, \tau)$ & $\mbox{N}\left(0, \tausq/\zeta_h \right)$ & $\mbox{Gamma}(1/2, a_{\zeta_j})$ &  $\mbox{Gamma}(1/2, 1)$ \\[0.5ex]
		$\mbox{NEG}(0, \tau, \lambda)$ & $\mbox{N}\left(0, \tausq/\zeta_h \right)$ & $\mbox{Inverse-}\chi^2(2, 2a_{\zeta_j})$ & $\mbox{Gamma}(\lambda, 1)$ \\[0.5ex] \hline
	\end{tabular}
	}
	\caption{\it Hierarchical formulation of the Laplace, Horseshoe and Negative-Exponential-Gamma (NEG) priors \eqref{eq:global_local_priors} following the general global-local representation \eqref{eq:globallocal}, for known $\tau$}.
	\label{tab:global_local} 
\end{table}
Each prior specification in \eqref{eq:global_local_priors} can be recasted into the scale mixture framework \eqref{eq:globallocal}, as summarized in Table \ref{tab:global_local}. For the Horseshoe and NEG priors, we use convenient hierarchical representations of $\pDens(\zeta_h)$ based on auxiliary variables $a_{\zeta_h}$, $1 \le h \le H$, involving tractable Gamma and $\text{Inverse-}\chi^2$ distributions. Details about the involved density functions and auxiliary variable representations are summarized in Section \ref{appsec:UsefulDistributions} of the supplementary material. 

A linear regression model with one of the three global-local priors in \eqref{eq:global_local_priors} is then expressible as:
\begin{align}
	\begin{array}{c}
	\by\vert\beta_0,\bbeta,\sigsq \sim \text{N}(\beta_0 \boldsymbol{1} + \bX \bbeta, \sigsq \bI),\\[1.5ex]
	\beta_0 \sim \mbox{N}(\mu_{\beta_0}, \sigsq_{\beta_0}), \qquad \bbeta \vert \bzeta,\tausq \sim \mbox{N}(\bzero, \tausq\:\diag(\bzeta)^{-1}),\\[1.5ex]
	\sigsq \vert \asigsq\sim\mbox{Inverse-}\chi^2(\nusigsq,1/\asigsq),\quad\asigsq\sim\mbox{Inverse-}\chi^2(1,1/(\nusigsq\ssigsq^2)),\\[1.5ex]
	\tausq \vert \atausq \sim \mbox{Inverse-}\chi^2(1, 1/\atausq), \quad \atausq \sim \mbox{Inverse-}\chi^2(1, 1/\stausq^2),\\[1.5ex]
	\zeta_h \vert a_{\zeta_h} \simind \pDens(\zeta_h \vert a_{\zeta_h}),\quad a_{\zeta_h} \simind \pDens(a_{\zeta_h}), \quad1 \le h \le H,
	\end{array}
\label{eq:globallocal_model}
\end{align} 
where $\beta_0$ is the model intercept (which is usually excluded from the selection procedure), $\boldsymbol{1}$ is a vector full of ones, $\bbeta = [\beta_1, \dots, \beta_H]^T$ is the vector of coefficients having global-local shrinkage priors, $\bX = [\bx_1, \dots, \bx_p]$ is the associated design matrix, $\bzeta = [\zeta_1, \dots, \zeta_H]^T$ and $\ba_{\bzeta} = [a_{\zeta_1}, \dots, a_{\zeta_H}]^T$. The common Gaussian and Half-t prior distributions are considered for $\beta_0$ and $\sigma$, respectively. Importantly, we specify a $\mbox{Half-}t(\stausq, 1)$ distribution with $\stausq>0$ for the global scale parameter $\tau$, allowing for weak prior informativeness about the global degree of sparseness when a large scale parameter $\stausq$ is used. Hence, we let the variable-selection procedure be free from hyperparameters that have to be manually tuned by the user. The densities $\pDens(\zeta_h \vert a_{\zeta_h})$ and $\pDens(a_{\zeta_h})$ vary according to the global-local prior specification adopted, as shown in Table \ref{tab:global_local}.

The model posterior density function $\pDens(\beta_0, \bbeta, \sigsq, \asigsq, \tausq, \atausq, \bzeta, \ba_{\bzeta} \vert \by)$ admits a tractable MFVB approximation when the following mean-field restriction is used:
$$
\qDens(\beta_0, \bbeta, \sigsq, \asigsq, \tausq, \atausq, \bzeta, \ba_{\bzeta}) = \qDens(\beta_0, \bbeta)\: \qDens(\sigsq)\: \qDens(\tausq)\: \qDens(\asigsq)\: \qDens(\atausq) \prod\limits_{h=1}^H \lbrace \qDens(\zeta_h)\: \qDens(a_{\zeta_h}) \rbrace.
$$
The optimal $\qDens$-density functions then results as follows:
\begin{equation}
\begin{array}{c}
\qDens^*(\beta_0, \bbeta) \mbox{ is a N}(\bmuq{\beta_0,\bbeta}, \bSigmaq{\beta_0,\bbeta}) \mbox{ density function},\\[1.5ex]
\qDens^*(\sigsq) \mbox{ is an Inverse-}\chi^2\big(\xiq{\sigsq}, \lambdaq{\sigsq}\big)\mbox{ density function},\\[1.5ex]
\qDens^*(\asigsq) \mbox{ is an Inverse-}\chi^2\big(\xiq{\asigsq}, \lambdaq{\asigsq}\big)\mbox{ density function},\\[1.5ex]
\qDens^*(\tausq) \mbox{ is an Inverse-}\chi^2(\xiq{\tausq}, \lambdaq{\tausq})\mbox{ density function},\\[1.5ex]
\qDens^*(\atausq) \mbox{ is an Inverse-}\chi^2(\xiq{\atausq}, \lambdaq{\atausq})\mbox{ density function},\\[1.5ex]
\qDens^*(\zeta_h)\mbox{ is }
\begin{cases}
	\mbox{an Inverse-Gaussian}(\muq{\zeta_h}, 1) \mbox{ density function} &\mbox{for a Laplace prior}\\
	\mbox{a Gamma}(1, \lambdaq{\zeta_h}) \mbox{ density function}&\mbox{for a Horseshoe prior}\\
	\mbox{an Inverse-Gaussian}(\muq{\zeta_h}, \lambdaq{\zeta_h}) \mbox{ density function} &\mbox{for a NEG prior}
\end{cases} \\[4ex]
\quad\mbox{and }\qDens^*(a_{\zeta_h})\mbox{ is }
\begin{cases}
	\qquad\qquad\qquad\qquad- &\mbox{for a Laplace prior}\\
	\mbox{a Gamma}(1, \lambdaq{a_{\zeta_h}}) \mbox{ density function} &\mbox{for a Horseshoe prior}\\
	\mbox{a Gamma}(\lambda + 1,\lambdaq{a_{\zeta_h}})\mbox{ density function} &\mbox{for a NEG prior,}
\end{cases}
\end{array}
\label{eq:qstargloballocal}
\end{equation}
for $1 \le h \le H$. When a Laplace prior is specified, the model does not include the $a_{\zeta_h}$ auxiliary variables and therefore $\qDens(a_{\zeta_h})$ is not included. The expressions for the parameter updates of these approximating densities, together with their full derivations, are reported in Section \ref{appsec:Derivations} of the supplementary material.

\subsection{From Shrinkage to Selection: the Signal Adaptive Variable Selector\label{subsec:SAVSAlgorithm}}

One limitation of continuous global-local shrinkage priors is the unavailability of direct information from the posterior of each $\beta_h$ for selecting relevant effects. Typically, the posterior distributions of less relevant coefficients arising from such priors are highly concentrated around zero, with marked peaks and negligible tails, although not having full mass at zero. Therefore, global-local shrinkage priors do not provide any sparse posterior solution. This issue becomes even more relevant when variational inference is used to fit the model because the approximate marginal posterior densities of $\bbeta$ are Gaussian, and so the peaks of the true marginal posterior densities are approximated by bell-shaped curves. 

Several heuristic methods have been developed for post-processing posterior distributions arising from global-local priors and determining whether the associated covariates have to be selected or not. A simple but possibly misleading solution is to select as relevant the covariates associated with coefficients whose posterior credible intervals do not contain the zero. Nonetheless, this approach usually exhibits poor performances due to the difficulty of accurately estimating the uncertainty in high dimensional problems, and depends on the chosen credible level. Carvalho \emph{et al.} (2010) define a local shrinkage factor which can take values in the unit interval and help determine whether each variable is suggested to be selected or not according to a pre-specified threshold, analogously to the classical posterior inclusion probability of Barbieri \myand Berger (2004). Bondell \myand Reich (2012) propose a method based on posterior credible regions, although its implementation and results rely upon the use of conjugate Normal priors. Zhang \myand Bondell (2018) extended the method to global-local priors and propose an intuitive approach to tune the prior hyperparameters based on minimizing a discrepancy measure between the induced distribution of $R^2$ from the prior and the desired distribution. 

All these methods and many others have a common issue, that is the dependence on the choice of one or more thresholds. Bhattacharya \emph{et al.} (2015) propose grouping the entries of posterior medians into null and non-null groups using 2-means clustering. While this approach does not require any tuning parameters, issues emerge when there are signals of varying strengths. Li \myand Pati (2017) propose a similar approach which is based on first obtaining a posterior distribution of the number of signals by clustering the signal and the noise coefficients and then estimating the signals from the posterior median. 

In this work, we opt for the signal adaptive variable selector (SAVS) partially motivated by Hahn \myand Carvalho (2015) and accurately developed by Ray \myand Bhattacharya (2018). The SAVS approach post-processes a point estimate from the posterior distribution of a coefficient having global-local prior distribution via \emph{soft-thresholding} to determine whether the associated covariate is assumed to be relevant or not. We adapt this procedure for usage in variational inference and propose Algorithm \ref{alg:savs} as an implementation of the SAVS approach based on the optimal approximate posterior densities $\qDens^*(\beta_h)$, $1 \le h \le H$. 
\begin{algorithm}[t!]
  \begin{center}
    \begin{minipage}[t]{155mm}
      	Inputs: $\mu_{\qDens^*(\beta_h)} \equiv E_{\qDens^*}(\beta_h)$ and $\mathbf{x}_h$ being the covariate vector corresponding to $\beta_h$, $1 \le h \le H$. \\[0.5ex]
      	If $\lVert \mathbf{x}_h \rVert^2 \le \vert \mu_{\qDens^*(\beta_h)} \vert^{-3}$:
		\begin{itemize}
			\item[] $\mu^*_{\qDens^*(\beta_h)} = 0$ \quad and\quad $\gamma_h = 0$;
		\end{itemize}
	else:
		\begin{itemize}
			\item[] $\mu^*_{\qDens^*(\beta_h)} = \text{sign}(\mu_{\qDens^*(\beta_h)}) \lVert \mathbf{x}_h \rVert^{-2} \left(\vert\mu_{\qDens^*(\beta_h)}\vert \: \lVert \mathbf{x}_h \rVert^2  - \mu_{\qDens^*(\beta_h)}^{-2} \right)$ \quad and\quad $\gamma_h = 1$.
		\end{itemize}
	Output: A sparse estimate $\mu^*_{\qDens^*(\beta_h)}$ for $\qDens^*(\beta_h)$ and the associated binary selector $\gamma_h$.
    \end{minipage}
    \vspace{-1em}
  \end{center}
  \caption{\it Signal Adaptive Variable Selector (SAVS) algorithm for performing variable selection using the optimal approximate density function $\qDens^*(\beta_h)$ of a generic coefficient with global-local prior. }
\label{alg:savs}
\end{algorithm}
The procedure takes the approximate posterior mean parameter of a generic coefficient subject to selection and the associated unstandardized covariate as inputs. It then returns a \emph{sparsified} approximate posterior summary estimate $\mu^*_{\qDens^*(\beta_h)}$, together with a binary variable $\gamma_h$ indicating whether the $h$th covariate is suggested to be selected or not. The attractiveness of this approach comes from the fact that  it is completely automated and does not require any tuning parameters. Ray \myand Bhattacharya (2018) provide a theoretical justification for the SAVS approach, noticing that its output can be obtained by solving an optimization problem closely related to the adaptive lasso of Zou (2006), and show it is highly competitive among alternative Bayesian selection procedures. 

\section{Linear Mixed Models with Global-Local Priors on Fixed Effects\label{sec:VariationalInferenceLMMwithGLPriors}}

Variational approximations for linear mixed models with two- and three-level random effects are described in Section \ref{sec:VariationalBayesInference}, using a generic $\bbeta \sim \text{N}(\bmu_{\bbeta}, \bSigma_{\bbeta})$ prior distribution for the fixed effects parameter vector. In this work, our interest is in developing variational approximations for generalizations of model \eqref{eq:gauss_heter_lmm} embedding prior specifications for fixed effects selection such as those discussed in Section \ref{sec:ApproximateVarSelShrinkagePriors}. 

In order to do so, we subdivide the $p$-dimensional fixed effects vector $\bbeta$ as follows: 
$$
\bbeta = \left[\begin{array}{c} \bbeta^\raneff \\ [0.5ex] \bbeta^\additional \\[0.5ex] \bbeta^\subjtosel \end{array}\right],
$$
where $\bbeta^\raneff$ is a $p_\raneff$-dimensional vector of fixed effects associated to the random (\textit{R}) effects component of the model, $\bbeta^{\additional}$ is a $p_{\additional}$-dimensional vector of additional (\textit{A}) fixed effects and $\bbeta^{\subjtosel}$ is a $p_{\subjtosel}$-dimensional vector of fixed effects which are subject to selection (\textit{S}). Here $p = p_\raneff + p_\additional + p_\subjtosel$, with $p_\raneff$, $p_\additional$ and $p_\subjtosel$ varying according to the application of interest. For the two- and three-level mixed models considered in this work,  $p_\raneff = q$ and $p_\raneff = \max(\qLone, \qLtwo)$ respectively. Typically, $q$, $\qLone$ and $\qLtwo$ are relatively small, while $p_\additional$ and $p_\subjtosel$ could take moderate to large values. 

Similarly, we subdivide the fixed effects design matrix as follows:
$$
\bX = \left[\: \bX^\raneff \:\Big\vert\: \bX^\additional \:\Big\vert\: \bX^\subjtosel \:\right],
$$ 
with $\bX^\subjtosel$ assumed to have columns with zero mean and unit variance, unless differently specified. The fixed effects linear contribution of model \eqref{eq:gauss_heter_lmm} then factorizes into $\bX \bbeta = \bX^\raneff \bbeta^\raneff + \bX^\additional \bbeta^\additional + \bX^\subjtosel \bbeta^\subjtosel$, and the same applies to the two-level mixed model \eqref{eq:gauss_heter_lmm_2level} and the three-level mixed model \eqref{eq:gauss_heter_lmm_3level} specifications. Notice that for the former specification $\bZ_i = \bX^\raneff_i$, for all $1 \le i \le m$.

We assume without loss of generality that $\bbeta^\raneff$, $\bbeta^\additional$ and $\bbeta^\subjtosel$ are a-priori independent from each other, and specify the following prior distributions:
$$
\bbeta^\raneff \sim \mbox{N}(\bmu_{\bbeta^\raneff}, \bSigma_{\bbeta^\raneff}), \qquad \bbeta^\additional \sim \mbox{N}(\bmu_{\bbeta^\additional}, \bSigma_{\bbeta^\additional}), \qquad \bbeta^\subjtosel \sim \pDens(\bbeta^\subjtosel) = \prod_{h=1}^{p_\subjtosel} \pDens(\beta^\subjtosel_h),
$$
with hyperparameters $\bmu_{\bbeta^\raneff} \in \mathbb{R}^{p_\raneff}$, $\bmu_{\bbeta^\additional} \in \mathbb{R}^{p_\additional}$, $\bSigma_{\bbeta^\raneff}$ and $\bSigma_{\bbeta^\additional}$ symmetric positive definite matrices. The prior specification for $\bbeta^\subjtosel$ assumes a-priori independence among all the coefficients subject to selection, with $\pDens(\beta_h^{\subjtosel})$ taking one of the three different global-local prior distributions treated in Section \ref{sec:ApproximateVarSelShrinkagePriors} for $1 \le h \le p_\subjtosel$, specifically:
$$
\beta^\subjtosel_h \vert \tau \simind \text{Laplace}(0, \tau), \quad \beta^\subjtosel_h \vert \tau \simind \text{Horseshoe}(0, \tau) \quad \text{or} \quad\beta^\subjtosel_h \vert \tau \simind \text{NEG}(0, \tau, \lambda).
$$
The resulting linear mixed model is a generalization of \eqref{eq:gauss_heter_lmm} that accounts for global-local prior specification over a subset of the fixed effects, and can be expressed as:
\begin{equation}
\label{eq:gauss_heter_lmm_penalized}
\begin{array}{c}
\by \vert \bbeta^\raneff, \bbeta^\additional, \bbeta^\subjtosel, \bu, \sigsq \sim \text{N}(\bX^\raneff \bbeta^\raneff + \bX^\additional \bbeta^\additional + \bX^\subjtosel \bbeta^\subjtosel + \bZ \bu, \sigsq \bI), \quad \bu \vert \bG \sim \text{N}(\bzero, \bG),\\[1.5ex] 
\left[\begin{array}{c} \bbeta^\raneff \\ \bbeta^\additional \\ \bbeta^\subjtosel \end{array}\right] \Bigg\vert \:\bzeta, \tausq \sim \mbox{N}\left(\left[\begin{array}{c} \bmu_{\bbeta^\raneff} \\ \bmu_{\bbeta^\additional} \\ \bzero \end{array}\right], \left[\begin{array}{ccc} \bSigma_{\bbeta^\raneff} & \bO & \bO \\ \bO & \bSigma_{\bbeta^\additional} & \bO \\ \bO & \bO & \tausq\diag(\bzeta)^{-1} \end{array} \right] \right), \\[5ex] 
\sigsq\vert \asigsq \sim \text{Inverse-}\chi^2(\nusigsq, 1/\asigsq), \quad \asigsq \sim \text{Inverse-}\chi^2(1, 1/(\nusigsq \ssigsq^2)),  \\[1ex]
\zeta_h \vert a_{\zeta_h} \simind \begin{cases} 
	\:\mbox{Inverse-}\chi^2(2,1) \qquad & \mbox{for a Laplace prior}\\
	\:\mbox{Gamma}(1/2, a_{\zeta_h}) \qquad & \mbox{for a Horseshoe prior}\\
	\:\mbox{Inverse-}\chi^2(2, 2a_{\zeta_h}) \qquad & \mbox{for a NEG prior,}
\end{cases} \\[4ex]
a_{\zeta_h} \simind \begin{cases} 
	\qquad\quad - \qquad & \mbox{for a Laplace prior}\\
	\mbox{Gamma}(1/2, 1) \qquad & \mbox{for a Horseshoe prior}\\
	\mbox{Gamma}(\lambda, 1) \qquad & \mbox{for a NEG prior,}
\end{cases}\\[4ex]
\text{for } 1 \le h \le p_\subjtosel,\\[1ex]
\tausq \vert \atausq \sim \mbox{Inverse-}\chi^2(1, 1/\atausq), \quad \atausq \sim \mbox{Inverse-}\chi^2(1, 1/{\stausq^2}),\\[1.5ex]
 \bG\vert \AG \sim \pDens(\bG \vert \AG), \quad \AG \sim \pDens(\AG).
\end{array}
\end{equation}
Here $\bzeta \equiv [\zeta_1, \ldots, \zeta_{p_\subjtosel}]^T$ and $\ba_{\bzeta} \equiv [a_{\zeta_1}, \ldots, a_{\zeta_{p_\subjtosel}}]^T$. 
This model can be fitted via MFVB assuming that the full posterior density function is approximated as
\begin{equation}
	\pDens(\bbeta, \bu, \sigsq, \asigsq, \bzeta, \ba_{\bzeta}, \tausq, \atausq, \bG, \AG \vert \by) \approx \qDens(\bbeta, \bu, \sigsq, \asigsq, \bzeta, \ba_{\bzeta}, \tausq, \atausq, \bG, \AG)
\label{eq:MFVBapproxim}
\end{equation}
and a tractable solution arises with the following mean-field restriction:
\begin{equation}
\begin{array}{c}
	\qDens(\bbeta, \bu, \sigsq, \asigsq, \bzeta, \ba_{\bzeta}, \tausq, \atausq, \bG, \AG) = \\[1ex] \qDens(\bbeta, \bu)\: \qDens(\sigsq)\: \qDens(\asigsq) \left\lbrace \displaystyle{\prod_{h=1}^{p_\subjtosel}}
 \qDens(\zeta_h)\: \qDens(a_{\zeta_h}) \right\rbrace \qDens(\tausq)\: \qDens(\atausq)\: \qDens(\bG)\: \qDens(\AG).
\end{array}
\label{eq:prodrestr_globallocal}
\end{equation}
Arguments similar to those given in Section \ref{subsec:NaiveUpdatesTwoLevelThreeLevelLMM} and \ref{subsec:VariationalInferenceLMwithGLPriors} lead to the optimal approximating densities being:
\begin{equation}
\hspace{-0.1cm}
\begin{array}{c}
\qDens^*(\bbeta, \bu) \text{ is a } \text{N}(\bmuq{\bbeta, \bu}, \bSigmaq{\bbeta, \bu})\text{ density function},\\[1.5ex]
\qDens^*(\sigsq) \text{ is an }\mbox{Inverse-}\chi^2(\xiq{\sigsq}, \lambdaq{\sigsq})\text{ density function},\\[1.5ex]
\qDens^*(\asigsq) \text{ is an }\mbox{Inverse-}\chi^2(\xiq{\asigsq}, \lambdaq{\asigsq})\text{ density function},\\[1.5ex]
\qDens^*(\zeta_h)\mbox{ is }
\begin{cases}
	\mbox{an Inverse-Gaussian}(\muq{\zeta_h}, 1) \mbox{ density function} &\mbox{ for a Laplace prior}\\
	\mbox{a Gamma}(1, \lambdaq{\zeta_h}) \mbox{ density function}&\mbox{ for a Horseshoe prior}\\
	\mbox{an Inverse-Gaussian}(\muq{\zeta_h}, \lambdaq{\zeta_h}) \mbox{ density function} &\mbox{ for a NEG prior,}
\end{cases}\\[4ex]
\qDens^*(a_{\zeta_h})\mbox{ is }
\begin{cases}
	\qquad\qquad\qquad\qquad- &\mbox{ for a Laplace prior}\\
	\mbox{a Gamma}(1, \lambdaq{a_{\zeta_h}}) \mbox{ density function} &\mbox{ for a Horseshoe prior}\\
	\mbox{a Gamma}(\lambda + 1,\lambdaq{a_{\zeta_h}})\mbox{ density function} &\mbox{ for a NEG prior,}
\end{cases} \\[4ex]
\text{for } 1 \le h \le p_\subjtosel,\\[1.5ex]
\qDens^*(\tausq) \mbox{ is an Inverse-}\chi^2(\xiq{\tausq}, \lambdaq{\tausq})\mbox{ density function }\\[1.5ex]
\mbox{and }\qDens^*(\atausq) \mbox{ is an Inverse-}\chi^2(\xiq{\atausq}, \lambdaq{\atausq})\mbox{ density function}.
\end{array}
\label{eq:qstar_gauss_heter_lmm_globallocal}
\end{equation}
The optimal approximating densities for the matrices $\bG$ and $\AG$ vary according to the adopted random effect structure, as explained in Section \ref{subsec:NaiveUpdatesTwoLevelThreeLevelLMM}. Notice that \eqref{eq:prodrestr_globallocal} jointly approximates $\bbeta^\raneff$, $\bbeta^\additional$, $\bbeta^\subjtosel$ and $\bu$, allowing all the fixed effects to share posterior dependence with the random effects. Also, the
$\qDens^*(\beta^\subjtosel_h)$ approximating densities are Gaussian, although different global-local prior specifications may lead to marginal posterior density functions $\pDens(\beta^\subjtosel_h \vert \by)$ having shapes around zero that are different from the typical bell-shaped behavior, especially those of fixed effects associated to irrelevant covariates. Using MFVB we sacrifice some degree of accuracy, yet obtaining a substantial computational time advantage over standard MCMC sampling procedures. Section \ref{sec:Assessments} investigates the quality of these approximations and the fixed effects selection performances.

\subsection{Na{\"i}ve Updates\label{subsec:NaiveMFVBGLPriors}}

The updates of the parameters of the $\qDens$-densities in \eqref{eq:qstar_gauss_heter_lmm_globallocal} can be derived combining and adapting the results discussed in Sections \ref{subsec:NaiveUpdatesTwoLevelThreeLevelLMM} and \ref{subsec:VariationalInferenceLMwithGLPriors}, and references therein. In particular, those related to the $\qDens^*(\bbeta, \bu)$ optimal density function are:
\begin{equation}
\begin{array}{c}
	\bSigmaq{\bbeta,\bu} \longleftarrow \left(\muq{1/\sigsq} \:\bC^T \bC + \left[\begin{array}{cccc} \bSigma_{\bbeta^\raneff}^{-1} & \bO & \bO & \bO \\[1ex]
	\bO & \bSigma_{\bbeta^\additional}^{-1} & \bO & \bO \\[1ex]
	\bO & \bO & \muq{1/\tausq}\: \diag(\bmuq{\bzeta}) & \bO \\[1ex]
	\bO & \bO & \bO & E_{\qDens}(\bG^{-1}) \end{array}\right]\right)^{-1} \\[6.5ex]
	\mbox{and\quad}\bmuq{\bbeta,\bu} \longleftarrow \bSigmaq{\bbeta,\bu} \left(\muq{1/\sigsq}\: \bC^T \by + \left[\begin{array}{c} \bSigma_{\bbeta^\raneff}^{-1} \bmu_{\bbeta^\raneff} \\[1ex] \bSigma_{\bbeta^\additional}^{-1} \bmu_{\bbeta^\additional} \\[1ex] \bzero \\[1ex] \bzero \end{array}\right] \right).
\label{eq:SigmaMu_updates_naive}
\end{array}
\end{equation}
The update for $E_\qDens(\bG^{-1})$ is given by \eqref{eq:EGinv_twolevel} or \eqref{eq:EGinv_threelevel} depending on whether we are considering a two-level or a three-level mixed model specification, respectively. The main difference with the expressions given in \eqref{eq:baseModelUpdates} is that a global-local prior specification on each element of $\bbeta^\subjtosel$ introduces the diagonal matrix $\muq{1/\tausq}\: \diag(\bmuq{\bzeta})$ of dimension $p_\subjtosel \times p_\subjtosel$ inside the update expression for $\bSigmaq{\bbeta,\bu}$. This diagonal matrix is updated at each iteration of the MFVB algorithm employing the updated values of $\muq{1/\tausq}$ and $\bmuq{\bzeta}$, accordingly to the global-local prior adopted. 

The updates for the parameters of $\qDens^*(\sigsq)$ and $\qDens^*(\asigsq)$ are identical to those in \eqref{eq:baseModelUpdates}. The updates for the parameters of the optimal approximating $\qDens$-densities of $\bG$ and $\AG$ are identical to those described in Section \ref{subsec:NaiveUpdatesTwoLevelThreeLevelLMM}. 
The updates for the parameters of $\qDens^*(\zeta_h)$ and $\qDens^*(a_{\zeta_h})$, $1 \le h \le p_\subjtosel$, $\qDens^*(\tausq)$ and $\qDens^*(\atausq)$ follow with minor modifications from Section \ref{subsec:VariationalInferenceLMwithGLPriors}, replacing $H$ with $p_\subjtosel$. 

Nevertheless, updates \eqref{eq:SigmaMu_updates_naive} suffer from the same problematics elucidated in Section \ref{subsec:NaiveUpdatesTwoLevelThreeLevelLMM} for large $m$ values. Therefore, an appropriate streamlined enhancement for efficient implementations is required.

\subsection{Streamlined Updates\label{subsec:StreamlinedMFVBGLPriors}}

Results 1 and 4 from Nolan \emph{et al.} (2020) can be extended to derive a streamlined MFVB algorithm for efficiently updating the parameters of the densities in \eqref{eq:qstar_gauss_heter_lmm_globallocal} and, in particular, for exploiting the sparse matrix structures presented in updates \eqref{eq:SigmaMu_updates_naive}. Such results are based on the two- and three-level \emph{sparse matrix least squares problems} defined in Nolan \myand Wand (2020), and make use of the associated $\SolveTwoLevelSparseMatrix$ and $\SolveThreeLevelSparseMatrix$ routines, which are recalled in Section \ref{appsec:RoutinesNolanWand2020} of the supplementary material. The following results explain how to efficiently compute $\bmuq{\bbeta,\bu}$ and the relevant sub-blocks of $\bSigmaq{\bbeta,\bu}$ which are necessary for finding the optimal $\qDens$-density listed in \eqref{eq:qstar_gauss_heter_lmm_globallocal}.
\begin{result}[Result 1 of Nolan \emph{et al.} (2020), revisited]
\label{res:res_streamlined_2level}
	The MFVB updates of $\bmuq{\bbeta,\bu}$ and each of the sub-blocks of $\bSigmaq{\bbeta,\bu}$ that are relevant for variational inference concerning model \eqref{eq:qstar_gauss_heter_lmm_globallocal} with a two-level random effects specification are expressible as a two-level sparse matrix problem of the form $\bA \bmuq{\bbeta,\bu} = \ba$, where $\ba$ and the non-zero sub-blocks of $\bA$ are, according to the notation in Section \ref{subsec:StreamlinedUpdatesTwoLevelThreeLevelLMM}:
	\begin{equation*}
	\begin{array}{c}
		\AL{11} = \muq{1/\sigsq} \displaystyle{\sum_{i=1}^m} (\bX_i^T \bX_i) + \left[\begin{array}{ccc} \bSigma_{\bbeta^\raneff}^{-1} & \bO & \bO \\[1ex]
	\bO & \bSigma_{\bbeta^\additional}^{-1} & \bO \\[1ex]
	\bO & \bO & \muq{1/\tausq}\: \diag(\bmuq{\bzeta}) \end{array}\right],\\[5ex]
	\ba_1 = \muq{1/\sigsq} \displaystyle{\sum_{i=1}^m} (\bX_i^T \by_i) + \left[\begin{array}{c} \bSigma_{\bbeta^\raneff}^{-1} \bmu_{\bbeta^\raneff} \\[1ex] \bSigma_{\bbeta^\additional}^{-1} \bmu_{\bbeta^\additional} \\[1ex] \bzero \end{array}\right],\\[5ex]
		\AL{22,i} = \muq{1/\sigsq} \bZ_i^T \bZ_i + \bM_{q(\bSigma^{-1})}, \quad \AL{12,i} = \muq{1/\sigsq} \bX_i^T \bZ_i,\quad \ba_{2,i} = \muq{1/\sigsq} \bZ_i^T \by_i,
	\end{array}
	\end{equation*}
for $1 \le i \le m$. Moreover, $\bX_i = [\: \bX_{i}^\raneff \:\vert\: \bX_{i}^\additional \:\vert\: \bX_{i}^\subjtosel \:]$. The $\SolveTwoLevelSparseMatrix$ routine efficiently solves the associated linear system and provides the solutions: 
$$ \bmuq{\bbeta} = \bx_1, \quad \bSigmaq{\bbeta} = \AU{11}$$ 
and 
$$ \bmuq{\bu_i} = \bx_{2,i}, \quad \bSigmaq{\bu_i} = \AU{22,i}, \quad E_\qDens\lbrace (\bbeta - \bmuq{\bbeta}) (\bu_i - \bmuq{\bu_i})^T \rbrace = \AU{12,i}, \quad 1 \le i \le m.$$
\end{result}
\begin{result}[Result 4 of Nolan \emph{et al.} (2020), revisited]
\label{res:res_streamlined_3level}
	The MFVB updates of $\bmuq{\bbeta,\bu}$ and each of the sub-blocks of $\bSigmaq{\bbeta,\bu}$ that are relevant for variational inference concerning model \eqref{eq:qstar_gauss_heter_lmm_globallocal} with a three-level random effects specification are expressible as a three-level sparse matrix problem of the form $\bA \bmuq{\bbeta,\bu} = \ba$, where $\ba$ and the non-zero sub-blocks of $\bA$ are, according to the notation in Section \ref{sec:threeLevSMP} of the supplementary material:
	\begin{equation*}
	\begin{array}{c}
		\AL{11} = \muq{1/\sigsq} \displaystyle{\sum_{i=1}^m\sum_{j=1}^{n_i}} (\bX_{ij}^T \bX_{ij}) + \left[\begin{array}{ccc} \bSigma_{\bbeta^\raneff}^{-1} & \bO & \bO \\[1ex] 
	\bO & \bSigma_{\bbeta^\additional}^{-1} & \bO \\[1ex]
	\bO & \bO & \muq{1/\tausq}\: \diag(\bmuq{\bzeta}) \end{array}\right],\\[5ex]
	\ba_1 = \muq{1/\sigsq} \displaystyle{\sum_{i=1}^m \sum_{j=1}^{n_i}} (\bX_{ij}^T \by_{ij} ) + \left[\begin{array}{c} \bSigma_{\bbeta^\raneff}^{-1} \bmu_{\bbeta^\raneff} \\[1ex] \bSigma_{\bbeta^\additional}^{-1} \bmu_{\bbeta^\additional} \\[1ex] \bzero \end{array}\right],\\[5ex]
		\AL{22,i} = \muq{1/\sigsq} \displaystyle{\sum_{j=1}^{n_i}} \left((\bZLone_{ij})^T \bZLone_{ij}\right) + \bM_{q(\bSigmaLoneMinusOne)}, \quad \AL{12,i} = \muq{1/\sigsq} \sum_{j=1}^{n_i} (\bX_{ij})^T \bZLone_{ij},\\[1.5ex]
		\ba_{2,i} = \muq{1/\sigsq} \displaystyle{\sum_{j=1}^{n_i}} (\bZLone_{ij})^T \by_{ij},\quad
\AL{22,ij} = \muq{1/\sigsq} (\bZLtwo_{ij})^T \bZLtwo_{ij} + \bM_{q(\bSigmaLtwoMinusOne)},\\[1.5ex]
	\AL{12,i,j} = \muq{1/\sigsq} (\bZLone_{ij})^T \bZLtwo_{ij}, \quad \AL{12,ij} = \muq{1/\sigsq} \bX_{ij}^T \bZLtwo_{ij}, \quad \ba_{2,ij} = \muq{1/\sigsq} (\bZLtwo_{ij})^T \by_{ij},
	\end{array}
	\end{equation*}
for $1 \le i \le m$ and $1 \le j \le n_i$. Moreover, $\bX_{ij} = [\: \bX_{ij}^\raneff \:\vert\: \bX_{ij}^\additional \:\vert\: \bX_{ij}^\subjtosel \:]$. The $\SolveThreeLevelSparseMatrix$ routine efficiently solves the associated linear system, and provides the solutions: 
$$ \bmuq{\bbeta} = \bx_1, \quad \bSigmaq{\bbeta} = \bA^{11},$$ 
$$ \bmuq{\buLone_i} = \bx_{2,i}, \quad \bSigmaq{\buLone_i} = \AU{22,i}, \quad E_\qDens\lbrace (\bbeta - \bmuq{\bbeta}) (\buLone_i - \bmuq{\buLone_i})^T \rbrace = \AU{12,i}, \quad 1 \le i \le m,$$ 
and 
$$ \bmuq{\buLtwo_{ij}} = \bx_{2,ij}, \quad \bSigmaq{\buLtwo_{ij}} = \AU{22,ij}, \quad E_\qDens\lbrace (\bbeta - \bmuq{\bbeta}) (\buLtwo_{ij} - \bmuq{\buLtwo_{ij}}^T \rbrace = \AU{12,ij},$$ $$E_\qDens\lbrace (\buLone_i - \bmuq{\buLone_i}) (\buLtwo_{ij} - \bmuq{\buLtwo_{ij}})^T \rbrace = \AU{12,i,j}, \quad1\le i \le m,\quad 1 \le j \le n_i.$$
\end{result}
We employ these two results to derive streamlined MFVB algorithms for determining the optimal parameters of the densities in \eqref{eq:qstar_gauss_heter_lmm_globallocal} for multilevel models having two-level and three-level random effects. The former specification is accommodated by Algorithm \ref{alg:streamlinedMFVB_twolevel_globallocal}, the latter by Algorithm \ref{alg:streamlinedMFVB_threelevel_globallocal}. Notice all the sub-blocks of $\bA$ and components of $\ba$ described in Results \ref{res:res_streamlined_2level} and \ref{res:res_streamlined_3level} can be updated simply performing linear transformations of matrices involving multiplications of sub-vectors of $\by$ and sub-matrices of $\bX$ and $\bZ$ which need to be computed only once instead of at each iteration of the associated algorithms.

Similar results are provided in Nolan \emph{et al.} (2020) in terms of \emph{sparse least squares problems} of the type $\lVert \bb - \bB \bmuq{\bbeta,\bu} \rVert^2$, after exploiting the equalities $\bA = \bB^T \bB$ and $\ba = \bB^T \bb$. This class of problems can be solved via efficient QR-decompositions, instead of sparse matrix problems of the type $\bA \bmuq{\bbeta,\bu} = \ba$ which rely upon matrix inversion routines.
Section 2.1 of Nolan \myand Wand (2020) claims that the former class of problems and associated $\SolveTwoLevelSparseLeastSquares$ and $\SolveThreeLevelSparseLeastSquares$ routines proposed in Appendix A of Nolan \emph{et al.} (2020) are numerically preferred to the latter, since QR-decomposition methods are more computationally stable. However, streamlined MFVB approximations based on the QR-decomposition require an additional set of matrices to be used at each algorithm iteration. Efficient QR-decomposition routines and functions for performing matrix multiplications with the associated $\bQ$ and $\bR$ matrices are not available in all standard computing environments and programming languages. Furthermore, the sub-blocks of $\bB$ and components of $\bb$ become particularly sparse for large $p$, as for the cases considered in this work, entailing onerous memory consumption and compromising efficiency of the QR-decomposition. For these reasons, we opt for routines based on sparse matrix linear system problems such as those of Results \ref{res:res_streamlined_2level} and \ref{res:res_streamlined_3level}.

The convergence of Algorithms \ref{alg:streamlinedMFVB_twolevel_globallocal} and   \ref{alg:streamlinedMFVB_threelevel_globallocal} can be assessed in several ways. One way is to monitor the relative increment of the variational lower bound to the marginal log-likelihood having general expression given by the first addend of \eqref{eq:logliksplit}. The algorithm can be stopped when the increment falls below a pre-specified threshold. The drawback is that much tedious algebra is required to derive an explicit lower bound expression. An alternative way is to measure the absolute relative increments among all the parameters updated at each algorithm iteration and stopping when the maximum increment falls below a pre-specified threshold. However, both the approaches can be computationally expensive, especially for large $m$ values, and these convergence checks may significantly slow down the overall streamlined iterations.
Therefore, we suggest letting Algorithms \ref{alg:streamlinedMFVB_twolevel_globallocal} and   \ref{alg:streamlinedMFVB_threelevel_globallocal} run for a reasonable number of iterations fixed in advance and increase the number of iterations if convergence checks suggest that the desired level of convergence has not been achieved. 

A code bundle with \textsf{R} and \textsf{C++} implementations of our algorithms is publicly available in the GitHub repository \textsf{DMTWcode} at \texttt{github.com/lucamaestrini/DMTWcode}.

\begin{algorithm}[H]
  \begin{center}
    \begin{minipage}[t]{155mm}
      \begin{small}
          Data Inputs: $\by_i(o_i\times1), \bX^\raneff_i(o_i\times p_\raneff), \bX^\additional_i(o_i\times p_\additional), \bX^\subjtosel_i(o_i\times p_\subjtosel), \bZ_i (o_i \times q), \ 1\le i\le m$. \\[0.5ex]
          \phantom{Data Inputs:} Build $\bX_i = [\: \bX_i^\raneff \:\vert\: \bX_i^\additional \:\vert\: \bX_i^\subjtosel \:] (o_i \times p)$. \\[0.75ex]
          Global-local prior type choice: \textsf{Laplace}, \textsf{Horseshoe} or \textsf{NEG}.\\[0.75ex]
          Hyperparameter Inputs: $\bmu_{\bbeta^\raneff} (p_\raneff \times 1), \bmu_{\bbeta^\additional} (p_\additional \times 1), \bSigma_{\bbeta^\raneff} (p_\raneff \times p_\raneff) \mbox{ and } \bSigma_{\bbeta^\additional} (p_\additional \times p_\additional)$ \\[0.5ex] \phantom{Hyperparameter Inputs:} both symmetric and positive definite, $\nusigsq, \ssigsq, \stausq, \nuSigma, \sSigmaOne,\ldots,\sSigmaq > 0$. \\ [0.5ex] \phantom{Hyperparameter Inputs:} If \textsf{NEG}: $\lambda>0$.\\[0.75ex]
          Initialize: $\muq{1/\sigsq} > 0, \muq{1/\asigsq} > 0, \muq{1/\tausq} > 0, \muq{1/\atausq} > 0, \bmuq{\bzeta} (p_\subjtosel \times 1), \bmuq{\ba_{\bzeta}} (p_\subjtosel \times 1),$ \\[0.5ex] $\phantom{Initialize} \bMq{\bSigma^{-1}} (q \times q), \bMq{\ASigma^{-1}} (q \times q)$ both symmetric and positive definite.\\[0.75ex]
          $\xiq{\sigsq}\longleftarrow\nusigsq+\sum_{i=1}^m o_i; \quad \xiq{\bSigma}\longleftarrow \nuSigma + m + 2 q -2; \quad \xiq{\asigsq}\longleftarrow\nusigsq+1; \quad \xiq{\ASigma}\longleftarrow\nuSigma+q;$\\[0.75ex]
          $\xiq{\tausq} \longleftarrow p_\subjtosel + 1; \quad \xiq{\atausq} \longleftarrow 2; \quad \bLambda_{\ASigma} \longleftarrow \{\nuSigma\:\diag(\sSigmaOne^2,\ldots,\sSigmaq^2)\}^{-1}.$\\[0.75ex]
          Cycle until convergence:
          \begin{itemize}
          	\item[] Compute $\ba_1, \AL{11}, \lbrace \ba_{2,i}, \AL{22,i}, \AL{12,i} : 1 \le i \le m\rbrace$ with expressions from Result \ref{res:res_streamlined_2level}.
		\item[] $\SscAlgStreamlinedMFVBTwoLevel\longleftarrow\SolveTwoLevelSparseMatrix\Big(\ba_1, \AL{11}, \lbrace \ba_{2,i}, \AL{22,i}, \AL{12,i} : 1 \le i \le m\rbrace\Big).$
		\item[] $\bmuq{\bbeta}\longleftarrow \bx_1 \mbox{ component of } \SscAlgStreamlinedMFVBTwoLevel; \quad \bSigmaq{\bbeta}\longleftarrow \AU{11} \mbox{ component of } \SscAlgStreamlinedMFVBTwoLevel;$
		\item[] $\bmuq{(\bbeta^\subjtosel)^2}\longleftarrow \mbox{diagonal}\left(\bSigmaq{\bbeta^\subjtosel} + \bmuq{\bbeta^\subjtosel} \bmuq{\bbeta^\subjtosel}^T\right);$
		\item[] $\lambdaq{\sigsq}\longleftarrow\muq{1/\asigsq}; \quad \bLambdaq{\bSigma}\longleftarrow\bMq{\ASigma^{-1}}.$
		\item[] For $i = 1,\ldots, m$:
		\begin{itemize}
			\item[] $\bmuq{\bu_i}\longleftarrow \bx_{2,i} \mbox{ component of } \SscAlgStreamlinedMFVBTwoLevel; \quad \bSigmaq{\bu_i}\longleftarrow \AU{22,i} \mbox{ component of } \SscAlgStreamlinedMFVBTwoLevel;$
			\item[] $E_\qDens\lbrace(\bbeta-\bmuq{\bbeta})(\bu_i-\bmuq{\bu_i})^T\rbrace\longleftarrow \AU{12,i} \mbox{ component of }\SscAlgStreamlinedMFVBTwoLevel$;
			\item[] $\lambdaq{\sigsq}\longleftarrow\lambdaq{\sigsq} + \big\lVert \by_i - \bX_i \bmuq{\bbeta} - \bZ_i \bmuq{\bu_i} \big\rVert^2 + \tr(\bX_i^T \bX_i \bSigmaq{\bbeta}) + \tr(\bZ_i^T \bZ_i \bSigmaq{\bu_i}) \\[0.5ex] 
			\phantom{\lambda_{\qDens(1/\sigsq)}\longleftarrow} + 2\tr[\bZ_i^T \bX_i E_\qDens\lbrace(\bbeta-\bmuq{\bbeta})(\bu_i-\bmuq{\bu_i})^T\rbrace];$
			\item[] $\bLambdaq{\bSigma}\longleftarrow \bLambdaq{\bSigma} + \bmuq{\bu_i}\bmuq{\bu_i}^T + \bSigmaq{\bu_i};$
		\end{itemize}
		\item[] $\muq{1/\sigsq} \longleftarrow \xiq{\sigsq}/\lambdaq{\sigsq}; \quad \bMq{\bSigma^{-1}} \longleftarrow (\xiq{\bSigma} - q + 1) \bLambdaq{\bSigma}^{-1};$
		\item[] $\lambdaq{\asigsq}\longleftarrow\muq{1/\sigsq}+1/(\nusigsq\ssigsq^2); \quad \muq{1/\asigsq} \longleftarrow \xiq{\asigsq}/\lambdaq{\asigsq};$ 
		\item[] $\lambdaq{\tausq}\longleftarrow\muq{1/\atausq} + \bmuq{\bzeta}^T \bmuq{(\bbeta^\subjtosel)^2}; \quad \muq{1/\tausq}\longleftarrow\xiq{\tausq}/\lambdaq{\tausq};$
		\item[] $\lambdaq{\atausq}\longleftarrow\muq{1/\tausq} + 1/\stausq^2; \quad \muq{1/\atausq}\longleftarrow\xiq{\atausq}/\lambdaq{\atausq};$
		\item[] $\boldsymbol{g}\longleftarrow \frac{1}{2} \muq{1/\tausq} \bmuq{(\bbeta^\subjtosel)^2};$
		\item[] If \textsf{Laplace}: $\bmuq{\bzeta} \longleftarrow \sqrt{\boldsymbol{1}/(2\boldsymbol{g})};$
		\item[] If \textsf{Horseshoe}: $\boldsymbol{\lambda}_{\qDens(\bzeta)} \longleftarrow \bmuq{\ba_{\bzeta}} + \boldsymbol{g}; \quad \bmuq{\bzeta}\longleftarrow \boldsymbol{1}/\boldsymbol{\lambda}_{\qDens(\bzeta)}; \\[0.5ex]\phantom{If \emph{HorseShoe}\:} \boldsymbol{\lambda}_{\qDens(\ba_{\bzeta})} \longleftarrow \bmuq{\bzeta} + \boldsymbol{1}; \quad \bmuq{\ba_{\bzeta}}\longleftarrow \boldsymbol{1}/\boldsymbol{\lambda}_{\qDens(\ba_{\bzeta})};$
		\item[] If \textsf{NEG}: $\boldsymbol{\lambda}_{\qDens(\bzeta)} \longleftarrow 2\bmuq{\ba_{\bzeta}}; \quad \bmuq{\bzeta}\longleftarrow \sqrt{\boldsymbol{\lambda}_{\qDens(\bzeta)} / (2\boldsymbol{g})}; \quad \bmuq{\boldsymbol{1}/\bzeta} \longleftarrow \boldsymbol{1}/\bmuq{\bzeta} + \boldsymbol{1}/(2\bmuq{\ba_{\bzeta}}); \\[0.5ex]\phantom{If \emph{NEG}\:} \boldsymbol{\lambda}_{\qDens(\ba_{\bzeta})} \longleftarrow \bmuq{\boldsymbol{1}/\bzeta} + \boldsymbol{1}; \quad \bmuq{\ba_{\bzeta}}\longleftarrow (\lambda+1)(\boldsymbol{1}/\boldsymbol{\lambda}_{\qDens(\ba_{\bzeta})});$
		\item[] $\bLambdaq{\ASigma}\longleftarrow 
\diag\big\{\mbox{diagonal}\big(\bMq{\bSigma^{-1}}\big)\big\}+\bLambda_{\ASigma};\quad \bMq{\ASigma^{-1}}\longleftarrow\xiq{\ASigma}\bLambda_{\qDens(\ASigma)}^{-1}.$
	\end{itemize}
	Outputs: $\bmuq{\bbeta}, \bSigmaq{\bbeta}, \left\lbrace \bmuq{\bu_i}, \bSigmaq{\bu_i}, E_\qDens\lbrace (\bbeta - \bmuq{\bbeta})(\bu_i - \bmuq{\bu_i})^T \rbrace : 1 \le i \le m \right\rbrace,$\\[0.5em]
	\phantom{Outputs:} $\xiq{\sigsq}, \lambdaq{\sigsq}, \xiq{\asigsq}, \lambdaq{\asigsq}, \xiq{\tausq}, \lambdaq{\tausq}, \xiq{\atausq}, \lambdaq{\atausq},$\\[0.5em]
	\phantom{Outputs:} $\mbox{if \textsf{Laplace}: } \bmuq{\bzeta}, \mbox{ if \textsf{Horseshoe}: } \left\lbrace\boldsymbol{\lambda}_{\qDens(\bzeta)}, \boldsymbol{\lambda}_{\qDens(\ba_{\bzeta})}\right\rbrace, \mbox{ if \textsf{NEG}: } \left\lbrace\bmuq{\bzeta}, \boldsymbol{\lambda}_{\qDens(\bzeta)}, \boldsymbol{\lambda}_{\qDens(\ba_{\bzeta})}\right\rbrace,$\\[0.5em]
	\phantom{Outputs:} $\xiq{\bSigma}, \bLambdaq{\bSigma}, \xiq{\ASigma}, \bLambda_{\qDens(\ASigma)}.$
        \end{small}
      \end{minipage}
    \end{center}
 \caption{\it Streamlined algorithm for obtaining the mean field variational Bayes approximate posterior density functions \eqref{eq:qstar_gauss_heter_lmm_globallocal} for the parameters of the linear mixed model \eqref{eq:gauss_heter_lmm_penalized} with the two-level random effects specification \eqref{eq:gauss_heter_lmm_2level}. The approximation is based on the mean-field density restriction \eqref{eq:prodrestr_globallocal}.}
\label{alg:streamlinedMFVB_twolevel_globallocal}
\end{algorithm}
\begin{algorithm}[H]
  \begin{center}
    \begin{minipage}[t]{155mm}
      \begin{small}
          Data Inputs: $\by_{ij}(o_{ij}\times1), \bX^\raneff_{ij}(o_{ij}\times p_\raneff), \bX^\additional_{ij}(o_{ij}\times p_\additional), \bX^\subjtosel_{ij}(o_{ij}\times p_\subjtosel), \\[0.5ex]
          \phantom{Data Inputs:} \bZLone_{ij} (o_{ij} \times \qLone), \bZLtwo_{ij} (o_{ij} \times \qLtwo), \ 1\le i\le m, \ 1 \le j \le n_i$. \\[0.5ex]
          \phantom{Data Inputs:} Build $\bX_{ij} = [\: \bX_{ij}^\raneff \:\vert\: \bX_{ij}^\additional \:\vert\: \bX_{ij}^\subjtosel \:] (n_i \times p)$. \\[0.75ex]
          Global-local prior type choice: \textsf{Laplace}, \textsf{Horseshoe} or \textsf{NEG}.\\[0.75ex]
          Hyperparameter Inputs: $\bmu_{\bbeta^\raneff} (p_\raneff \times 1), \bmu_{\bbeta^\additional} (p_\additional \times 1), \bSigma_{\bbeta^\raneff} (p_\raneff \times p_\raneff) \mbox{ and } \bSigma_{\bbeta^\additional} (p_\additional \times p_\additional)$ \\[0.5ex] \phantom{Hyperparameter Inputs:} both symmetric and positive definite, $\nusigsq, \ssigsq, \stausq, \nuSigmaLone, \nuSigmaLtwo,\\ [0.5ex] \phantom{Hyperparameter Inputs:}  \sSigmaOneLone,\ldots,\sSigmaqLone, \sSigmaOneLtwo,\ldots,\sSigmaqLtwo > 0$. If \textsf{NEG}: $\lambda>0$.\\[0.75ex]
          Initialize: $\muq{1/\sigsq} > 0, \muq{1/\asigsq} > 0, \muq{1/\tausq} > 0, \muq{1/\atausq} > 0, \bmuq{\bzeta} (p_\subjtosel \times 1), \bmuq{\ba_{\bzeta}} (p_\subjtosel \times 1),$ \\[0.5ex] 
          \phantom{Initialize:} $\bMq{\bSigmaLoneMinusOne} (\qLone \times \qLone), \bMq{\bSigmaLtwoMinusOne} (\qLtwo \times \qLtwo), \bMq{\ASigmaLone^{-1}} (\qLone \times \qLone), \bMq{\ASigmaLtwo^{-1}} (\qLtwo \times \qLtwo)$ \\
          \phantom{Initialize:} all symmetric and positive definite.\\[0.75ex]
          $\xiq{\sigsq}\longleftarrow\nusigsq+\sum_{i=1}^m\sum_{j=1}^{n_i} o_{ij}; \quad \xiq{\bSigmaLone}\longleftarrow \nuSigmaLone + m + 2 \qLone -2; \quad \xiq{\bSigmaLtwo}\longleftarrow \nuSigmaLtwo + \sum_{i=1}^m n_i + 2 \qLtwo -2;\\[0.75ex]
          \xiq{\asigsq}\longleftarrow\nusigsq+1; \quad \xiq{\ASigmaLone}\longleftarrow\nuSigmaLone+\qLone; \quad \xiq{\ASigmaLtwo}\longleftarrow\nuSigmaLtwo+\qLtwo;$\\[0.75ex]
          $\xiq{\tausq} \longleftarrow p_\subjtosel + 1; \quad \xiq{\atausq} \longleftarrow 2; \\[0.75ex] \bLambda_{\ASigmaLone} \longleftarrow \{\nuSigmaLone\:\diag(\sSigmaOneLone^2,\ldots,\sSigmaqLone^2)\}^{-1}; \quad \bLambda_{\ASigmaLtwo} \longleftarrow \{\nuSigmaLtwo\:\diag(\sSigmaOneLtwo^2,\ldots,\sSigmaqLtwo^2)\}^{-1}.$\\[0.75ex]
          Cycle until convergence:
          \begin{itemize}
          	\item[] Compute $\ba_1, \AL{11}, \lbrace \ba_{2,i}, \AL{22,i}, \AL{12,i} : 1 \le i \le m\rbrace,
		\lbrace \ba_{2,ij}, \AL{22,ij}, \AL{12,ij}, \AL{12,i,j} : \\[0.5ex] 1 \le i \le m, 1 \le j \le n_i \rbrace$ with expressions from Result \ref{res:res_streamlined_3level}.
		\item[] $\SscAlgStreamlinedMFVBThreeLevel\longleftarrow\SolveThreeLevelSparseMatrix\Big(\ba_1, \AL{11}, \lbrace \ba_{2,i}, \AL{22,i}, \AL{12,i} : 1 \le i \le m\rbrace,$\\[-2ex]
		\item[] $\qquad\qquad\lbrace \ba_{2,ij}, \AL{22,ij}, \AL{12,ij}, \AL{12,i,j} : 1 \le i \le m, 1 \le j \le n_i \rbrace\Big).$
		\item[] $\bmuq{\bbeta}\longleftarrow \bx_1 \mbox{ component of } \SscAlgStreamlinedMFVBThreeLevel; \quad \bSigmaq{\bbeta}\longleftarrow \AU{11} \mbox{ component of } \SscAlgStreamlinedMFVBThreeLevel;$
		\item[] $\bmuq{(\bbeta^\subjtosel)^2}\longleftarrow \mbox{diagonal}\left(\bSigmaq{\bbeta^\subjtosel} + \bmuq{\bbeta^\subjtosel} \bmuq{\bbeta^\subjtosel}^T\right);$
		\item[] $\lambdaq{\sigsq}\longleftarrow\muq{1/\asigsq}; \quad \bLambdaq{\bSigmaLone}\longleftarrow\bMq{\ASigmaLone^{-1}}; \quad \bLambdaq{\bSigmaLtwo}\longleftarrow\bMq{\ASigmaLtwo^{-1}}.$
		\item[] For $i = 1,\ldots, m$:
		\begin{itemize}
			\item[] $\bmuq{\buLone_i}\longleftarrow \bx_{2,i} \mbox{ component of } \SscAlgStreamlinedMFVBThreeLevel; \quad \bSigmaq{\buLone_i}\longleftarrow \AU{22,i} \mbox{ component of } \SscAlgStreamlinedMFVBThreeLevel;$
			\item[] $E_\qDens\lbrace(\bbeta-\bmuq{\bbeta})(\buLone_i-\bmuq{\buLone_i})^T\rbrace\longleftarrow \AU{12,i} \mbox{ component of }\SscAlgStreamlinedMFVBThreeLevel;$
			\item[] $\bLambdaq{\bSigmaLone}\longleftarrow\bLambdaq{\bSigmaLone} + \bmuq{\buLone_i}\bmuq{\buLone_i}^T + \bSigmaq{\buLone_i}.$
			\item[] For $j=1,\ldots,n_i$:
			\begin{itemize}		
				\item[] $\bmuq{\buLtwo_{ij}}\longleftarrow \bx_{2,ij} \mbox{ component of } \SscAlgStreamlinedMFVBThreeLevel; \quad \bSigmaq{\buLtwo_{ij}}\longleftarrow \AU{22,ij} \mbox{ component of } \SscAlgStreamlinedMFVBThreeLevel;$
				\item[] $E_\qDens\lbrace(\bbeta-\bmuq{\bbeta})(\buLtwo_{ij}-\bmuq{\buLtwo_{ij}})^T\rbrace\longleftarrow \AU{12,ij} \mbox{ component of }\SscAlgStreamlinedMFVBThreeLevel;$
				\item[] $E_\qDens\lbrace(\buLone_i-\bmuq{\buLone_i})(\buLtwo_{ij}-\bmuq{\buLtwo_{ij}})^T\rbrace\longleftarrow \AU{12,i,j} \mbox{ component of }\SscAlgStreamlinedMFVBThreeLevel;$
				\item[] $\lambdaq{\sigsq}\longleftarrow\lambdaq{\sigsq} + \big\lVert \by_{ij} - \bX_{ij} \bmuq{\bbeta} - \bZLone_{ij} \bmuq{\buLone_i} - \bZLtwo_{ij} \bmuq{\buLtwo_{ij}} \big\rVert^2 \\[0.5ex]
				 \phantom{\lambdaq{\sigsq}\longleftarrow} + \tr(\bX_{ij}^T \bX_{ij} \bSigmaq{\bbeta}) + \tr((\bZLone_{ij})^T \bZLone_{ij} \bSigmaq{\buLone_i}) + \tr((\bZLtwo_{ij})^T \bZLtwo_{ij} \bSigmaq{\buLtwo_{ij}}) \\[0.5ex]
				 \phantom{\lambdaq{\sigsq}\longleftarrow} + 2\tr[(\bZLone_{ij})^T \bX_{ij} E_\qDens\lbrace(\bbeta-\bmuq{\bbeta})(\buLone_i-\bmuq{\buLone_i})^T\rbrace] \\[0.5ex]
				 \phantom{\lambdaq{\sigsq}\longleftarrow} + 2\tr[(\bZLtwo_{ij})^T \bX_{ij} E_\qDens\lbrace(\bbeta-\bmuq{\bbeta})(\buLtwo_{ij}-\bmuq{\buLtwo_{ij}})^T\rbrace] \\[0.5ex]
				 \phantom{\lambdaq{\sigsq}\longleftarrow} + 2\tr[(\bZLone_{ij})^T \bZLtwo_{ij} E_\qDens\lbrace(\buLone_i-\bmuq{\buLone_i})(\buLtwo_{ij}-\bmuq{\buLtwo_{ij}})^T\rbrace];$
				\item[] $\bLambdaq{\bSigmaLtwo}\longleftarrow\bLambdaq{\bSigmaLtwo} + \bmuq{\buLtwo_{ij}}\bmuq{\buLtwo_{ij}}^T + \bSigmaq{\buLtwo_{ij}}.$
				\item[] \emph{continued on a subsequent page \ldots}
			\end{itemize}
		\end{itemize}
	\end{itemize}
	\end{small}
      \end{minipage}
    \end{center}
 \caption{\it Streamlined algorithm for obtaining the mean field variational Bayes approximate posterior density functions \eqref{eq:qstar_gauss_heter_lmm_globallocal} for the parameters of the linear mixed model \eqref{eq:gauss_heter_lmm_penalized} with the three-level random effects specification \eqref{eq:gauss_heter_lmm_3level}. The approximation is based on the mean-field density restriction \eqref{eq:prodrestr_globallocal}. The algorithm description requires more than one page and is continued on a subsequent page.}
\label{alg:streamlinedMFVB_threelevel_globallocal}
\end{algorithm}
\setcounter{algorithm}{2}
\begin{algorithm}[H]
  \begin{center}
    \begin{minipage}[t]{155mm}
      \begin{small}
      	\begin{itemize}
		\item[] $\muq{1/\sigsq} \longleftarrow \xiq{\sigsq}/\lambdaq{\sigsq};$
		\item[] $\bMq{\bSigmaLoneMinusOne}\longleftarrow(\xiq{\bSigmaLone}-\qLone+1) \bLambda^{-1}_{\qDens(\bSigmaLone)}; \quad \bMq{\bSigmaLtwoMinusOne}\longleftarrow(\xiq{\bSigmaLtwo}-\qLtwo+1) \bLambda^{-1}_{\qDens(\bSigmaLtwo)};$
		\item[] $\lambdaq{\asigsq}\longleftarrow\muq{1/\sigsq}+1/(\nusigsq\ssigsq^2); \quad \muq{1/\asigsq} \longleftarrow \xiq{\asigsq}/\lambdaq{\asigsq};$ 
		\item[] $\lambdaq{\tausq}\longleftarrow\muq{1/\atausq} + \bmuq{\bzeta}^T \bmuq{(\bbeta^\subjtosel)^2}; \quad \muq{1/\tausq}\longleftarrow\xiq{\tausq}/\lambdaq{\tausq};$
		\item[] $\lambdaq{\atausq}\longleftarrow\muq{1/\tausq} + 1/\stausq^2; \quad \muq{1/\atausq}\longleftarrow\xiq{\atausq}/\lambdaq{\atausq};$
		\item[] $\boldsymbol{g}\longleftarrow \frac{1}{2} \muq{1/\tausq} \bmuq{(\bbeta^\subjtosel)^2};$
		\item[] If \textsf{Laplace}: $\bmuq{\bzeta} \longleftarrow \sqrt{\boldsymbol{1}/(2\boldsymbol{g})};$
		\item[] If \textsf{Horseshoe}: $\boldsymbol{\lambda}_{\qDens(\bzeta)} \longleftarrow \bmuq{\ba_{\bzeta}} + \boldsymbol{g}; \quad \bmuq{\bzeta}\longleftarrow \boldsymbol{1}/\boldsymbol{\lambda}_{\qDens(\bzeta)}; \\[0.5ex]\phantom{If \textsf{HorseShoe}\:} \boldsymbol{\lambda}_{\qDens(\ba_{\bzeta})} \longleftarrow \bmuq{\bzeta} + \boldsymbol{1}; \quad \bmuq{\ba_{\bzeta}}\longleftarrow \boldsymbol{1}/\boldsymbol{\lambda}_{\qDens(\ba_{\bzeta})};$
		\item[] If \textsf{NEG}: $\boldsymbol{\lambda}_{\qDens(\bzeta)} \longleftarrow 2\bmuq{\ba_{\bzeta}}; \quad \bmuq{\bzeta}\longleftarrow \sqrt{\boldsymbol{\lambda}_{\qDens(\bzeta)} / (2\boldsymbol{g})}; \quad \bmuq{\boldsymbol{1}/\bzeta} \longleftarrow \boldsymbol{1}/\bmuq{\bzeta} + \boldsymbol{1}/(2\bmuq{\ba_{\bzeta}}); \\[0.5ex]\phantom{If \textsf{NEG}\:} \boldsymbol{\lambda}_{\qDens(\ba_{\bzeta})} \longleftarrow \bmuq{\boldsymbol{1}/\bzeta} + \boldsymbol{1}; \quad \bmuq{\ba_{\bzeta}}\longleftarrow (\lambda+1)(\boldsymbol{1}/\boldsymbol{\lambda}_{\qDens(\ba_{\bzeta})});$
		\item[] $\bLambdaq{\ASigmaLone}\longleftarrow 
\diag\big\{\mbox{diagonal}\big(\bMq{\bSigmaLoneMinusOne}\big)\big\}+\bLambda_{\ASigmaLone};\quad \bMq{(\ASigmaLone)^{-1}}\longleftarrow\xiq{\ASigmaLone}\bLambdaq{\ASigmaLone}^{-1}.$
		\item[] $\bLambdaq{\ASigmaLtwo}\longleftarrow 
\diag\big\{\mbox{diagonal}\big(\bMq{\bSigmaLtwoMinusOne}\big)\big\}+\bLambda_{\ASigmaLtwo};\quad \bMq{(\ASigmaLtwo)^{-1}}\longleftarrow\xiq{\ASigmaLtwo}\bLambdaq{\ASigmaLtwo}^{-1}.$
	\end{itemize}
	Outputs: $\bmuq{\bbeta}, \bSigmaq{\bbeta}, \left\lbrace \bmuq{\buLone_i}, \bSigmaq{\buLone_i}, E_\qDens\lbrace (\bbeta - \bmuq{\bbeta})(\buLone_i - \bmuq{\buLone_i})^T \rbrace : 1 \le i \le m \right\rbrace,$ \\[0.5em]
	\phantom{Outputs:} $\left\lbrace \bmuq{\buLtwo_{ij}}, \bSigmaq{\buLtwo_{ij}}, E_\qDens\lbrace (\bbeta - \bmuq{\bbeta})(\buLtwo_{ij} - \bmuq{\buLtwo_{ij}})^T \rbrace : 1 \le i \le m, 1 \le j \le n_i \right\rbrace,$ \\[0.5em]
	\phantom{Outputs:} $\xiq{\sigsq}, \lambdaq{\sigsq}, \xiq{\asigsq}, \lambdaq{\asigsq}, \xiq{\tausq}, \lambdaq{\tausq}, \xiq{\atausq}, \lambdaq{\atausq},$\\[0.5em]
	\phantom{Outputs:} $\mbox{ if \textsf{Laplace}: } \bmuq{\bzeta}, \mbox{if \textsf{Horseshoe}: } \left\lbrace\boldsymbol{\lambda}_{\qDens(\bzeta)}, \boldsymbol{\lambda}_{\qDens(\ba_{\bzeta})}\right\rbrace, \mbox{ if \textsf{NEG}: } \left\lbrace\bmuq{\bzeta}, \boldsymbol{\lambda}_{\qDens(\bzeta)}, \boldsymbol{\lambda}_{\qDens(\ba_{\bzeta})}\right\rbrace,$\\[0.5em]
	\phantom{Outputs:} $\xiq{\bSigmaLone}, \bLambdaq{\bSigmaLone}, \xiq{\bSigmaLtwo}, \bLambdaq{\bSigmaLtwo}, \xiq{\ASigmaLone}, \bLambdaq{\ASigmaLone}, \xiq{\ASigmaLtwo}, \bLambdaq{\ASigmaLtwo}.$
        \end{small}
      \end{minipage}
    \end{center}
\caption{\textbf{continued.} \it This is a continuation of the description of this algorithm that commences on a preceding page.}
\end{algorithm}

\newpage

\section{Simulation Study Investigations\label{sec:Assessments}}

We discuss the results of a simulation study conducted to assess: 
\begin{enumerate}
	\item the accuracy of the optimal $\qDens^*$-density approximations compared to the marginal posterior density functions obtained via MCMC, when global-local priors are specified;
	\item fixed effects selection performances via the SAVS procedure for effectively discriminating the relevant fixed effects from those being irrelevant;
	\item computational timings and memory storage requirements for both na{\"i}ve and  streamlined algorithm implementations, especially when the number of parameters increases. 
\end{enumerate}
Notice that the accuracy and variable selection performances are not affected by the use of streamlined updates in place of na{\"i}ve counterparts, given that both the implementations converge to the same solution.

The simulation study was performed on a \textsf{MacBook Pro} laptop with a 1.4 gigahertz processor and 8 gigabytes of random access memory. To allow for maximal speed, both the MFVB algorithms and corresponding MCMC sampling schemes were implemented in \textsf{C++} employing the \textsf{Armadillo} library (Sanderson \myand Curtin, 2016) and executed into an \textsf{R} environment using the \textsf{RcppArmadillo} package (Eddelbuettel \myand Sanders, 2014).

The simulation study focused on three-level random effect models, which give rise to more complex three-level sparse matrix structures. We simulated 50 datasets according to model specification \eqref{eq:gauss_heter_lmm_penalized}  with $m=100$ groups, each with $n_i = 15$ sub-groups, each having $o_{ij}=20$ units, for $1 \le i \le 100, 1 \le j \le 15$. We included a random intercept and a random slope for both the group and sub-group levels $(\qLone = \qLtwo = p_\raneff = 2)$ with true parameter values being $\bbeta^\raneff = [0.58\,\, 1.98]^T$, and $p_\additional = 3$ additional fixed effects having true parameter values $\bbeta^\additional = [0.7\,\, -0.9\,\, 1.8]^T$. We also considered a sparse design setting with $p_\subjtosel = 50$ fixed effects such that the first 10 of them were $\beta_{1}^\subjtosel = 1.91, \beta_{2}^\subjtosel = 1.96, \beta_{3}^\subjtosel = -0.10, \beta_{4}^\subjtosel = 1.62, \beta_{5}^\subjtosel = -1.45, \beta_{6}^\subjtosel = -1.53, \beta_{7}^\subjtosel = 0.24, \beta_{8}^\subjtosel = 1.76, \beta_{9}^\subjtosel = 1.79$ and $\beta_{10}^\subjtosel = -0.15$, while the remaining 40 were assumed to be irrelevant and hence the corresponding parameters had true value equal to zero $(\beta_h^\subjtosel = 0 \text{ for each } 11 \le h \le 50)$. The true variance parameter $\sigsq$ was fixed to 0.7. The random effects vectors $\buLone_i$ and $\buLtwo_{ij}$ were respectively generated independently from $\mbox{N}(\bzero, \bSigmaLone)$ and $\mbox{N}(\bzero, \bSigmaLtwo)$ distributions, having
$$
\bSigmaLone = \left[\begin{array}{cc} 0.42 & -0.09 \\ -0.09 & 0.52 \end{array}\right] \qquad\mbox{and}\qquad \bSigmaLtwo = \left[\begin{array}{cc} 0.80 & -0.24 \\ -0.24 & 0.75 \end{array}\right].
$$
For each data replication, the slope-associated column of $\bX^{\raneff}$ was generated from a standard Gaussian distribution, while the rows of $\bX^\additional$ and $\bX^\subjtosel$ were generated from two multivariate Gaussian distributions having zero mean vector and covariance matrices generated from $\mbox{Wishart}(p_\additional, \bI)$ and $\mbox{Wishart}(p_\subjtosel, \bI)$ distributions, respectively. This strategy produces covariates with different variability and non-zero correlations that better mimic a real-data scenario. 
The setting embeds a fixed effects design matrix $\bX$ of size 30,000 $\times$ 55, and a sparse random effects design matrix $\bZ$ of size 30,000 $\times$ 3,200 with 96 millions cells, of which $99.875\%$ are zeros.

We proceed fitting model \eqref{eq:gauss_heter_lmm_penalized} for each data replication using diffuse priors with hyperparameters $\bmu_{\bbeta^\raneff} = \bmu_{\bbeta^\additional} = \bzero$, $\bSigma_{\bbeta^\raneff} = \bSigma_{\bbeta^\additional} = 10^{10}\bI$, $\nusigsq = 1$, $\nuSigmaLone = \nuSigmaLtwo = 2$, $\ssigsq = s_{\mbox{\tiny{$\bSigmaLone,1$}}} = s_{\mbox{\tiny{$\bSigmaLone,2$}}} = s_{\mbox{\tiny{$\bSigmaLtwo,1$}}} = s_{\mbox{\tiny{$\bSigmaLtwo,2$}}} = 10^5$. Without loss of generality, a Horseshoe prior for all the elements of $\bbeta^\subjtosel$ have been specified, with $\stausq=10^5$ to limit prior information about the degree of sparsity.
Along all the simulation study, the MFVB approximations were obtained by running 200 iterations of the streamlined or na{\"i}ve algorithms, i.e. four times the number of iterations used in the numerical experiments of Nolan \textit{et al.} (2020).
This number of MFVB iterations was chosen to guarantee appropriate convergence of the variational algorithms, as well as a fair computational time comparison between na{\"i}ve and streamlined MFVB. Maestrini and Wand (2021) suggest running variational algorithms until the relative change in the variational parameters reaches a certain threshold. In our numerical experiments, the 200 MFVB iterations guaranteed that the relative change in the variational parameter estimates was smaller that $10^{-3}$ and for this or smaller values the accuracy and fixed effect selection performance of MFVB was not impacted.
Variational algorithm convergence could also be assessed by monitoring the lower bound at every iteration or after a certain number of iterations; however, deriving the variational lower bound is subject to calculation errors and its computation can be time consuming.

\subsection{Accuracy Assessment\label{subsec:AccuracyAssessment}}

We measured the quality of the variational approximations using the accuracy index proposed in Section 3.4 of Faes \emph{et al.} (2011). For a generic univariate parameter $\theta \in \Theta$, this is defined as
\begin{equation}
\mbox{Accuracy}(\theta) \equiv \left(1 -\dfrac{1}{2} \int_{\Theta} \left\vert \qDens^*(\theta) - \pDens(\theta \vert \by) \right\vert \mathrm{d}\theta \right)\%,
\label{eq:accuracy}
\end{equation}
where $\pDens(\theta \vert \by)$ is the marginal posterior density of $\theta$ and $\qDens^*(\theta)$ is the associated optimal approximating density function obtained via MFVB. This index takes values between 0\% and 100\%, with a score of 100\% indicating perfect matching between $\qDens^*(\theta)$ and $\pDens(\theta\vert\by)$, and 0\% if the densities have no overlapping mass. In practice, computation of the true $\pDens(\theta \vert \by)$ is numerically challenging, so we employed binned kernel density estimation with direct plug-in bandwidth selection (e.g. Section 3.6.1 of Wand \myand Jones, 1995) and applied it to the MCMC samples using the \textsf{R} package \textsf{KernSmooth} (Wand, 2020). The MCMC samplings were performed using a warmup of length 5,000 followed by 100,000 iterations, to which we applied a thinning value of 20. 
The integration in \eqref{eq:accuracy} was then accurately performed via trapezoidal numerical quadrature.
\begin{figure}[b!]
	\centering
	\includegraphics[width=0.95\textwidth]{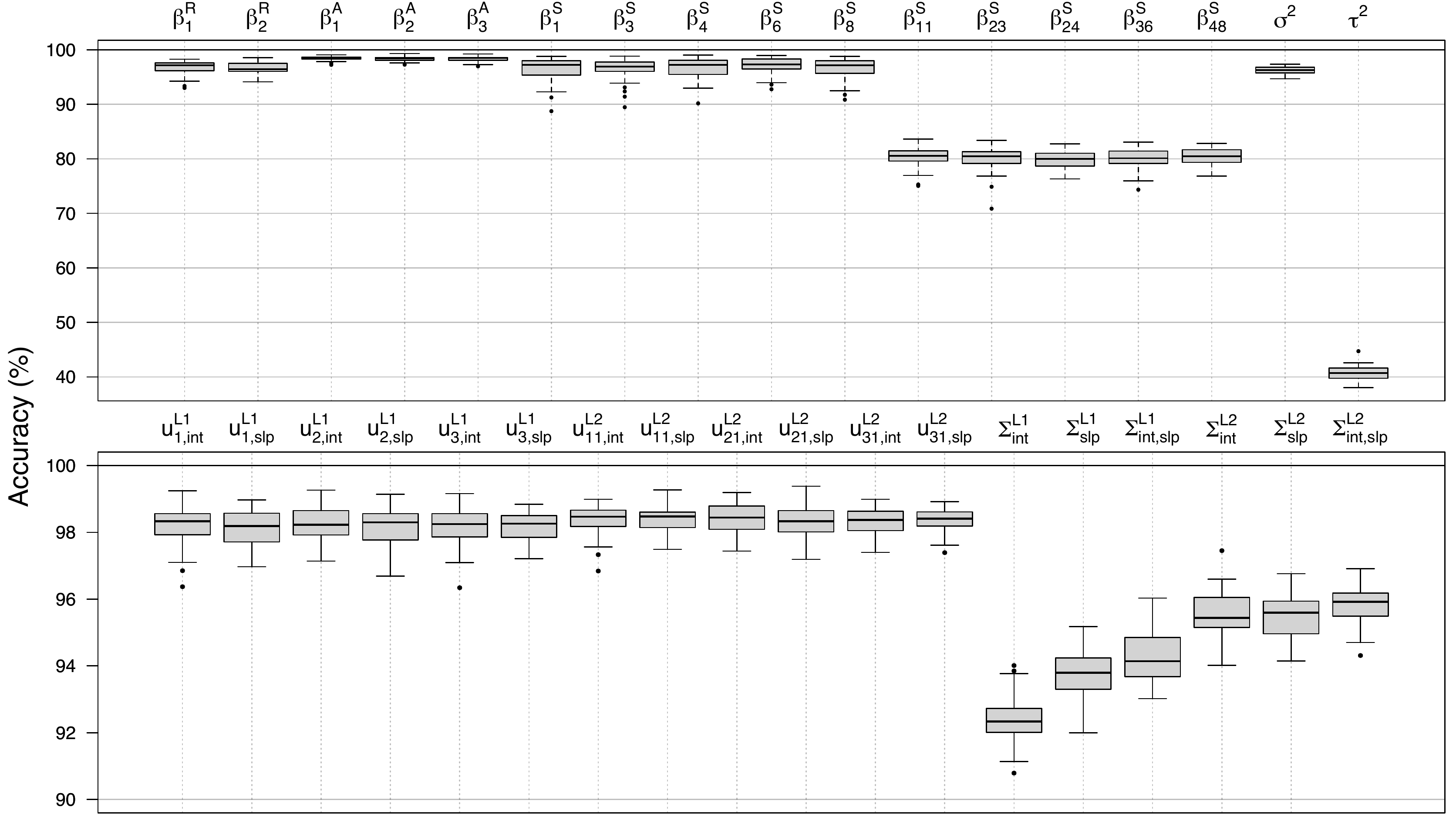}
	\caption{\it Side-by-side boxplots of the accuracy scores from the simulation study for a selection of model parameters and random effects. Outliers are displayed as solid points.}
	\label{fig:accuracy_boxplots}
\end{figure} 
Figure \ref{fig:accuracy_boxplots} shows the accuracy results through boxplots for the 50 data replicates. The boxplots refer to all the entries of $\bbeta^\raneff$ and $\bbeta^\additional$, to 10 elements of $\bbeta^\subjtosel$ chosen such that 5 of them have non-zero values ($\beta^\subjtosel_1$, $\beta^\subjtosel_3$, $\beta^\subjtosel_4$, $\beta^\subjtosel_6$ and $\beta^\subjtosel_8$) and 5 are null ($\beta^\subjtosel_{11}$, $\beta^\subjtosel_{23}$, $\beta^\subjtosel_{24}$, $\beta^\subjtosel_{36}$ and $\beta^\subjtosel_{48}$), the intercept- and slope-associated elements of $\buLone_i$ and $\buLtwo_{ij}$, for $1 \le i \le 3$ and $j=1$, the entries of $\bSigmaLone$ and $\bSigmaLtwo$, $\sigsq$, and $\tausq$. The intercept- and slope-associated parameters are identified by the subscripts ``int'' and ``slp'', respectively. 

Variational approximations showed high accuracy scores for all the model parameters considered and across the different data replications. All the fixed effects subject to selection and having non-zero true values exhibited accuracy scores greater than 90\%. In contrast, those having true values equal to zero had lower accuracy scores between 75\% and 85\% due to the spiky marginal posterior densities being approximated by Gaussian variational densities. All the other fixed effects parameters, random effects and variance parameters showed accuracy scores greater than 90\%. The accuracy scores of the global variance parameter $\tausq$ were under 50\%. Despite not directly shown, all the $\qDens^*(\zeta_h)$'s had accuracy scores between 75\% and 80\%. Similar results were obtained for the alternative global-local prior specifications.
\begin{figure}[t!]
	\centering
	\includegraphics[width=0.95\textwidth]{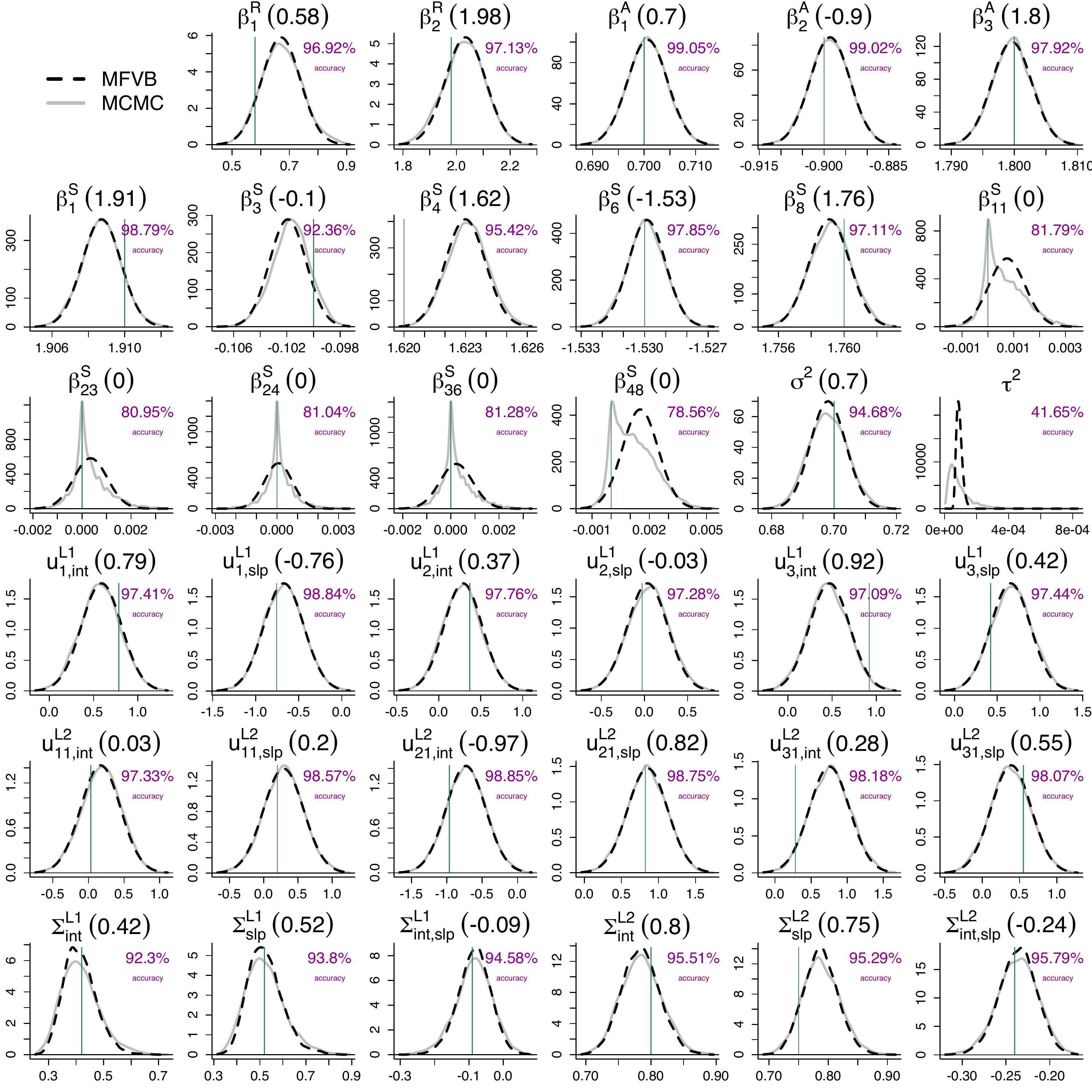}
	\caption{\it Approximate posterior density functions of some of the three-lever random effects model parameters obtained from the first replication of the simulation study. Each plot shows the optimal approximate posterior density function $\qDens^*(\theta)$ obtained via MFVB (black dashed curves) and the MCMC-based $\pDens(\theta\vert\by)$ densities (grey curves). A vertical line indicates the parameter true value. The percentages of accuracy are also provided.}
\label{fig:accuracy_densities}
\end{figure}

Figure \ref{fig:accuracy_densities} displays the MFVB and MCMC approximate posterior densities, together with the associated accuracy scores, obtained for the first data replication of the simulation study. 
These plots allow to visually assess the quality of the approximation and show that the approximate posterior density functions are generally concentrated around the true parameter values.
Notice that the global-local prior shrinks the MCMC marginal posterior densities of the $\beta^\subjtosel$-parameters having null true value towards zero. The corresponding $\qDens^*$-densities are Gaussian and, although they are not able to capture peak and tail behaviors, they provide satisfactory approximations. Notice also that the $\tausq$ approximate posterior densities are concentrated around a very small value, as expected for the sparse design setting we considered in the simulation study.

\subsection{Fixed Effects Selection Assessment\label{subsec:VSAssessment}}

We assessed fixed effects selection performances by running Algorithm \ref{alg:streamlinedMFVB_threelevel_globallocal} on the same 50 data replications, also experimenting the other two global-local priors considered in this work: the Laplace and Normal-Exponential-Gamma with $\lambda = 0.25$. We also considered the Gaussian prior specification described in Nolan \emph{et al.} (2020) with hyperparameters $\bmu_{\bbeta^\subjtosel} = \bzero$ and $\bSigma_{\bbeta^\subjtosel} = 10^{10} \bI$.
Then, for each data replication and each of the different four prior specifications considered we used the approximating densities corresponding to fixed effects subject to selection to perform the SAVS procedure of Algorithm \ref{alg:savs}. We employed this procedure also for the approximate posteriors obtained from the Gaussian prior to provide a comparison with a prior that does not belong to the global-local family. 

Let TP (true positives) denote the number of selected fixed effects having true value different from zero and TN (true negatives) denote the number of unselected irrelevant fixed effects having true value equal to zero. We measured the fixed effects selection performance for each prior choice using the $F_1$-score (Van Rijsbergen, 1979), which is defined as:
\begin{equation}
\label{eq:f1_score}
F_1 \equiv \left(\frac{2 \times \text{precision} \times \text{recall}}{\text{precision} + \text{recall}}\right) \times 100 \%,
\end{equation}
with $\text{precision} \equiv \text{TP}/(\text{TP} + \text{FP})$ and $\text{recall} \equiv \text{TP}/(\text{TP} + \text{FN})$, where FP and FN denote the number of false positives (incorrectly selected fixed effects) and false negatives (relevant fixed effects that have not been selected), respectively. This index takes values between 0\% and 100\%, with higher values to be preferred.

The median $F_1$-score was 63.5\% (1st quartile: 43.5\%; 3rd quartile: 94.2\%) for the Gaussian prior and 95.24\% (1st quartile: 80\%; 3rd quartile: 100\%) for the Laplace prior. Both the Horseshoe and Negative-Exponential-Gamma priors exhibited a $F_1$-score equal to 100\% for all the 50 data replications of the simulation study, meaning that perfect selection ($\text{TP}=10$ and $\text{TN}=40$) was always achieved. From this simplified simulated sparse data scenario it is apparent that global-local priors effectively provide better variable selection performances than the Gaussian prior. The lower performances of the Gaussian and Laplace priors are due to the SAVS procedure being applied to optimal approximate $\qDens^*$-densities that have not been properly shrunk towards zero and so irrelevant fixed effects tend to be selected. Nonetheless, all the four different priors gave $\text{FP}=0$, meaning that they did not incorrectly select irrelevant fixed effects.

\subsection{Speed and Memory Saving Assessment\label{subsec:SpeedAssessment}}

Streamlined variational inference has been conceived to obtain efficient implementations of variational algorithms. Hence, the assessment of speed and memory savings is another important aspect. We assessed speed and memory performances of the streamlined variational Algorithm \ref{alg:streamlinedMFVB_threelevel_globallocal} and its na{\"ive} counterpart, whose updates are described in Section \ref{subsec:NaiveMFVBGLPriors}. Four different group numbers and three different lengths for the vector of fixed effects to be selected were considered, namely $m \in \lbrace 10, 50, 100, 200 \rbrace$ and $p_\subjtosel \in \lbrace 25, 100, 200 \rbrace$, aiming to explore scalability of the streamlined methodology to high dimensions. For each combination of $m$ and $p_\subjtosel$, we simulated 10 data replications from model \eqref{eq:gauss_heter_lmm_penalized} with a random intercept and one slope for a single continuous predictor, choosing the sub-group dimensions $n_i$ uniformly on the discrete set $\lbrace 10, \ldots, 20 \rbrace$ and the sub-group specific unit dimensions $o_{ij}$ uniformly on the discrete set $\lbrace 20, \ldots, 30 \rbrace$. This setting allows to test models with heterogeneous dimensions but the same number of groups $m$. We considered a Horseshoe prior for the fixed effects subject to selection, and all the other model dimensions and hyperparameters specifications were the same as those described before. 

For each simulated dataset, we collected the computational timings and the total size of the input data required for performing both the streamlined and na{\"i}ve algorithms. We also ran MCMC (warmup of length 5,000 followed by 25,000 iterations, to which we applied a thinning value of 5) and recorded its computational timings, although MFVB and MCMC do not admit a genuine comparison due to the fact that both depend upon different convergence requirements, as explained in Section 5.1 of Ormerod \emph{et al.} (2017). Moreover, our efficient implementation of MCMC is based on independent resampling from the full conditional densities of $\bbeta$, $\buLone_i$ and $\buLtwo_{ij}$, whereas MFVB uses a joint approximation $\qDens(\bbeta,\bu)$ for all these vectors. Nonetheless, MCMC timings are also reported to provide an intuition on the computational effort required for sampling from the posterior distribution when $m$ and $p_\subjtosel$ increase.
\begin{table}[t!]
\centering
\resizebox{\textwidth}{!}{%
\begin{tabular}{cc|cccc|ccc}
\hline \\[-1em]
 & \multicolumn{1}{c}{} & \multicolumn{4}{c}{\underline{Total runtime of the algorithm (seconds)}} & \multicolumn{3}{c}{\underline{Total size of the required data inputs (megabytes)}} \\[0.5em]
$m$ \qquad& $p_\subjtosel$ \qquad& \emph{Streamlined MFVB} & \emph{Na{\"i}ve MFVB} & $\frac{\text{\it Na{\"i}ve MFVB}}{\text{\it Streamlined MFVB}}$ & \emph{MCMC} & \emph{Streamlined MFVB} & \emph{Na{\"i}ve MFVB} & $\frac{\text{\it Na{\"i}ve MFVB}}{\text{\it Streamlined MFVB}}$ \\[0.5em] \hline\\[-1.2em]
10 & 25 & $0.70_{(0.06)}$ & $3.86_{(0.50)}$ & $5.54$ & $13.03_{(0.96)}$ & $1.39_{(0.07)}$ & $11.06_{(1.00)}$ & $7.97$ \\[0.15em]
 & 100 & $3.90_{(0.23)}$ & $6.51_{(0.82)}$ & $1.67$ & $46.99_{(1.47)}$ & $3.68_{(0.24)}$ & $12.73_{(1.34)}$ & $3.46$ \\[0.15em]
 & 200 & $9.71_{(1.22)}$ & $12.09_{(2.04)}$ & $1.24$ & $176.22_{(9.37)}$ & $6.60_{(0.52)}$ & $15.09_{(1.89)}$ & $2.29$ \\[0.15em] \hline\\[-1.2em]
50 & 25 & $3.29_{(0.11)}$ & $365.07_{(37.10)}$ & $111.01$ & $63.28_{(3.76)}$ & $6.69_{(0.20)}$ & $240.40_{(13.92)}$ & $35.91$ \\[0.15em]
 & 100 & $19.66_{(0.87)}$ & $415.29_{(51.66)}$ & $21.12$ & $138.42_{(5.72)}$ & $18.37_{(0.81)}$ & $254.04_{(20.84)}$ & $13.83$ \\[0.15em]
 & 200 & $49.25_{(1.47)}$ & $501.33_{(39.76)}$ & $10.18$ & $305.25_{(6.51)}$ & $34.00_{(1.04)}$ & $269.34_{(14.90)}$ & $7.92$ \\[0.15em] \hline\\[-1.2em]
100 & 25 & $6.77_{(0.32)}$ & $2877.33_{(177.77)}$ & $424.72$ & $151.05_{(10.48)}$ & $13.56_{(0.34)}$ & $967.34_{(42.15)}$ & $71.33$ \\[0.15em]
 & 100 & $40.42_{(1.66)}$ & $3172.66_{(116.83)}$ & $78.50$ & $266.71_{(10.06)}$ & $36.77_{(1.29)}$ & $1010.25_{(26.94)}$ & $27.47$ \\[0.15em]
 & 200 & $97.81_{(2.07)}$ & $3403.55_{(204.38)}$ & $34.80$ & $494.15_{(9.34)}$ & $67.53_{(1.31)}$ & $1013.90_{(48.67)}$ & $15.01$ \\[0.15em] \hline\\[-1.2em]
200 & 25 & $13.30_{(0.30)}$ & $>$ 5 hours & $> 1355$ & $328.87_{(20.04)}$ & $26.90_{(0.56)}$ & $3817.06_{(113.85)}$ & $141.88$ \\[0.15em]
 & 100 & $79.97_{(1.18)}$ & $>$ 5 hours & $> 225$ & $578.88_{(13.45)}$ & $73.74_{(1.01)}$ & $3941.15_{(97.90)}$ & $53.45$ \\[0.15em]
 & 200 & $197.20_{(4.13)}$ & $>$ 5 hours & $> 95$ & $904.64_{(15.77)}$ & $135.84_{(2.73)}$ & $3991.66_{(102.48)}$ & $29.38$ \\[0.15em] \hline\\[-1em]
\end{tabular}%
}
\caption{\it Average (standard deviation of) elapsed computing times in seconds and average (standard deviation of) total size of required data inputs in megabytes for fitting model \eqref{eq:gauss_heter_lmm_penalized} with three-level random effects specification and $p_\subjtosel$ fixed effects subject to penalization via Horseshoe prior. Results are shown for different group sizes $m$ and different values for $p_\subjtosel$.}
\label{tab:computational_timings}
\end{table}

The tabulated results are shown in Table \ref{tab:computational_timings}. The ``ratio" columns help understand the gain obtained employing the streamlined MFVB methodology over its na{\"i}ve counterpart.
For increasing $m$, streamlined MFVB reached convergence faster than na{\"i}ve MFVB. Notice that in the biggest scenario under examination (last row of the table), the streamlined MFVB algorithm ran in less then 4 minutes on average, while na{\"i}ve MFVB required more than 5 hours. Bigger scenarios are computationally demanding for the na{\"i}ve implementation and may fail to run due to excessive storage demand, whilst we did not experience these issues with Algorithm \ref{alg:streamlinedMFVB_threelevel_globallocal}. Similar comments apply to the huge saving of memory allocation for the required input data provided by the streamlined MFVB implementation.

We conclude the discussion by noticing that our methodology takes the approximate joint posterior dependence between the fixed effects and random effects parameters into account through the multivariate Gaussian approximating density $\qDens(\bbeta,\bu)$. This choice allows to better capture the \emph{a posteriori} covariance structure between $\bbeta$ and $\bu$ and ensures better results in terms of approximation accuracy. Nevertheless, alternative and less restrictive factorizations can be considered, especially if the quality of the approximation can be sacrificed and faster algorithms are desired. For instance, the additional factorization $\qDens(\bbeta,\bu) = \qDens(\bbeta^\raneff, \bu) \qDens(\bbeta^\additional, \bbeta^\subjtosel)$ could remarkably speed up the computations if $p_\raneff$ is small, regardless of the size of $p_\additional$ and $p_\subjtosel$. For implementing this and any other additional independence constraints, mean field restrictions such as \eqref{eq:prodrestr_globallocal} and the Algorithms proposed in Section \ref{subsec:StreamlinedMFVBGLPriors} need to be modified accordingly.

\section{Application to Data from a Perinatal Study\label{sec:Application}}

We present an application of the methodology and algorithms proposed in Section \ref{sec:VariationalInferenceLMMwithGLPriors} to the National Collaborative Perinatal Project data (Klebanoff, 2009), a multisite prospective cohort study which took place in the United States of America between 1959 and 1974. This study was designed to identify the effects of complications during pregnancy or the perinatal period on birth and child outcomes. The data are publicly available from the U.S. National Archives with identifier \textsf{606622}. Many online resources already employed this dataset or subset of it for several analysis and over the years the dataset has become a high-quality reference for biomedical and behavioral research in many areas such as obstetrics, perinatology, pediatrics, and developmental psychology.

The same data were examined in Nolan \emph{et al.} (2020) and we expressly account for the same model specification to experiment a suitable fixed effects selection for a moderately large set of regressors that have been excluded from their analysis and that may have a relevant impact onto explaining the response variable. A full-blown analysis goes beyond the scope of this paper and we focused on predicting the height-for-age z-score for 37,257 infants followed longitudinally over their first year of life, following indications from Taylor (1980). The height-for-age z-score is a standardized measure of the World Health Organization for the height of children after accounting for age; see World Health Organization (2006) for insights on how notable discrepancies from this index standard reference values constitutes an alarm signal for malnutrition symptoms. 

We performed a Bayesian analysis that accounts for the heterogeneity of the evolution of such index across infants. Our analysis was performed through a two-level random effects linear model having a random intercept, and linear and quadratic slopes ($q = 3$) to account for the quadratic evolution of that score over time. All the fixed effects regressors, excluding the intercept, age of the infant (in days since birth) and its square, were subject to selection; these include characteristics of the infant at birth (e.g. weight, length, head circumference, sex and Apgar scores), and characteristics of the mother, father and family. We also accounted for possible interactions between some infant characteristics and sex, for a total of 38 candidate predictors subject to selection.

The model we fitted respects the general specification \eqref{eq:gauss_heter_lmm_penalized} and can be expressed for the generic $i$th infant as follows:
\begin{equation}
\begin{array}{c}
	\by_i \vert \bbeta, \bu_i, \sigsq \simind \mbox{N}\left(\bX^\raneff_i\bbeta^\raneff + \bX^\subjtosel_i\bbeta^\subjtosel + \bZ_i \bu_i, \sigsq\bI\right), \quad \bu_i \vert \Sigma \simind \mbox{N}_3(\bzero, \bSigma),\quad1 \le i \le 37,257\:, \\[1.5ex]
	\left[\begin{array}{c} \bbeta^\raneff \\[1ex] \bbeta^\subjtosel \end{array}\right] \Bigg\vert \:\bzeta, \tausq \sim \mbox{N}\left(\left[\begin{array}{c} \bzero_3 \\[1ex] \bzero_{38} \end{array}\right], \left[\begin{array}{cc} 10^{10}\bI_3 & \bO \\[1ex] \bO & \tausq\:\diag(\bzeta)^{-1} \end{array} \right] \right), \\[3.5ex] 
	\sigsq\vert \asigsq \sim \text{Inverse-}\chi^2(1, 1/\asigsq), \quad \asigsq \sim \text{Inverse-}\chi^2(1, 10^{-10}), \\[1.5ex]
	\bSigma\vert \ASigma \sim \text{Inverse-G-Wishart}(\Gfull, 6,\ASigma^{-1}),\\[1.5ex]
\ASigma \sim \text{Inverse-G-Wishart}(\Gdiag,1, 2 \times 10^{-10}\bI_3),\\[1.5ex]
	\zeta_h \vert a_{\zeta_h} \simind \begin{cases} 
	\:\mbox{Inverse-}\chi^2(2,1) \qquad & \mbox{for a Laplace prior}\\
	\:\mbox{Gamma}(1/2, a_{\zeta_h}) \qquad & \mbox{for a Horseshoe prior}\\
	\:\mbox{Inverse-}\chi^2(2, 2a_{\zeta_h}) \qquad & \mbox{for a NEG prior,}
\end{cases} \\[4ex]
a_{\zeta_h} \simind \begin{cases} 
	\qquad\quad - \qquad & \mbox{for a Laplace prior}\\
	\mbox{Gamma}(1/2, 1) \qquad & \mbox{for a Horseshoe prior}\\
	\mbox{Gamma}(\lambda, 1) \qquad & \mbox{for a NEG prior,}
\end{cases}\\[4ex]
\text{for } 1 \le h \le 38,\\[1ex]
\tausq \vert \atausq \sim \mbox{Inverse-}\chi^2(1, 1/\atausq), \quad \atausq \sim \mbox{Inverse-}\chi^2(1, 10^{-10}).
\end{array}
\end{equation}
For the $i$th infant, $o_i$ time-point measurements were recorded, ranging from one to four in number. The $\bX_i^\raneff$ matrix has size $n_i \times 3$ with the first column being a vector of ones, the second one consists of the time-point measurements for the $i$th infant and the third column containing the square of the elements of the second one. We set $\bX_i^\additional = \bO$, while the $\bX_i^\subjtosel$ matrix of size $n_i \times 38$ consisted of all the considered predictors subject to selection. Moreover, $\bZ_i = \bX_i^\raneff$ by definition. Uninformative priors were placed over all the model parameters. We fitted the model using the three global-local priors treated in this work, and the Gaussian prior for $\bbeta^\subjtosel$ treated in Nolan \emph{et al.} (2020). 

Streamlined MFVB and MCMC were used for model fitting. The former was performed running Algorithm \ref{alg:streamlinedMFVB_twolevel_globallocal} and stopping it after 200 iterations, while the latter was performed running 25,000 iterations to which a thinning factor of 5 was applied after discarding 5,000 burnin iterations. The whole input data required approximately 200 megabytes of memory storage, while a na{\"ive} MFVB procedure would necessitate several gigabytes of memory to entirely store the $\bZ$ matrix, that is composed by 11,693,705,562 cells of which the 99.997\% are zeros. All the covariates excluding the binary ones were standardized and the estimates were rescaled back to the original scale before presenting the results. The streamlined MFVB algorithms took 2 to 3 minutes to run for each prior specification, while the associated MCMC samplers took more than 35 minutes. 
\begin{figure}[t!]
	\centering
	\includegraphics[width=\textwidth]{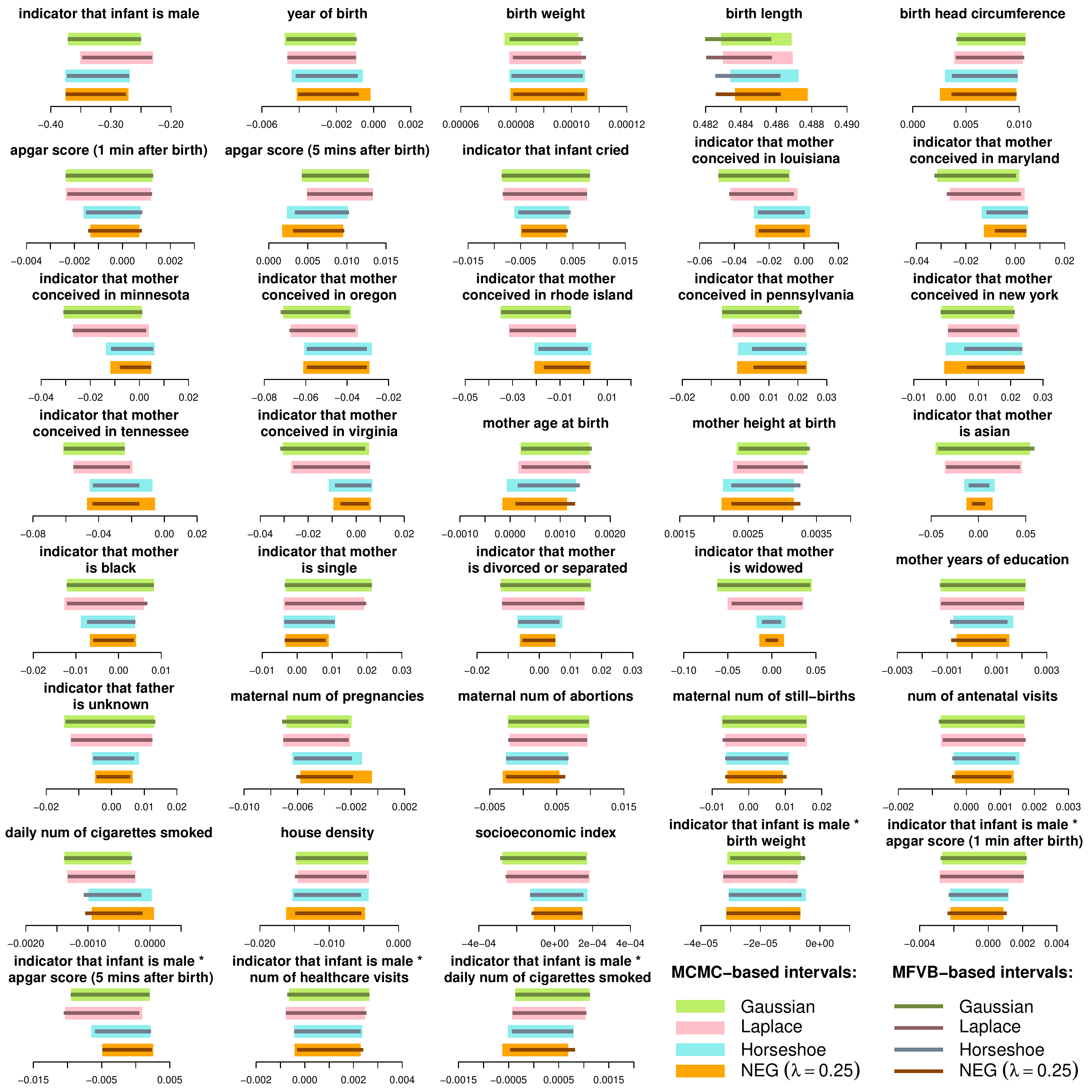}
	\caption{\it The $90\%$ high posterior density credible intervals for the fixed effects subject to selection in the real data application. The four different priors for $\bbeta^\subjtosel$ are represented with different colors. For each fixed effect, the thicker lines correspond to the intervals obtained from the MCMC approximate marginal posterior densities, while the thinner lines represent those obtained from the streamlined MFVB approximating densities.} 
\label{fig:cppdata_hpd}
\end{figure}

We omit the presentation of model interpretation, goodness-of-fit analysis, accuracy of the approximations, convergence of the MCMC chains and visualization of the fitted height-for-age z-score trajectories over time. Instead, in Figure \ref{fig:cppdata_hpd} we present  90\% high posterior density credible intervals for all the fixed effects subject to selection. The thicker lines represent the intervals obtained from the MCMC posterior samples through the \textsf{emp.hpd} function of the \textsf{TeachingDemos} package (Snow, 2020), while the thinner lines represent those calculated from the determined optimal MFVB approximate posteriors. The two lines are superimposed to facilitate immediate comparison, for each different prior specification.

Overall, MFVB provided very accurate high posterior density credible intervals when compared to MCMC. The MCMC chain associated to the \textsf{birth length} fixed effect showed some convergence problems, probably due to its moderate correlation with the response variable and this reduced the overlap with MFVB. Importantly, the Laplace global-local prior produced results very similar to those provided by the Gaussian prior specification. On the other hand, both the Horseshoe and Normal Exponential Gamma priors seem to have effectively shrunk most of the fixed effects towards zero, especially those associated to dummy variables. 

The SAVS procedure applied to the approximate posterior densities classifies the following fixed effects as relevant: indicator that the infant is male, infant's year of birth, birth length, birth head circumference, the Apgar score assessed 5 minutes after the infant's birth and the mother height. 
%We have few knowledge about the considered topic to confirm whether this results are in line with literature references, but we are confident that sex, length of both the infant and the mother and some additional infant-related covariates are effectively relevant for predicting the height-for-age z-score. 
Some effects such as the ethnicity of the mother, the place where she conceived, her marital status and informations on previous pregnancies were not indicated as relevant by the SAVS procedure, albeit some of their associated high posterior density credible intervals are far from zero. Notice also that, for some fixed effects, global-local priors drag the intervals towards the origin, with different intensities probably depending on the variability and correlation of the covariates. More ad-hoc analyses are firmly suggested, although these go beyond the scope of the current work.

\section{Conclusions}\label{sec:Conclusions}

In this work, we developed streamlined mean-field variational Bayes procedures for Gaussian response linear mixed models having nested random effects structures and admitting fixed effects prior specifications alternative to the classical Gaussian one. The priors we considered are amenable to automated and hyperparameter-free fixed effects selection procedures. Simulated and real data examples showed how streamlined variational inference can provide impressive benefits in terms of computational time and memory saving when compared to inefficient implementations of variational approximations. Albeit the marginal posterior densities of fixed effects subject to selection are approximated by bell-shaped curves, our studies showed high performances of the automated selection procedure.
It is also worth mentioning that the more restrictive mean-field approximations discussed at the end of Section \ref{sec:Assessments} can sensibly improve the benefit of streamlined variational inference for large $p$ scenarios and, in general, when speed is more important than approximation accuracy.

Numerous ramifications of the proposed methodology can be envisaged. These include the treatment of models with unit-specific errors, heteroskedastic covariance structures for groups and sub-groups, higher levels of nesting or crossed random effects and the extension to generalized linear mixed models. Ormerod and Wand (2010), Wand \textit{et al.} (2011), Nolan and Wand (2017), Maestrini \myand Wand (2018) and McLean \myand Wand (2019) provide variational inference algorithms for models with a variety of response distributions such as Bernoulli with probit or logit link, Poisson, Negative Binomial, $t$, Asymmetric Laplace, Skew Normal, Skew $t$ and Finite Normal Mixtures. The variational algorithms presented in these references have been derived using data-augmentation representations of the response distributions. The same strategy can be adopted to derive MFVB algorithms for non-Gaussian response mixed models and allow for the implementation of streamlined variational inference through a relatively straightforward adaptation of the algorithms presented in this work. Maestrini (2019) also provides explicit streamlined variational inference algorithms for two-level mixed models with binary-logistic and Poisson responses, and generic Gaussian priors for the fixed effects coefficients.

Regarding the selection of fixed effects, alternative shrinkage priors belonging to the global-local family can be accounted for with minor modifications to the proposed algorithms. Following the prescriptions of Neville \emph{et al.} (2014), it is possible to study global-local prior specifications that are not based on auxiliary variables and investigate whether this could lead to better approximation accuracies, although at the cost of more computationally intensive variational updates. We also mention the possibility of admitting spike-and-slab prior formulations following, for example, a variational inference approach similar to that of Carbonetto \myand Stephens (2012). The drawback is that spike-and-slab priors require more involved algebra for implementing streamlined variational inference. Using the Bayesian Adaptive Lasso (BaLasso) prior proposed by Leng \emph{et al.} (2014) it is also possible to extend our individual fixed-effects selection approach to account for (ordered) group selection by imposing different shrinkage levels to different coefficients. A variational Bayes approach for generalized linear models with priors of this type has been explored by Tung \emph{et al.} (2019), where the variational parameter updates of their VBGLMM algorithm could be streamlined using the framework studied in our work.

One last possible direction to explore concerns streamlined variational inference for models with priors for selecting random effects. Consider for simplicity the two-level random effects case in which $\boldsymbol{u}_i \simind \mbox{N}(\boldsymbol{0}, \boldsymbol{\Sigma})$ for $1 \le i \le m$. If the covariance matrix $\boldsymbol{\Sigma}$ is supposed to be diagonal, i.e. $\mbox{Cov}(u_{ih}, u_{ik}) = 0$ for each $1 \le h, k \le q$ with $h \neq k$ and $1 \le i \le m$, and a global-local prior distribution is imposed to each random effect, then explicit streamlined MFVB updates can be obtained with straightforward manipulation of the results presented in the supplementary material and references therein. This applies to three-level models in a similar way.
However, assuming that $\boldsymbol{\Sigma}$ is diagonal may sometimes be quite restrictive and the highest computational advantage of streamlined MFVB over its na{\"i}ve counterpart is achieved when the random effects covariance matrix is non-diagonal.
Furthermore, it is typically easier to provide an interpretation to the selection of fixed effects than explaining why certain random effects are relevant only for a subset of clusters or sub-clusters. For these reasons the present work provides a first extension of the streamlined MFVB approach of Nolan \emph{et al.} (2020) focusing on fixed-effects selection. Another relevant extension is the derivation of (streamlined) MFVB algorithms with random effects selection procedures admitting non-diagonal structures for $\boldsymbol{\Sigma}$. The covariance matrix decomposition approach of Chen \& Dunson (2003) and related methods discussed in Section 1 can be exploited for this extension.

\section*{Acknowledgments}

The research presented in this article was supported by the Australian Research Council Discovery Project DP180100597 and the Australian Research Council Center of Excellence for Mathematical and Statistical Frontiers (grant CE140100049).

\section*{References}

\bib
Andrews, D. F. \myand Mallows, C. L. (1974). Scale mixtures of normal distributions. \emph{Journal of the Royal Statistical Society B}, {\bf 36}, 99--102.

\bib
Armagan, A. \myand Dunson, D. B. (2011). Sparse variational analysis of linear mixed models for large data sets. \emph{Statistics \myand Probability Letters}, {\bf 81}, 1056--1062.

\bib
Armagan, A., Dunson, D. B. \myand Lee, J. (2013). Generalized double Pareto shrinkage. \emph{Statistica Sinica}, {\bf 23}, 119--143.

\bib
Atay-Kayis, A. \myand Massam, H. (2005). A Monte Carlo method for computing marginal likelihood in nondecomposable Gaussian graphical models. \emph{Biometrika}, {\bf 92}, 317--335.

\bib
Baltagi, B. H. (2013). \emph{Econometric Analysis of Panel Data, Fifth Edition}. Chichester, U.K.: John Wiley \myand Sons.

\bib
Barbieri, M. M. \myand Berger, J. O. (2004). Optimal predictive model selection. \emph{The Annals of Statistics}, {\bf 32}, 870--897.

\bib
Bhadra, A., Datta, J., Polson, N. G. \myand Willard, B. (2017). The horseshoe+ estimator of ultra-sparse signals. \emph{Bayesian Analysis}, {\bf 12}, 1105--1131.

\bib
Bhadra, A., Datta, J., Polson, N. G. \myand Willard, B. (2019). Lasso meets horseshoe: a survey. \emph{Statistical Science}, {\bf 34}, 405--427.

\bib
Bhattacharya, A., Chakraborty, A. \myand Mallick, B. K. (2016). Fast sampling with Gaussian scale mixture priors in high-dimensional regression. \emph{Biometrika}, {\bf 103}, 985--991.

\bib
Bhattacharya, A., Pati, D., Pillai, N. S. \myand Dunson, D. B. (2015). Dirichlet-Laplace priors for optimal shrinkage. \emph{Journal of the American Statistical Association}, {\bf 110}, 1479--1490.

\bib
Bishop, C. M. (2006). \emph{Pattern Recognition and Machine Learning}. New York: Springer.

\bib
Blei, D. M., Kucukelbir, A. \myand McAuliffe, J. D. (2017). Variational inference: a review for statisticians. \emph{Journal of the American Statistical Association}, {\bf 112}, 859--877.

\bib
Bogdan, M., Chakrabarti, A., Frommlet, F. \myand Ghosh, J. K. (2011). Asymptotic Bayes-optimality under sparsity of some multiple testing procedures. \emph{The Annals of Statistics}, {\bf 39}, 1551--1579.

\bib
Bondell, H. D. \myand Reich, B. J. (2012). Consistent high-dimensional Bayesian variable selection via penalized credible regions. \emph{Journal of the American Statistical Association}, {\bf 107}, 1610--1624.

\bib
Boyd, S. \myand Vandenberghe, L. (2004). \emph{Convex Optimization}, Cambridge, U.K.: Cambridge University Press.

\bib
Brown, H. \myand Prescott, R. (2014). \emph{Applied Mixed Models in Medicine, Third Edition}, Chichester, U.K.: John Wiley \& Sons. 

\bib
B{\"u}rkner, P.-C. (2017). \textsf{brms}: an \textsf{R} package for Bayesian multilevel models using \textsf{Stan}. \emph{Journal of Statistical Software}, {\bf 80}, 1--28.

\bib
Carbonetto, P. \myand Stephens, M. (2012). Scalable variational inference for Bayesian variable selection in regression, and its accuracy in genetic association studies. \emph{Bayesian Analysis}, {\bf 7}, 73--108.

\bib
Carpenter, B., Gelman, A., Hoffman, M., Lee, D., Goodrich, B., Betancourt, M., Brubaker, M., Guo, J., Li, P. \myand Riddell, A. (2017). \textsf{Stan}: a probabilistic programming language. \emph{Journal of Statistical Software}, {\bf 76}, 1--32.

\bib
Carvalho, C. M., Polson, N. G. \myand Scott, J. G. (2009). Handling sparsity via the horseshoe. \emph{Journal of Machine Learning Research}, {\bf 5}, 73--80.

\bib
Carvalho, C. M., Polson, N. G. \myand Scott, J. G. (2010). The horseshoe estimator for sparse signals. \emph{Biometrika}, {\bf 97}, 465--480.

\bib
Chen, Z. \myand Dunson, D. B. (2003). Random effects selection in linear mixed models. \emph{Biometrics}, {\bf 59}, 762--769.

\bib
Eddelbuettel, D. \myand Sanderson, C. (2014). \textsf{RcppArmadillo}: accelerating \textsf{R} with high-performance \textsf{C++} linear algebra. \emph{Computational Statistics and Data Analysis}, {\bf 71}, 1054--1063. 

\bib
Efron, B. (2008). Microarrays, empirical Bayes and the two-groups model. \emph{Statistical Science}, {\bf 23}, 1--22.

\bib
Faes, C., Ormerod, J. \myand Wand, M.P. (2011). Variational Bayesian inference for parametric and nonparametric regression with missing data. \emph{Journal of the American Statistical Association}, {\bf 106}, 959--971.

\bib
Fan, Y. \myand Li, R. (2012). Variable selection in linear mixed effects models. \emph{The Annals of Statistics}, {\bf 40}, 2043--2068.

\bib
Fitzmaurice, G. , Davidian, M., Verbeke, G. \myand Molenberghs, G. (2008). \emph{Longitudinal Data Analysis}. Boca Raton, Florida: Chapman \myand Hall/CRC.

\bib
Frank, I. E. \myand Friedman, J. H. (1993). A statistical view of some chemometrics regression tools. \emph{Technometrics}, {\bf 35}, 109--148.

\bib
Gelman, A. (2006). Prior distributions for variance parameters in hierarchical models. \emph{Bayesian Analysis}, \textbf{1}, 515--533.

\bib
George, E. I. \myand McCulloch, R. E. (1997). Approaches for Bayesian variable selection. \emph{Statistica Sinica}, {\bf 7}, 339--373.

\bib
Goldstein, H. (2010). \emph{Multilevel Statistical Models, Fourth Edition}. Chichester, U.K.: John Wiley \& Sons.

\bib
Griffin, J. E. \myand Brown, P. J. (2010). Inference with normal-gamma prior distributions in regression problems. \emph{Bayesian Analysis}, {\bf 5}, 171--188.

\bib
Griffin, J. E. \myand Brown, P. J. (2011). Bayesian hyper-lassos with non-convex penalization. \emph{Australian and New Zealand Journal of Statistics}, {\bf 53}, 423--442.

\bib
Groll, A. \myand Tutz, G. (2012). Variable selection for generalized linear mixed models by $l_1$-penalized estimation. \emph{Statistics and Computing}, {\bf 24}, 137--154.

\bib
Hahn, P. R. \myand Carvalho, C. M. (2015). Decoupling shrinkage and selection in Bayesian linear models: a posterior summary perspective. \emph{Journal of the American Statistical Association}, {\bf 110}, 435--448.

\bib
Hoerl, A. E. \myand Kennard, R. W. (1970). Ridge regression: biased estimation for nonorthogonal problems. \emph{Technometrics}, {\bf 12}, 55--67.

\bib
Huang, A. \myand Wand, M. P. (2013). Simple marginally noninformative prior distributions for covariance matrices. \emph{Bayesian Analysis}, {\bf 8}, 439--452.

\bib
Hughes, D. M., Garc{\'i}a-Fina{\~n}a, M. \myand Wand, M. P. (2021). Fast approximate inference for multivariate longitudinal data. \emph{Biostatistics} (volume and page numbers pending). 

\bib
Hui, F. K. C., M{\"u}ller, S. \myand Welsh, A. H. (2017). Joint selection in mixed models using regularized PQL. \emph{Journal of the American Statistical Association}, {\bf 112}, 1323--1333.

\bib
Ishwaran, H. \myand Rao, J. S. (2005). Spike and slab variable selection: frequentist and Bayesian strategies. \emph{The Annals of Statistics}, {\bf 33}, 730--773.

\bib
Johnstone, I. M. \myand Silverman, B. W. (2005). Bayes selection of wavelet thresholds. \emph{The Annals of Statistics}, {\bf 33}, 1700--1752. 

\bib
Kinney, S. K. \myand Dunson, D. B. (2007). Fixed and random effects selection in linear and logistic models. \emph{Biometrics}, {\bf 63}, 690--698.

\bib
Klebanoff M. A. (2009). The collaborative perinatal project: a 50-year retrospective. \emph{Paediatric and Perinatal Epidemiology}, {\bf 23}, 2--8. 

\bib
Korte, A., Vilhjálmsson, B., Segura, V., Platt, A., Long, Q. \myand Nordborg, M. (2012). A mixed-model approach for genome-wide association studies of correlated traits in structured populations. \emph{Nature Genetics}, {\bf 44}, 1066--1071.

\bib
Lee, C. Y. Y. \myand Wand, M. P. (2016). Streamlined mean field variational Bayes for longitudinal and multilevel data analysis. \emph{Biometrical Journal} {\bf 58}, 868--895.

\bib
Leng, C., Tran, M.N. \myand Nott, D. (2014). Bayesian adaptive Lasso. \emph{Annals of the Institute of Statistical Mathematics}, {\bf 66}, 221--244.

\bib
Li, H. \myand Pati, D. (2017). Variable selection using shrinkage priors. \emph{Computational Statistics and Data Analysis}, {\bf 107}, 107--119.

\bib
Li, J., Wang, Z., Li, R. \myand Wu, R. (2015). Bayesian group Lasso for nonparametric varying-coefficient models with application to functional genome-wide association studies. \emph{The Annals of Applied Statistics}, {\bf 9}, 640--664.

\bib
Lindner, C. C. \myand Rodger, C. A. (1997). \emph{Design Theory}. Boca Raton, Florida: Chapman \myand Hall/CRC.

\bib
Li, Y., Wang, S., Song, P. X.--K., Wang, N., Zhou, L. \myand Zhu, J. (2018). Doubly regularized estimation and selection in linear mixed-effects models for high-dimensional longitudinal data. \emph{Statistics and Its Interface}, {\bf 11}, 721--737.

\bib
Luts, J., Broderick, T. \myand Wand, M. P. (2014). Real-time semiparametric regression. \emph{Journal of Computational and Graphical Statistics}, {\bf 23}, 589--615.

\bib
Maestrini, L. \myand Wand, M.P. (2018). Variational message passing for skew t regression. \emph{Stat}, {\bf 7}, 1--11.

\bib
Maestrini, L. (2019). \emph{On Variational Approximations for Frequentist and Bayesian Inference}. Doctor of Philosophy thesis, Università degli Studi di Padova, Italy.

\bib
Maestrini, L. \myand Wand, M. P. (2021). The inverse G-Wishart distribution and variational message passing. \emph{Australian and New Zealand Journal of Statistics}, {\bf 63}, 517--541.

\bib
McLean, M.W. \myand Wand, M.P. (2019). Variational message passing for elaborate response regression models. \textit{Bayesian Analysis}, {\bf 14}, 371--398.

\bib
Menictas, M., Di Credico, G. \myand Wand, M. P. (2022). Streamlined variational inference for linear mixed models with crossed random effects. \emph{Journal of Computational and Graphical Statistics} (volume and page numbers pending).

\bib
Menictas, M., Nolan, T.H., Simpson, D. G. \myand Wand, M. P. (2021). Streamlined variational inference for higher level group-specific curve models. \emph{Statistical Modelling}, {\bf 21}, 479--519. 

\bib
Minka, T., Winn, J., Guiver, J., Zaykov, Y., Fabian, D. \myand Bronskill, J. (2018). \textsf{Infer.NET 0.3}, Microsoft Research Cambridge, Cambridge, U.K. \emph{http://dotnet.github.io/infer}.

\bib
Mitchell, T. J. \myand Beauchamp, J. J. (1988). Bayesian variable selection in linear regression. \emph{Journal of the American Statistical Association}, {\bf 83}, 1023--1032.

\bib
Neville, S. E., Ormerod, J. T \myand Wand, M. P. (2014). Mean field variational Bayes for continuous sparse signal shrinkage: pitfalls and remedies. \emph{Electronic Journal of Statistics}, {\bf 8}, 1113--1151.

\bib
Nolan, T. H., Menictas, M. \myand Wand, M. P. (2020). Streamlined computing for variational inference with higher level random effects. \emph{Journal of Machine Learning Research}, {\bf 21}, 1--62.

\bib
Nolan, T.H. \myand Wand, M.P. (2017). Accurate logistic variational message passing: algebraic and numerical details. \emph{Stat}, {\bf 6}, 102--112.

\bib
Nolan, T. H. \myand Wand, M. P. (2020). Streamlined solutions to multilevel sparse matrix problems. \emph{ANZIAM Journal}, {\bf 62}, 18--41.

\bib
O'Hara, R. B. \myand Sillanp{\"a}{\"a}, M. J. (2009). A review of Bayesian variable selection methods: what, how and which. \emph{Bayesian Analysis}, {\bf 4}, 85--118.

\bib
Ormerod, J. T. \myand Wand, M. P. (2010). Explaining variational approximations. \emph{The American Statistician}, {\bf 64}, 140--153.

\bib
Ormerod, J. T., You, C. \myand M{\"u}ller, S. (2017). A variational Bayes approach to variable selection. \emph{Electronic Journal of Statistics}, {\bf 11}, 3549--3594.

\bib
Park, T. \myand Casella, G. (2008). The Bayesian lasso. \emph{Journal of the American Statistical Association}, {\bf 103}, 681--686.

\bib
Pinheiro, J. C. \myand Bates, D. M. (2006). \emph{Mixed-Effects Models in} \textsf{S} \textit{and} \textsf{S-PLUS}. New York: Springer Science \& Business Media.

\bib
Polson, N. G. \myand Scott, J. G. (2011). Shrink globally, act locally: sparse Bayesian regularization and prediction. \emph{Bayesian Statistics}, {\bf 9}, 501--538.

\bib
Rao, J. N. K. \myand Molina,  I. (2015). \emph{Small Area Estimation, Second Edition}. Hoboken, New Jersey: John Wiley \myand Sons.

\bib
Ray, P., \myand Bhattacharya, A. (2018). Signal Adaptive Variable Selector for the horseshoe prior. Unpublished manuscript available at \emph{https://arxiv.org/abs/1810.09004}.

\bib
R Core Team. (2022). \textsf{R}: A language and environment for statistical computing. R Foundation for Statistical Computing, Vienna, Austria. \emph{https://www.R-project.org/}.

\bib
Ruppert, D., Wand, M. P. \myand Carroll, R. (2003). \emph{Semiparametric Regression}. Cambridge, U.K.: Cambridge University Press.

\bib
Sanderson, C. \myand Curtin, R. (2016). \textsf{Armadillo}: a template-based \textsf{C++} library for linear algebra. \emph{Journal of Open Source Software}, {\bf 1}, 26.

\bib
Schelldorger, J., B{\"u}hlmann, P. \myand Van De Geer, S. (2011). Estimation for high-dimensional linear mixed-effects models using $l_1$-penalization. \emph{Scandinavian Journal of Statistics}, {\bf 38}, 197--214.

\bib
Sikorska, K., Rivadeneira, F., Groenen, P. J., Hofman, A., Uitterlinden, A. G., Eilers, P. H. \myand Lesaffre, E. (2012). Fast linear mixed model computations for genome-wide association studies with longitudinal data. \emph{Statistics in Medicine}, {\bf 32}, 165--180.

\bib
Snow, G. (2020). \textsf{TeachingDemos}: demonstrations for teaching and learning. \textsf{R} package version 2.12. \emph{https://CRAN.R-project.org/package=TeachingDemos}.

\bib
Tang, X., Xu, X., Ghosh, M. \myand Ghosh, P. (2018). Bayesian variable selection and estimation based on global-local shrinkage priors. \emph{Sankhya A}, {\bf 80}, 215--246.

\bib
Taylor, P. M. (1980). The first year of life: the collaborative perinatal project of the National Institute of Neurological and Communicative Disorders and Stroke. \emph{Journal of the American Medical Association}, {\bf 244}, 1503. 

\bib
Tibshirani, R. (1996). Regression shrinkage and selection via the lasso. \emph{Journal of the Royal Statistical Society B}, {\bf 58}, 267--288.

\bib
Tung, D. T., Tran, M.-N. \myand Cuong, T. M. (2019). Bayesian adaptive lasso with variational Bayes for variable selection in high-dimensional generalized linear mixed models. \emph{Communications in Statistics - Simulation and Computation}, {\bf 48}, 530--543.

\bib
Van Rijsbergen, C. J. (1979). \emph{Information Retrieval (2nd ed.)}. London: Butterworths.

\bib
Verbeke, G. \myand Molenberghs, G. (2000). \emph{Linear Mixed Models for Longitudinal Data}. New York: Springer.

\bib
Vonesh, E. F. \myand Chinchilli, V. G. (1997). \emph{Linear and Nonlinear Models for the Analysis of Repeated Measurements}. London: Chapman and Hall.

\bib
Wand, M. P. (2020). \textsf{KernSmooth}: functions for kernel smoothing supporting Wand \myand Jones (1995). \textsf{R} package version 2.23-18. \emph{https://CRAN.R-project.org/package=KernSmooth}.

\bib
Wand, M. P. \myand Jones, M.C. (1995). \emph{Kernel Smoothing}. London: Chapman and Hall.

\bib
Wand, M. P., Ormerod, J. T., Padoan, S. A., \myand Fr{\"u}hwirth, R. (2011). Mean field variational Bayes for elaborate distributions. \emph{Bayesian Analysis}, {\bf 6}, 847--900.

\bib
Wand, M. P. \myand Ormerod, J. T. (2011). Penalized wavelets: embedding wavelets into semiparametric regression. \emph{Electronic Journal of Statistics}, {\bf 5}, 1654--1717.

\bib
Wang, S. S. J. \myand Wand,  M. P. (2011). Using \textsf{Infer.NET} for statistical analyses. \emph{The American Statistician}, {\bf 65}, 115--126.

\bib
West, M. (1987). On scale mixtures of normal distributions. \emph{Biometrika}, {\bf 74}, 646--648. 

\bib
Whittaker, E. T. \myand Watson, G. N. (1990). \emph{A Course in Modern Analysis}. Cambridge, U.K.: Cambridge University Press.

\bib
World Health Organization (2006). WHO child growth standards: length/height-for-age, weight-for-age, weight-for-length, weight-for-height and body mass index-for-age: methods and development. Available at \emph{https://apps.who.int/iris/handle/10665/43413}.

\bib
Yang, M. (2013). Bayesian nonparametric centered random effects models with variable selection. \emph{Biometrical Journal}, {\bf 55}, 217--230.

\bib
Yang, M., Wang, M. \myand Dong, G. (2020). Bayesian variable selection for mixed effects model with shrinkage prior. \emph{Computational Statistics}, {\bf 35}, 227--243.

\bib
Zhang, Y. \myand Bondell, H. D. (2018). Variable selection via penalized credible regions with Dirichlet-Laplace global-local shrinkage priors. \emph{Bayesian Analysis}, {\bf 13}, 823--844.

\bib
Zou, H. (2006). The adaptive lasso and its oracle properties. \emph{Journal of the American Statistical Association}, {\bf 101}, 1418--1429.

\bib
Zou, H. \myand Hastie, T. (2005). Regularization and variable selection via the elastic net. \emph{Journal of the Royal Statistical Society Series B}, {\bf 67}, 301--320.

\null\vfill\eject

%%%%%%%%%%%%%%%%%%%%%%%%%%%%%%%%%%%%%%%%%%%%%%%%%%%%%%%%%%%%%%%%%%%
%%%%%%%%%%%%%%%%%%%%%%%%%%%%%%%%%%%%%%%%%%%%%%%%%%%%%%%%%%%%%%%%%%%
%%%%%%%%%%%%%%%%%%%%%%%%%%%%%%%%%%%%%%%%%%%%%%%%%%%%%%%%%%%%%%%%%%%
%
% START OF SUPPLEMENTARY MATERIAL
%
%%%%%%%%%%%%%%%%%%%%%%%%%%%%%%%%%%%%%%%%%%%%%%%%%%%%%%%%%%%%%%%%%%%
%%%%%%%%%%%%%%%%%%%%%%%%%%%%%%%%%%%%%%%%%%%%%%%%%%%%%%%%%%%%%%%%%%%
%%%%%%%%%%%%%%%%%%%%%%%%%%%%%%%%%%%%%%%%%%%%%%%%%%%%%%%%%%%%%%%%%%%
%%%%%%%%%%%%%%%%%%%%%%%%%%%%%%%%%%%%%%%%%%%%%%%%%%%%%%%%%%%%%%%%%%%
\appendix
%%%%%%%%%%%%%%%%%%%%%%%%%%%%%%%%%%%%%%%%%%%%%%%%%%%%%%%%%%%%%%%%%%%
%%%%%%%%%%%%%%%%%%%%%%%%%%%%%%%%%%%%%%%%%%%%%%%%%%%%%%%%%%%%%%%%%%%
%%%%%%%%%%%%%%%%%%%%%%%%%%%%%%%%%%%%%%%%%%%%%%%%%%%%%%%%%%%%%%%%%%%
\renewcommand{\thealgorithm}{S.\arabic{algorithm}}
\setcounter{algorithm}{0}
\renewcommand{\thesection}{S.\arabic{section}}
\setcounter{section}{0}
%%%%%%%%%%%%%%%%%%%%%%%%%%%%%%%%%%%%%%%%%%%%%%%%%%%%%%%%%%%%%%%%%%%

%%%%%%%%%%%%%%%%%%%%%%%%%%%%%%%%%%%%%%%%%%%%%%%%%%%%%%%%%%%

%\centerline{\Large\textbf{Appendix}}
\centerline{\Large Supplementary Material for:}
\vskip5mm
\centerline{\Large\bf Sparse Linear Mixed Model Selection via}
\vskip2mm
\centerline{\Large\bf Streamlined Variational Bayes}
\vskip5mm
\centerline{\normalsize\sc By Emanuele Degani$\null^{\dag,\natural}$, Luca Maestrini$\null^\ddag$,}
\centerline{\normalsize\sc Dorota Toczydłowska$\null^\sharp$ \myand Matt P. Wand$\null^\sharp$}
\vskip5mm
\centerline{\textit{Università degli Studi di Padova$\null^\dag$, Banca d'Italia -- Eurosystem$\null^\natural$,}}
\vskip1mm
\centerline{\textit{The Australian National University$\null^\ddag$ and University of Technology Sydney$\null^\sharp$}}

\section{Distributions and Associated Useful Results\label{appsec:UsefulDistributions}}

Many probability distributions are used throughout the paper. We provide details on their probability density functions together with additional useful results.

\subsection{Inverse-Gaussian Distribution\label{appsubsec:InverseGaussian}}

A continuous random variable $x$ has an Inverse Gaussian distribution with mean parameter $\mu>0$ and rate parameter $\lambda > 0$, written $x \sim \mbox{Inverse-Gaussian}(\mu,\lambda)$, if the density function of $x$ is
$$
\pDens(x) = \lambda^{1/2} (2\pi x^3)^{-1/2} \exp\left\lbrace -\frac{\lambda (x-\mu)^2}{2\mu^2 x}\right\rbrace, \quad x > 0.
$$

\subsection{Laplace Distribution\label{appsubsec:Laplace}}

A continuous random variable $x$ has a Laplace distribution (also known as \emph{Double-Exponential} or \emph{two-tailed Exponential} distribution) with mean parameter $\mu$ and scale parameter $\sigma > 0$, written $x \sim \mbox{Laplace}(\mu,\sigma)$, if the density function of $x$ is
$$
\pDens(x) = \frac{1}{2\sigma}\exp\left( -\frac{\vert x-\mu\vert}{\sigma}\right), \quad x\in\mathbb{R}.
$$
A useful result from Andrews \myand Mallows (1974) and West (1987) shows that
$$
\begin{array}{c}
\mbox{if} \quad x \vert b \sim \text{N}(\mu, \sigsq/b) \quad \mbox{and} \quad b \sim \mbox{Inverse-Gamma}(1,1/2),\\[1ex]
\mbox{then} \quad x \sim \mbox{Laplace}(\mu, \sigma).
\end{array}
$$

\subsection{Horseshoe Distribution\label{appsubsec:Horseshoe}}

A continuous random variable $x$ has a Horseshoe distribution with mean parameter $\mu$ and scale parameter $\sigma > 0$, written $x \sim \mbox{Horseshoe}(\mu,\sigma)$, if the density function of $x$ is
$$
\pDens(x) = (2\pi^3)^{-1/2} \sigma^{-1} \exp\left\{\frac{(x-\mu)^2}{2\sigsq}\right\} E_1\left\{\frac{(x-\mu)^2}{2\sigsq}\right\}, \quad x\in\mathbb{R}
$$
where $E_1(x) \equiv \int_x^\infty t^{-1} e^{-t} \mathrm{d}t$, with $x \neq 0$, is the exponential integral function of order 1. 
A useful result from Carvalho \emph{et al.} (2010) shows that
$$
\begin{array}{c}
\mbox{if} \quad x \vert b \sim \text{N}(\mu, \sigsq/b), \quad b\vert c \sim \mbox{Gamma}(1/2, c) \quad \mbox{and} \quad c \sim \mbox{Gamma}(1/2,1),\\[1ex]
\mbox{then} \quad x \sim \mbox{Horseshoe}(\mu, \sigma).
\end{array}
$$

\subsection{Normal-Exponential-Gamma Distribution\label{appsubsec:NEG}}

A continuous random variable $x$ has a Normal-Exponential-Gamma distribution with mean parameter $\mu$, scale parameter $\sigma > 0$ and shape parameter $\lambda > 0$, written $x \sim \mbox{NEG}(\mu,\sigma,\lambda)$, if the density function of $x$ is
$$
\pDens(x) = \pi^{-1/2} \sigma^{-1} \lambda 2^\lambda \Gamma(\lambda + 1/2) \exp\left\{\frac{(x-\mu)^2}{4\sigsq}\right\} D_{-2\lambda-1}\left(\left\vert \frac{x-\mu}{\sigma}  \right\vert\right), \quad x\in\mathbb{R},
$$
where $D_\nu(x) \equiv 2^{\nu/2+1/4} W_{\nu/2+1/4, -1/4}(x^2/2)/\sqrt{x}, \: x > 0$ is the parabolic cylinder function of order $\nu \in \mathbb{R}$ and $W_{k,m}$ is a confluent hypergeometric function or order $k$ and $m$, as defined by Whittaker \myand Watson (1990). 
A useful result from Griffin \myand Brown (2011) shows that
$$
\begin{array}{c}
\mbox{if} \quad x \vert b \sim \text{N}(\mu, \sigsq/b), \quad b\vert c \sim \mbox{Inverse-Gamma}(1, c) \quad \mbox{and} \quad c \sim \mbox{Gamma}(\lambda/2,1),\\[1ex]
\mbox{then} \quad x \sim \mbox{NEG}(\mu, \sigma, \lambda).
\end{array}
$$

\subsection{Gamma Distribution\label{appsubsec:Gamma}}

A continuous random variable $x$ has a Gamma distribution with shape parameter $\alpha>0$ and scale parameter $\beta>0$, written $x \sim \mbox{Gamma}(\alpha, \beta)$, if the density function of $x$ is
$$
\pDens(x) = \frac{\beta^\alpha}{\Gamma(\alpha)}x^{
\alpha-1} \exp\lbrace -\beta x \rbrace, \quad x > 0.
$$

\subsection{Inverse-$\chi^2$ and Inverse-Gamma Distributions\label{appsubsec:InvChi2-InvGamma}}

A continuous random variable $x$ has an Inverse-$\chi^2$ distribution with shape parameter $\xi>0$ and scale parameter $\lambda>0$, written $x \sim \mbox{Inverse-}\chi^2(\xi, \lambda)$, if the density function of $x$ is
$$
\pDens(x) = \frac{\left(\lambda/2\right)^{\xi/2}}{\Gamma(\xi/2)}x^{-(\xi/2)-1} \exp\lbrace -(\lambda/2)/x \rbrace, \quad x > 0.
$$
A continuous random variable $x$ has an Inverse-Gamma distribution with shape parameter $\alpha>0$ and scale parameter $\beta>0$, written $x \sim \mbox{Inverse-Gamma}(\alpha, \beta)$, if the density function of $x$ is
$$
\pDens(x) = \frac{\beta^\alpha}{\Gamma(\alpha)}x^{-\alpha-1} \exp\lbrace -\beta/x \rbrace, \quad x > 0.
$$
Notice that 
$$
\quad x \sim \mbox{Inverse-}\chi^2(\xi, \lambda) \quad \mbox{if and only if} \quad x \sim \mbox{Inverse-Gamma}(\xi/2, \lambda/2)
$$
and, equivalently,
$$
\quad x \sim \mbox{Inverse-Gamma}(\alpha,\beta) \quad \mbox{if and only if} \quad x \sim \mbox{Inverse-}\chi^2(2\alpha, 2\beta).
$$

\subsection{Half-Cauchy Distribution}

A continuous random variable $x$ has a Half-Cauchy distribution with scale parameter $\sigma>0$, written $x\sim\mbox{Half-Cauchy}(\sigma)$, if the density function of $x$ is
\begin{equation*}
p(x)=2/[\sigma\pi\{1+(x/\sigma)^2\}], \quad x > 0.
\end{equation*}

\subsection{Half-t Distribution\label{appsubsec:Halft}}

A continuous random variable $x$ has a $\mbox{Half-}t$ distribution with $\nu > 0$ degrees of freedom and scale parameter $\sigma > 0$, written $x \sim \mbox{Half-}t(\sigma, \nu)$, if the density function of $x$ is
$$
\pDens(x) = \frac{2 \Gamma(\frac{\nu+1}{2})}{\sqrt{\pi\nu}\Gamma(\nu/2)\sigma \lbrace 1 + (x/\sigma)^2/\nu \rbrace^{\frac{\nu+1}{2}}}, \quad x > 0.
$$
If $\nu = 1$, then $x \sim \mbox{Half-Cauchy}(\sigma)$. Wand \emph{et al.} (2011) show that
$$
\begin{array}{c}
\mbox{if} \quad x\vert a \sim \mbox{Inverse-Gamma}(\nu/2, \nu/a) \quad \mbox{and} \quad a \sim \mbox{Inverse-Gamma}(1/2, 1/A^2),\\[1ex]
\mbox{then} \quad \sqrt{x} \sim \mbox{Half-}t(A, \nu).
\end{array}
$$
The same result can be equivalently formulated with:
$$
x\vert a \sim \mbox{Inverse-Gamma}(\nu/2, \nu/(2a))\quad \mbox{and} \quad a \sim \mbox{Inverse-Gamma}(1/2, 1/(2A^2)),
$$
or with:
$$
x\vert a \sim \mbox{Inverse-}\chi^2(\nu, 1/a)\quad \mbox{and}\quad a \sim \mbox{Inverse-}\chi^2(1, 1/(\nu A^2)).
$$

\subsection{Inverse-G-Wishart Distribution\label{appsubsec:InvGWishart}}

The $\mbox{Inverse-G-Wishart}$ distribution arises from the inverse of matrices with a \emph{G-Wishart} distribution (e.g. Atay-Kayis \myand Massam, 2005;
Maestrini \myand Wand, 2021). For any positive integer $d$, let $G$ be
an undirected graph with $d$ nodes labeled $1,\ldots,d$ and
set $E$ consisting of sets of pairs of nodes that are connected
by an edge. We say that the symmetric $d\times d$ matrix $\bM$ \emph{respects}
$G$ if
$$\bM_{ij}=0\quad\mbox{for all}\quad \{i,j\}\notin E.$$
A $d\times d$ random matrix $\bX$ has an Inverse G-Wishart distribution
with graph $G$ and parameters $\xi>0$ and symmetric $d\times d$
matrix $\bLambda$, written $\bX\sim\mbox{Inverse-G-Wishart}(G,\xi,\bLambda)$, if and only if the density function of $\bX$ satisfies
$$\pDens(\bX)\propto |\bX|^{-(\xi+2)/2}\exp\left\lbrace-\frac{1}{2}\tr(\bLambda\,\bX^{-1})\right\rbrace$$
over arguments $\bX$ such that $\bX$  is symmetric and positive definite
and $\bX^{-1}$ respects $G$. Two important special cases are
$$G=\Gfull\equiv\mbox{totally connected $d$-node graph},$$
for which the Inverse G-Wishart distribution coincides with the ordinary
Inverse Wishart distribution $\bX \sim \mbox{Inverse-Wishart}(\xi - d + 1, \bLambda)$, and
$$G=\Gdiag\equiv\mbox{totally disconnected $d$-node graph},$$
for which the Inverse G-Wishart distribution coincides with
a product of independent Inverse Chi-Squared distributions.
The subscripts of $\Gfull$ and $\Gdiag$ reflect the
fact that $\bX^{-1}$ is a full matrix and $\bX^{-1}$ is
a diagonal matrix in each special case.
In the $d=1$ special case, the Inverse G-Wishart distribution coincides with the
Inverse Chi-Squared distribution.  

\subsection{Huang-Wand Distribution\label{appsubsec:HuangWand}}

A $d \times d$ symmetric positive definite matrix $\bSigma$ has a $\mbox{Huang-Wand}$ distribution with $\nu > 0$ degrees of freedom and scale parameters $s_1, \ldots, s_d$, written $\bSigma \sim \mbox{Huang-Wand}(\nu; s_1, \ldots, s_d)$, if and only if the following augmented representation holds:
$$
\begin{array}{c}
\bSigma\vert\bA \sim \mbox{Inverse-G-Wishart}(\Gfull, \nu + 2d - 2, \bA^{-1})\\[1ex]
\bA \sim \mbox{Inverse-G-Wishart}(\Gdiag, 1, \lbrace \nu\, \diag(s_1^2, \ldots, s_d^2) \rbrace^{-1}).
\end{array}
$$
This distribution is defined in Huang \myand Wand, 2013). The formulation displayed here is provided in Maestrini \myand Wand (2021). 

If $\nu=2$, a marginally noninformative $\mbox{Huang-Wand}$ prior specification for $\bSigma$ can be obtained for arbitrarily large scale parameters. This corresponds to the standard deviation parameters $\sigma_j \equiv (\bSigma)_{jj}^{1/2}, 1 \leq j \leq d$, having $\mbox{Half-}t$ distributions with $\nu$ degrees of freedom and scale parameter given by $s_j$, and the correlation parameters $\rho_{jj'} \equiv (\bSigma)_{jj'}^{1/2}/(\sigma_j \sigma_{j'})$, $1 \leq j, j' \leq d$, having a Uniform distribution on the interval $(-1, 1)$.

\section{Multilevel Sparse Matrix Problem Algorithms\label{appsec:RoutinesNolanWand2020}}

Algorithms \ref{alg:streamlinedMFVB_twolevel_globallocal} and \ref{alg:streamlinedMFVB_threelevel_globallocal} described in Section \ref{sec:VariationalInferenceLMMwithGLPriors} rely upon
two matrix algebraic routines for efficiently solving the two-level
and three-level versions of the \emph{multilevel sparse matrix problems} defined in Nolan \myand Wand (2020). They correspond to the $\SolveTwoLevelSparseMatrix$ and $\SolveThreeLevelSparseMatrix$ algorithms that we list hereafter as Algorithms \ref{alg:SolveTwoLevelSparseMatrix} and \ref{alg:SolveThreeLevelSparseMatrix}, respectively. 

We briefly describe two-level and three-level sparse matrix structures, and give explicit definition of the aforementioned routines for efficiently solving the associated linear system problems, following Appendix A of Nolan \emph{et al.} (2020).

\subsection{Two-Level Sparse Matrix Problems\label{sec:twoLevSMP}}

Two-level sparse matrix problems are summarized in Section \ref{subsec:StreamlinedUpdatesTwoLevelThreeLevelLMM}. Such problems are efficiently solved by the \SolveTwoLevelSparseMatrix\ routine, which is here listed as 
Algorithm \ref{alg:SolveTwoLevelSparseMatrix} and is justified by 
Theorem 2.2 of Nolan \myand Wand (2020).

\begin{algorithm}[!th]
\begin{center}
\begin{minipage}[t]{165mm}
\begin{small}
\begin{itemize}
\setlength\itemsep{4pt}
\item[] Inputs: $\ba_1(p\times1),\AL{11}(p\times p),\,
\big\{\big(\ba_{2,i}(q\times 1),\AL{22,i}(q\times q),\AL{12,i}(p\times q)\big):\ 1\le i\le m\big\}$
\item[] $\bomega\longleftarrow\ba_1$\ \ \ ;\ \ \ $\bOmega\longleftarrow\bA_{11}$
\item[] For $i=1,\ldots,m$:
\begin{itemize}
\setlength\itemsep{4pt}
\item[]$\bomega\longleftarrow\bomega-\AL{12,i}\AL{22,i}^{-1}\ba_{2,i}$
\ \ \ ;\ \ \ 
$\bOmega\longleftarrow\bOmega-\AL{12,i}\AL{22,i}^{-1}\ALT{12,i}$
\end{itemize}
\item[] $\AU{11}\longleftarrow\bOmega^{-1}$
\ \ \ ;\ \ \ $\bx_1\longleftarrow\AU{11}\bomega$
\item[] For $i=1,\ldots,m$:
\begin{itemize}
\setlength\itemsep{4pt}
\item[] $\bx_{2,i}\longleftarrow\AL{22,i}^{-1}(\ba_{2,i}-\ALT{12,i}\bx_1)$\ \ \ ;\ \ \ 
$\AU{12,i}\longleftarrow\,-(\AL{22,i}^{-1}\ALT{12,i}\AU{11})^T$
\item[] $\AU{22,i}\longleftarrow\,\AL{22,i}^{-1}\big(\bI-\ALT{12,i}\bA^{12,i}\big)$ 
\end{itemize}
\item[] Outputs: $\bx_1,\AU{11},\big\{\big(\bx_{2,i},\AU{22,i},\AU{12,i}):\ 1\le i\le m\big\}$.
\end{itemize}
\end{small}
\end{minipage}
\end{center}
\caption{\textit{The} \SolveTwoLevelSparseMatrix\ \textit{algorithm for solving the two-level sparse
matrix problem $\bx=\AL{}^{-1}\ba$ and sub-blocks of $\AL{}^{-1}$ corresponding to the non-zero
sub-blocks of $\AL{}$. }}
\label{alg:SolveTwoLevelSparseMatrix} 
\end{algorithm}
%%%%%%%%%%%%%%%%%%%%%%%%%%%%%%%%%%%%%%%%%%%%%%%%%%%%%%%%%%%%%%%%%%%%%%%%%%%%%%%%%%%%%%%%

\subsection{Three-Level Sparse Matrix Problems\label{sec:threeLevSMP}}

Three-level sparse matrix problems are described in Section 3 of 
Nolan \myand Wand (2020). An illustrative three-level sparse matrix example is
\begin{equation*}
\bA =
\left[ \arraycolsep=2pt\def\arraystretch{1.5} 
   \begin{array}{cccccccc}
   \setstretch{4}
   \AL{11} & \AL{12,1} & \AL{12,11} & \AL{12,12} & \AL{12,2} & \AL{12,21} & \AL{12,22} & \AL{12,23} \\
   \ALT{12,1} & \AL{22,1} & \AL{12,1,1} & \AL{12,1,2} & \bO & \bO & \bO & \bO \\
   \ALT{12,11} & \ALT{12,1,1} & \AL{22,11} & \bO & \bO & \bO & \bO & \bO \\
   \ALT{12,12} & \ALT{12,1,2} & \bO & \AL{22,12} & \bO & \bO & \bO & \bO \\
   \ALT{12,2} & \bO & \bO & \bO & \AL{22,2} & \AL{12,2,1} & \AL{12,2,2} & \AL{12,2,3} \\
   \ALT{12,21} & \bO & \bO & \bO & \ALT{12,2,1} & \AL{22,21} & \bO & \bO \\
   \ALT{12,22} & \bO & \bO & \bO & \ALT{12,2,2} & \bO & \AL{22,22} & \bO \\
   \ALT{12,23} & \bO & \bO & \bO & \ALT{12,2,3} & \bO & \bO & \AL{22,23}
\end{array} \right],
\end{equation*}
which corresponds to level 2 group sizes $n_1=2$ and $n_2=3$,
and a level 3 group size $m=2$. A generic three-level sparse matrix $\bA$ consists of the following components:
\begin{itemize}
\item A $p \times p$ matrix $\AL{11}$, which is assigned the $(1,1)$-block position.
\item A set of partitioned matrices 
$\left\{\left[\begin{array}{c|c|c|c} \AL{12,i} & \AL{12,ij} 
& \dots & \AL{12,in_i} \end{array} \right]: 1 \le i \le m \right\}$, 
which is assigned the $(1, 2)$-block position. For each $1 \le i \le m$, $\AL{12,i}$ 
is $p \times q_1$, and for each $1 \le j \le n_i$, $\AL{12,ij}$ is $p \times q_2$.
\item A $(2,1)$-block, which is simply the transpose of the $(1,2)$-block.
\item A block diagonal structure along the $(2,2)$-block position, where each sub-block 
is a two-level sparse matrix, 
as defined in Section \ref{subsec:StreamlinedUpdatesTwoLevelThreeLevelLMM}. For each $1 \le i \le m$, $\AL{22,i}$ is $q_1 \times q_1$, and for each 
$1 \le j \le n_i$, $\AL{12,i,j}$ is $q_1 \times q_2$ and $\AL{22,ij}$ is $q_2 \times q_2$.
\end{itemize}
The three-level sparse matrix problem arising from the illustrative example above is defined as finding the vector $\bx$ such that $\bA\,\bx=\ba$, where:
\begin{equation*}
   \ba \equiv
      \left[ \arraycolsep=2pt\def\arraystretch{1.5}
         \begin{array}{c}
         \setstretch{4}
         \ba_{1} \\
         \ba_{2,1} \\
         \ba_{2,11} \\
         \ba_{2,12} \\
         \ba_{2,2} \\
         \ba_{2,21} \\
         \ba_{2,22} \\
         \ba_{2,23}
      \end{array} \right] \qquad
      \text{and} 
\qquad
   \bx \equiv
      \left[ 
         \arraycolsep=2pt\def\arraystretch{1.5}
         \begin{array}{c}
         \setstretch{4}
         \bx_{1} \\
         \bx_{2,1} \\
         \bx_{2,11} \\
         \bx_{2,12} \\
         \bx_{2,2} \\
         \bx_{2,21} \\
         \bx_{2,22} \\
         \bx_{2,23}
      \end{array} \right],
\end{equation*}
and determining the sub-blocks
of $\bA^{-1}$ corresponding to the non-zero sub-blocks of $\bA$. The structure of $\bA^{-1}$ is 
\begin{equation*}
\bA^{-1} =
\left[\arraycolsep=2pt\def\arraystretch{1.5}
   \begin{array}{cccccccc}
 \setstretch{4}
   \AU{11} & \AU{12,1} & \AU{12,11} & \AU{12,12} & \AU{12,2} & \AU{12,21} & \AU{12,22} & \AU{12,23} \\
   \AUT{12,1} & \AU{22,1} & \AU{12,1,1} & \AU{12,1,2} & \bigX & \bigX & \bigX & \bigX \\
   \AUT{12,11} & \AUT{12,1,1} & \AU{22,11} & \bigX & \bigX & \bigX & \bigX & \bigX \\
   \AUT{12,12} & \AUT{12,1,2} & \bigX & \AU{22,12} & \bigX & \bigX & \bigX & \bigX \\
   \AUT{12,2} & \bigX & \bigX & \bigX & \AU{22,2} & \AU{12,2,1} & \AU{12,2,2} & \AU{12,2,3} \\
   \AUT{12,21} & \bigX & \bigX & \bigX & \AUT{12,2,1} & \AU{22,21} & \bigX & \bigX \\
   \AUT{12,22} & \bigX & \bigX & \bigX & \AUT{12,2,2} & \bigX & \AU{22,22} & \bigX \\
   \AUT{12,23} & \bigX & \bigX & \bigX & \AUT{12,2,3} & \bigX & \bigX & \AU{22,23}
\end{array} \right].
\end{equation*}
For a general three-level sparse linear system problem, $\ba_1$ and $\bx_1$ are $p \times 1$ vectors. 
For each $1 \le i \le m$, $\ba_{2,i}$ and $\bx_{2,i}$ are $q_1 \times 1$ vectors.
For each $1 \le i \le m$ and $1 \le j \le n_i$ the vectors  $\ba_{2,ij}$ and $\bx_{2,ij}$ 
have dimension $q_2 \times 1$.

Such problems are efficiently solved by the \SolveThreeLevelSparseMatrix\ routine, which is here listed as 
Algorithm \ref{alg:SolveThreeLevelSparseMatrix} and is justified by 
Theorem 3.2 of Nolan \myand Wand (2020).

\begin{algorithm}[!th]
\begin{center}
\begin{minipage}[t]{165mm}
\begin{small}
\begin{itemize}
\setlength\itemsep{4pt}
\item[] Input: $\ba_1(p\times1),\AL{11}(p\times p),\,
\big\{\big(\ba_{2,i}(q_1\times1),\AL{22,i}(q_1\times q_1),\AL{12,i}(p\times q_1):\ 1\le i\le m\big\}$,\\
$\qquad\qquad\big\{\ba_{2,ij}(q_2\times1),\AL{22,ij}(q_2\times q_2),\AL{12,ij}(p\times q_2),
\AL{12,i,j}(q_1\times q_2)\big):\ 1\le i\le m,\ 1\le j\le n_i\big\}$.
\item[]$\bomega\longleftarrow\ba_1$\ \ \ ;\ \ \ $\bOmega\longleftarrow\bA_{11}$
\item[] For $i=1,\ldots,m$:
\begin{itemize}
\setlength\itemsep{4pt}
\item[] $\bh_{2,i}\longleftarrow\ba_{2,i}$\ \ \ ;\ \ \ $\bH_{12,i}\longleftarrow \AL{12,i}$
\ \ \ ;\ \ \ $\bH_{22,i}\longleftarrow \AL{22,i}$
\item[] For $j=1,\ldots,n_i$:
\begin{itemize}
\setlength\itemsep{4pt}
\item[] $\bh_{2,i}\longleftarrow\bh_{2,i} - \AL{12,i,j}\AL{22,ij}^{-1}\ba_{2,ij}$\ \ \ ;\ \ \ 
$\bH_{12,i}\longleftarrow\bH_{12,i} - \AL{12,ij}\AL{22,ij}^{-1}\ALT{12,i,j}$\\[1ex]
$\bH_{22,i}\longleftarrow\bH_{22,i} - \AL{12,i,j}\AL{22,ij}^{-1}\ALT{12,i,j}$
\item[]$\bomega\longleftarrow\bomega-\AL{12,ij}\AL{22,ij}^{-1}\ba_{2,ij}$\ \ \ ;\ \ \  
$\bOmega\longleftarrow\bOmega-\AL{12,ij}\AL{22,ij}^{-1}\ALT{12,ij}$
\end{itemize}
\item[] $\bomega\longleftarrow\bomega-\bH_{12,i}\bH_{22,i}^{-1}\bh_{2,i}$\ \ \ ;\ \ \ 
$\bOmega\longleftarrow\bOmega-\bH_{12,i}\bH_{22,i}^{-1}\bH_{12,i}^T$
\end{itemize}
\item[] $\AU{11}\longleftarrow \bOmega^{-1}$\ \ \ ;\ \ \ $\bx_1\longleftarrow\AU{11}\bomega$
\item[] For $i=1,\ldots,m$:
\begin{itemize}
\item[] $\bx_{2,i}\longleftarrow\,\bH_{22,i}^{-1}(\bh_{2,i}-\bH_{12,i}^T\bx_1)$\ \ \ ;\ \ \ 
$\AU{12,i}\longleftarrow\,-(\bH_{22,i}^{-1}\bH_{12,i}^T\AU{11})^T$
\item[] $\AU{22,i}\longleftarrow\,\bH_{22,i}^{-1}(\bI-\bH_{12,i}^T\AU{12,i})$ 
\item[] For $j=1,\ldots,n_i$:
\begin{itemize}
\item[]$\bx_{2,ij}\longleftarrow\AL{22,ij}^{-1}\big(\ba_{2,ij}-\AL{12,ij}^T\bx_1-\ALT{12,i,j}\bx_{2,i}\big)$
\item[]$\AU{12,ij}\longleftarrow\, -\big\{\AL{22,ij}^{-1}\big(\ALT{12,ij}\AU{11}+\ALT{12,i,j}\,\AUT{12,i}\big)\big\}^T$ 
\item[]$\AU{12,i,j}\longleftarrow\, -\big\{\AL{22,ij}^{-1}\big(\ALT{12,ij}\,\AU{12,i}+\ALT{12,i,j}\,\AU{22,i}\big)\big\}^T$ 
\item[]$\AU{22,ij}\longleftarrow \AL{22,ij}^{-1}\big(\bI - \ALT{12,ij}\AU{12,ij} - \ALT{12,i,j}\,\AU{12,i,j}\big)$ 
\end{itemize}
\end{itemize}
\item[] Outputs: $\bx_1,\AU{11},\big\{\big(\bx_{2,i},\AU{22,i},\AU{12,i}):\ 1\le i\le m\big\},$\\
$\null\qquad\qquad\big\{\big(\bx_{2,ij},\AU{22,ij},\AU{12,ij},\AU{12,i,j},
\big):\ 1\le i\le m,\ 1\le j\le n_i\big\}$.
\end{itemize}
\end{small}
\end{minipage}
\end{center}
\caption{\textit{The} \SolveThreeLevelSparseMatrix\ \textit{algorithm for solving the three-level sparse
matrix problem $\bx=\AL{}^{-1}\ba$ and sub-blocks of $\AL{}^{-1}$ corresponding to the non-zero
sub-blocks of $\AL{}$.}}
\label{alg:SolveThreeLevelSparseMatrix} 
\end{algorithm}
%%%%%%%%%%%%%%%%%%%%%%%%%%%%%%%%%%%%%%%%%%%%%%%%%%%%%%%%%%%%%%%%%%%%%%%%%%%%%%%%%%%%%%%%

\newpage

\section{Derivations\label{appsec:Derivations}}

We derive explicit updates for the parameters of the optimal density functions in \eqref{eq:qstargloballocal} using arguments similar to those provided in Neville \emph{et al.} (2014) with some adjustments. Such updates are then combined with the derivations in Appendix B of Nolan \emph{et al.} (2020) for deriving the streamlined MFVB algorithms for our two- and three- level linear mixed-effects models, i.e. Algorithms \ref{alg:streamlinedMFVB_twolevel_globallocal} and \ref{alg:streamlinedMFVB_threelevel_globallocal}. 

\subsection{Derivation of $\qDens^*(\beta_0, \bbeta)$ and Associated Parameter Updates}

Given the model formulation \eqref{eq:gauss_heter_lmm_penalized}, the full conditional density function for $(\beta_0, \bbeta)$ is:
\begin{align*}
\pDens(\beta_0, \bbeta\vert \text{rest}) &\propto \pDens(\by \vert \beta_0, \bbeta, \sigsq)\: \pDens(\beta_0)\: \pDens(\bbeta \vert \bzeta, \tausq) \\
&\propto \exp\Bigg\lbrace -\frac{1}{2} \Bigg[ \left[\begin{array}{c} \beta_0 \\ \bbeta \end{array}\right]^T \left(\frac{1}{\sigsq} \big[\:\boldsymbol{1} \:\big\vert\: \bX \:\big]^T \big[\:\boldsymbol{1} \:\big\vert\: \bX \:\big] + \left[\begin{array}{cc} \sigma^{-2}_{\beta_0} & \bzero^T \\[1ex] \bzero & \tau^{-2}\: \diag(\bzeta)\end{array} \right] \right) \left[\begin{array}{c} \beta_0 \\ \bbeta \end{array}\right]\\
&\qquad - 2 \left[\begin{array}{c} \beta_0 \\ \bbeta \end{array}\right]^T \left(\frac{1}{\sigsq} \big[\:\boldsymbol{1} \:\big\vert\: \bX \:\big]^T \by + \left[\begin{array}{cc} \sigma^{-2}_{\beta_0} & \bzero^T \\[1ex] \bzero & \tau^{-2}\: \diag(\bzeta)\end{array} \right] \left[\begin{array}{c} \mu_{\beta_0} \\ \bzero \end{array} \right] \right) \Bigg] \Bigg\rbrace
\end{align*}
and application of \eqref{eq:optimalqdensity} results in:
\begin{align*}
&\qDens^*(\beta_0, \bbeta) \propto \exp\left\lbrace E_\qDens\lbrace \log \pDens(\beta_0, \bbeta\vert \text{rest}) \rbrace \right\rbrace \\ &\propto \exp\Bigg\lbrace -\frac{1}{2} \Bigg[ \left[\begin{array}{c} \beta_0 \\ \bbeta \end{array}\right]^T \left(E_\qDens\lbrace \sigma^{-2} \rbrace \big[\:\boldsymbol{1} \:\big\vert\: \bX \:\big]^T \big[\:\boldsymbol{1} \:\big\vert\: \bX \:\big] + \left[\begin{array}{cc} \sigma^{-2}_{\beta_0} & \bzero^T \\[1ex] \bzero &\quad E_\qDens\lbrace\tau^{-2}\rbrace\: \diag(E_\qDens\lbrace\bzeta\rbrace)\end{array} \right] \right) \left[\begin{array}{c} \beta_0 \\ \bbeta \end{array}\right] \\
&\qquad - 2 \left[\begin{array}{c} \beta_0 \\ \bbeta \end{array}\right]^T \left(E_\qDens\lbrace \sigma^{-2} \rbrace \big[\:\boldsymbol{1} \:\big\vert\: \bX \:\big]^T \by + \left[\begin{array}{cc} \sigma^{-2}_{\beta_0} & \bzero^T \\[1ex] \bzero & E_\qDens\lbrace\tau^{-2}\rbrace\: \diag(E_\qDens\lbrace\bzeta\rbrace) \end{array} \right] \left[\begin{array}{c} \mu_{\beta_0} \\ \bzero \end{array} \right] \right)\Bigg] \Bigg\rbrace.
\end{align*}
After \emph{completion of the square} manipulations for the multivariate Gaussian distribution and standard algebraic manipulations, it follows immediately that:
$$
\qDens^*(\beta_0, \bbeta) \mbox{ is a N}\left(\bmuq{\beta_0, \bbeta}, \bSigmaq{\beta_0, \bbeta}\right) \text{ density function}
$$
with
$$
\bSigmaq{\beta_0, \bbeta} \longleftarrow \left(\muq{1/\sigsq} \:\big[\:\boldsymbol{1} \:\big\vert\: \bX \:\big]^T \big[\:\boldsymbol{1} \:\big\vert\: \bX \:\big] + \left[\begin{array}{cc} \sigma^{-2}_{\beta_0} & \bzero^T \\[1ex] \bzero & \muq{1/\tausq}\: \diag(\bmuq{\bzeta})\end{array} \right] \right)^{-1}
$$
and
$$
\bmuq{\beta_0, \bbeta} \longleftarrow \bSigmaq{\beta_0, \bbeta} \left(\muq{1/\sigsq}\: \big[\:\boldsymbol{1} \:\big\vert\: \bX \:\big]^T \by + \left[\begin{array}{c} \mu_{\beta_0}/\sigma^{2}_{\beta_0} \\ \bzero \end{array}\right] \right).
$$

\subsection{Derivation of $\qDens^*(\sigsq)$ and Associated Parameter Updates}

Given the model formulation \eqref{eq:gauss_heter_lmm_penalized}, the full conditional density function for $\sigsq$ is:
\begin{align*}
\pDens(\sigsq \vert \text{rest}) &\propto \pDens(\by \vert \beta_0, \bbeta, \sigsq)\: \pDens(\sigsq \vert \asigsq) \\
&\propto (\sigsq)^{-(\nusigsq + n)/2 - 1} \exp\left\lbrace -\frac{1}{\sigsq} \frac{\asigsq^{-1} + \lVert \by - \boldsymbol{1} \beta_0 - \bX \bbeta \rVert^2}{2} \right\rbrace
\end{align*}
and application of \eqref{eq:optimalqdensity} results in:
\begin{align*}
\qDens^*(\sigsq) &\propto \exp\left\lbrace E_\qDens\lbrace \log \pDens(\sigsq \vert \text{rest}) \rbrace \right\rbrace \\ &\propto (\sigsq)^{-(\nusigsq + n)/2 - 1} \exp\left\lbrace -\frac{1}{\sigsq} \frac{E_\qDens\lbrace\asigsq^{-1}\rbrace + E_\qDens\lbrace \lVert \by - \boldsymbol{1} \beta_0 - \bX \bbeta \rVert^2 \rbrace}{2} \right\rbrace.
\end{align*}
After standard algebraic manipulations, it follows immediately that:
$$
\qDens^*(\sigsq) \mbox{ is an Inverse-}\chi^2 \left(\xiq{\sigsq}, \lambdaq{\sigsq}\right) \text{ density function}
$$
with
$$
\xiq{\sigsq} \longleftarrow n + \nusigsq
$$
and
$$
\lambdaq{\sigsq}\longleftarrow\muq{1/\asigsq} + \left\lVert \by - \big[\:\boldsymbol{1} \:\big\vert\: \bX \:\big]\bmuq{\beta_0, \bbeta}\right\rVert^2 + \tr{\left\lbrace\bSigmaq{\beta_0, \bbeta} \big[\:\boldsymbol{1} \:\big\vert\: \bX \:\big]^T \big[\:\boldsymbol{1} \:\big\vert\: \bX \:\big]\right\rbrace}.
$$

\subsection{Derivation of $\qDens^*(\asigsq)$ and Associated Parameter Updates}

Given the model formulation \eqref{eq:gauss_heter_lmm_penalized}, the full conditional density function for $\asigsq$ is:
\begin{align*}
\pDens(\asigsq \vert \text{rest}) &\propto \pDens(\sigsq \vert \asigsq)\: \pDens(\asigsq) \\
&\propto (\asigsq)^{-(\nusigsq + 1)/2 - 1} \exp\left\lbrace -\frac{1}{\asigsq} \frac{\sigma^{-2} + (\nusigsq s_{\sigsq}^2)^{-1}}{2} \right\rbrace
\end{align*}
and application of \eqref{eq:optimalqdensity} results in:
\begin{align*}
\qDens^*(\asigsq) &\propto \exp\left\lbrace E_\qDens\lbrace \log \pDens(\asigsq \vert \text{rest}) \rbrace \right\rbrace \\ &\propto (\asigsq)^{-(\nusigsq + 1)/2 - 1} \exp\left\lbrace -\frac{1}{\asigsq} \frac{E_\qDens\lbrace \sigma^{-2} \rbrace + (\nusigsq s_{\sigsq}^2)^{-1}}{2} \right\rbrace.
\end{align*}
After standard algebraic manipulations, it follows immediately that:
$$
\qDens^*(\asigsq) \mbox{ is an Inverse-}\chi^2 \left(\xiq{\asigsq}, \lambdaq{\asigsq}\right) \text{ density function}
$$
with
$$
\xiq{\asigsq} \longleftarrow 1 + \nusigsq \quad \text{and} \quad \lambdaq{\asigsq}\longleftarrow\muq{1/\sigsq} +  (\nusigsq s_{\sigsq}^2)^{-1}.
$$

\subsection{Derivation of $\qDens^*(\tausq)$ and Associated Parameter Updates}

Given the model formulation \eqref{eq:gauss_heter_lmm_penalized}, the full conditional density function for $\tausq$ is:
\begin{align*}
\pDens(\tausq\vert \text{rest}) &\propto \left\lbrace \prod_{h=1}^{H} \pDens(\beta_h \vert \tausq) \right\rbrace \pDens(\tausq \vert \atausq) \\
&\propto (\tausq)^{-(H + 1)/2 - 1} \exp\left\lbrace -\frac{1}{\tausq} \frac{\atausq^{-1} + \sum_{h=1}^{H} \zeta_h (\beta_h)^2}{2}\right\rbrace
\end{align*}
and application of \eqref{eq:optimalqdensity} results in:
\begin{align*}
\qDens^*(\tausq) &\propto \exp\left\lbrace E_\qDens\lbrace \log \pDens(\tausq\vert \text{rest}) \rbrace \right\rbrace \\ &\propto (\tausq)^{-(H + 1)/2 - 1} \exp\left\lbrace -\frac{1}{\tausq} \frac{E_\qDens \lbrace \atausq^{-1} \rbrace + \sum_{h=1}^{H} E_\qDens\lbrace \zeta_h \rbrace E_\qDens \lbrace \beta_h^2 \rbrace}{2}\right\rbrace.
\end{align*}
After standard algebraic manipulations, it follows immediately that:
$$
\qDens^*(\tausq) \mbox{ is an Inverse-}\chi^2 \left(\xiq{\tausq}, \lambdaq{\tausq}\right) \text{ density function}
$$
with
$$
\xiq{\tausq} \longleftarrow H + 1\quad\text{and}\quad
\lambdaq{\tausq}\longleftarrow\muq{1/\atausq} + \bmuq{\bzeta}^T \bmuq{(\bbeta)^2}.
$$

\subsection{Derivation of $\qDens^*(\atausq)$ and Associated Parameter Updates}

Given the model formulation \eqref{eq:gauss_heter_lmm_penalized}, the full conditional density function for $\atausq$ is:
\begin{align*}
\pDens(\atausq\vert \text{rest}) &\propto \pDens(\tausq \vert \atausq) \pDens(\atausq) \\
&\propto (\atausq)^{-2} \exp\left\lbrace -\frac{1}{\atausq} \frac{\tau^{-2} + \stausq^{-2}}{2} \right\rbrace
\end{align*}
and application of \eqref{eq:optimalqdensity} results in:
\begin{align*}
\qDens^*(\atausq) \propto \exp\left\lbrace E_\qDens\lbrace \log\pDens(\atausq\vert \text{rest}) \rbrace \right\rbrace \propto (\atausq)^{-2} \exp\left\lbrace -\frac{1}{\atausq} \frac{E_\qDens\lbrace \tau^{-2} \rbrace + \stausq^{-2}}{2}\right\rbrace.
\end{align*}
After standard algebraic manipulations, it follows immediately that:
$$
\qDens^*(\atausq) \mbox{ is an Inverse-}\chi^2 \left(\xiq{\atausq}, \lambdaq{\atausq}\right) \text{ density function}
$$
with
$$
\xiq{\atausq} \longleftarrow 2\quad\text{and}\quad
\lambdaq{\atausq}\longleftarrow\muq{1/\tausq} + s^{-2}_{\tausq}.
$$

\subsection{Derivation of $\qDens^*(\zeta_h)$ and Associated Parameter Updates}

Given the model formulation  \eqref{eq:gauss_heter_lmm_penalized}, the full conditional density function for $\zeta_h$, $1 \le h \le H$, is:
$$
\pDens(\zeta_h \vert \text{rest}) \propto \pDens(\beta_h \vert \zeta_h, \tau)\: \pDens(\zeta_h \vert a_{\zeta_h}),
$$
where $\pDens(\beta_h \vert \zeta_h)$ is the density function of a $\mbox{N}(0, \tausq/\zeta_h)$ distribution, while $\pDens(\zeta_h\vert a_{\zeta_h})$ depends on the hierarchical specification given in Table \ref{tab:global_local} for each of the considered global-local priors.

\subsubsection{Laplace Prior Specification}

If $\beta_h\vert\tau \sim \mbox{Laplace}(0, \tau)$ then:
\begin{align*}
\pDens(\zeta_h \vert \text{rest}) \propto (\zeta_h)^{-3/2} \exp\left\lbrace -\zeta_h \frac{\beta_h^2}{2\tausq} - \frac{1}{2 \zeta_h} \right\rbrace
\end{align*}
and application of \eqref{eq:optimalqdensity} results in:
\begin{align*}
\qDens^*(\zeta_h) \propto \exp\left\lbrace E_\qDens\lbrace \log\pDens(\zeta_h \vert \text{rest}) \rbrace \right\rbrace \propto (\zeta_h)^{-3/2} \exp\left\lbrace -\zeta_h \frac{E_\qDens\lbrace \beta_h^2\rbrace E_\qDens\lbrace \tau^{-2} \rbrace}{2} - \frac{1}{2 \zeta_h} \right\rbrace.
\end{align*}
After standard algebraic manipulations, it follows immediately that:
$$
\qDens^*(\zeta_h) \mbox{ is an Inverse-Gaussian}(\muq{\zeta_h}, 1) \mbox{ density function }
$$
with
$$
\muq{\zeta_h} \longleftarrow \sqrt{1\Big/\left(\muq{1/\tausq} \muq{\beta_h^2}\right)}.
$$

\subsubsection{Horseshoe Prior Specification}

If $\beta_h^\vert\tau \sim \mbox{Horseshoe}(0, \tau)$ then:
\begin{align*}
\pDens(\zeta_h \vert \text{rest}) \propto \exp\left\lbrace -\zeta_h \left(\frac{\beta_h^2}{2\tausq} + a_{\zeta_h} \right) \right\rbrace
\end{align*}
and application of \eqref{eq:optimalqdensity} results in:
\begin{align*}
\qDens^*(\zeta_h) \propto \exp\lbrace E_\qDens\lbrace \log \pDens(\zeta_h \vert \text{rest}) \rbrace \rbrace \propto \exp\left\lbrace -\zeta_h \left(\frac{E_\qDens\lbrace \beta_h^2\rbrace E_\qDens\lbrace \tau^{-2} \rbrace}{2} + E_\qDens\lbrace a_{\zeta_h}\rbrace\right) \right\rbrace.
\end{align*}
After standard algebraic manipulations, it follows immediately that:
$$
\qDens^*(\zeta_h) \mbox{ is a Gamma}(1, \lambdaq{\zeta_h}) \mbox{ density function }
$$
with
$$
\lambdaq{\zeta_h} \longleftarrow \frac{\muq{1/\tausq} \muq{\beta_h^2}}{2} + \muq{a_{\zeta_h}}.
$$

\subsubsection{Normal-Exponential-Gamma Prior Specification}

If $\beta_h\vert\tau \sim \mbox{NEG}(0, \tau, \lambda)$ then:
\begin{align*}
\pDens(\zeta_h \vert \text{rest}) \propto (\zeta_h)^{-3/2} \exp\left\lbrace -\zeta_h \frac{\beta_h^2}{2\tausq} - \frac{a_{\zeta_h} }{\zeta_h} \right\rbrace
\end{align*}
and application of \eqref{eq:optimalqdensity} results in:
\begin{align*}
\qDens^*(\zeta_h) \propto \exp\left\lbrace E_\qDens\lbrace \log\pDens(\zeta_h \vert \text{rest}) \rbrace \right\rbrace \propto (\zeta_h)^{-3/2} \exp\left\lbrace -\zeta_h \frac{E_\qDens\lbrace \beta_h^2\rbrace E_\qDens\lbrace \tau^{-2} \rbrace}{2} - \frac{E_\qDens\lbrace a_{\zeta_h} \rbrace}{\zeta_h} \right\rbrace.
\end{align*}
After standard algebraic manipulations, it follows immediately that:
$$
\qDens^*(\zeta_h) \mbox{ is an Inverse-Gaussian}(\muq{\zeta_h}, \lambdaq{\zeta_h}) \mbox{ density function }
$$
with
$$
\muq{\zeta_h} \longleftarrow \sqrt{2\muq{a_{\zeta_h}}\Big/\left(\muq{1/\tausq} \muq{\beta_h^2}\right)}\quad\text{and}\quad
\lambdaq{\zeta_h} \longleftarrow 2 \muq{a_{\zeta_h}}.
$$

\subsection{Derivation of $\qDens^*(a_{\zeta_h})$ and Associated Parameter Updates}

Given the model formulation  \eqref{eq:gauss_heter_lmm_penalized}, the full conditional density function for $\zeta_h$, $1 \le h \le H$, is:
$$
\pDens(a_{\zeta_h} \vert \text{rest}) \propto \pDens(\zeta_h \vert a_{\zeta_h})\: \pDens(a_{\zeta_h}),
$$
where both $\pDens(\zeta_h\vert a_{\zeta_h})$ and $\pDens(a_{\zeta_h})$ depend on the hierarchical specification given in Table \ref{tab:global_local} for each of the considered global-local priors.

\subsubsection{Laplace Prior Specification}

According to Table \ref{tab:global_local}, for the Laplace prior it is not necessary to introduce the $a_{\zeta_h}$ auxiliary variables. Hence $\qDens(a_{\zeta_h})$ is not present in \eqref{eq:prodrestr_globallocal} and does not need to be determined.

\subsubsection{Horseshoe Prior Specification}

If $\beta_h\vert\tau \sim \mbox{Horseshoe}(0, \tau)$ then:
\begin{align*}
\pDens(a_{\zeta_h} \vert \text{rest}) \propto \exp\lbrace - (\zeta_h + 1) a_{\zeta_h} \rbrace
\end{align*}
and application of \eqref{eq:optimalqdensity} results in:
\begin{align*}
\qDens^*(a_{\zeta_h}) \propto \exp\lbrace E_\qDens\lbrace \log \pDens(a_{\zeta_h} \vert \text{rest}) \rbrace \rbrace \propto \exp\left\lbrace -a_{\zeta_h} \left(E_\qDens\lbrace \zeta_h \rbrace + 1\right) \right\rbrace.
\end{align*}
After standard algebraic manipulations, it follows immediately that:
$$
\qDens^*(a_{\zeta_h}) \mbox{ is a Gamma}(1,\lambdaq{a_{\zeta_h}}) \mbox{ density function }
$$
with
$$
\lambdaq{\zeta_h} \longleftarrow \muq{\zeta_h} + 1.
$$

\subsubsection{Normal-Exponential-Gamma Prior Specification}

If $\beta_h\vert\tau \sim \mbox{NEG}(0, \tau, \lambda)$ then:
\begin{align*}
\pDens(a_{\zeta_h} \vert \text{rest}) \propto (a_{\zeta_h})^{(\lambda+1)-1} \exp\lbrace a_{\zeta_h} (\zeta_h^{-1} + 1) \rbrace
\end{align*}
and application of \eqref{eq:optimalqdensity} results in:
\begin{align*}
\qDens^*(a_{\zeta_h}) \propto \exp\left\lbrace E_\qDens\lbrace \log\pDens(a_{\zeta_h} \vert \text{rest}) \rbrace \right\rbrace \propto (a_{\zeta_h})^{(\lambda+1)-1} \exp\lbrace a_{\zeta_h} (E_\qDens\lbrace \zeta_h^{-1} \rbrace + 1) \rbrace.
\end{align*}
After standard algebraic manipulations, it follows immediately that:
$$
\qDens^*(\zeta_h) \mbox{ is a Gamma}(\lambda + 1,\lambdaq{a_{\zeta_h}}) \mbox{ density function }
$$
with
$$
\lambdaq{\zeta_h} \longleftarrow \muq{1/\zeta_h} + 1.
$$

\end{document}